\documentclass[a4paper,fleqn,usenatbib]{mnras}

\usepackage{newtxtext,newtxmath}

\usepackage[T1]{fontenc}
\usepackage{ae,aecompl}

\usepackage{graphicx}	
\usepackage{amsmath}	
\usepackage{amssymb}	

\usepackage{xspace}     
\usepackage{enumitem}   
\usepackage{multirow}   
\usepackage{comment}    
\usepackage{verbatim}
\usepackage{ulem}

\newcommand{\Msun}{\,M$_{\odot}$\xspace}
\newcommand{\Lsun}{\,L$_{\odot}$\xspace}

\newcommand{\microns}{\,$\mu$m\xspace}

\newcommand{\gpercmsq}{\,g\,cm$^{-2}$\xspace}

\title[The evolutionary status of methanol masers]{The Evolutionary Status of Protostellar Clumps Hosting Class II Methanol Masers}

\author[B.~M. Jones et al.]{B.~M. Jones$^{1}$,\thanks{E-mail: bethany.jones@manchester.ac.uk (BMJ)}
G.~A. Fuller$^{1,2}$,
S.~L. Breen$^{3}$,
A. Avison$^{1,2}$,
J.~A. Green$^{4}$,
\newauthor
A. Traficante$^{5}$,
D. Elia$^{5}$,
S.~P. Ellingsen$^{6}$,
M.~A. Voronkov$^{7}$,
M. Merello$^{5,8}$,
\newauthor
S. Molinari$^{5}$,
E. Schisano$^{5}$.
\\
$^{1}$Jodrell Bank Centre for Astrophysics, Department of Physics and Astronomy, The School of Natural Sciences, The University of Manchester,\\ Manchester M13 9PL, UK\\
${^2}$UK ALMA Regional Centre Node, M13 9PL, UK \\
${^3}$Sydney Institute for Astronomy (SIfA), School of Physics, University of Sydney, NSW 2006, Australia\\
${^4}$CSIRO Astronomy and Space Science, 26 Dick Perry Avenue, Kensington, WA 6151, Australia\\
$^{5}$IAPS-INAF, Via Fosso del Cavaliere 100, 00133 Rome, Italy\\
${^6}$School of Natural Sciences, University of Tasmania, Private Bag 37, Hobart, Tasmania 7001, Australia\\
$^{7}$CSIRO Astronomy and Space Science, Australia Telescope National Facility, Box 76, Epping, NSW 1710, Australia\\
$^{8}$Universidade de S\~ ao Paulo, IAG Rua do Mat\~ ao, 1226, Cidade Universit\'aria, 05508-090, S\~ ao Paulo, Brazil\\
}

\date{Accepted XXX. Received YYY; in original form ZZZ}

\pubyear{2019}

\begin{document}
\label{firstpage}
\pagerange{\pageref{firstpage}--\pageref{lastpage}}
\maketitle

\begin{abstract}
The Methanol MultiBeam survey (MMB) provides the most complete sample of Galactic massive young stellar objects (MYSOs) hosting 6.7\,GHz class II methanol masers. 
We characterise the properties of these maser sources using dust emission detected by the \textit{Herschel} Infrared Galactic Plane Survey (Hi-GAL) to assess their evolutionary state. 
Associating 731 (73\%) of MMB sources with compact emission at four Hi-GAL wavelengths, we derive clump properties and define the requirements of a MYSO to host a 6.7\,GHz maser. 
The median far-infrared (FIR) mass and luminosity are 630\Msun and 2500\Lsun for sources on the near side of Galactic centre and 3200\Msun and 10000\Lsun for more distant sources. 
The median luminosity-to-mass ratio is similar for both at $\sim$4.2\Lsun/\Msun. 
We identify an apparent minimum 70\microns luminosity required to sustain a methanol maser of a given luminosity (with $L_{70} \propto L_{6.7}\,^{0.6}$). 
The maser host clumps have higher mass and higher FIR luminosities than the general Galactic population of protostellar MYSOs. 
Using principal component analysis, we find 896 protostellar clumps satisfy the requirements to host a methanol maser but lack a detection in the MMB. 
Finding a 70\microns flux density deficiency in these objects, we favour the scenario in which these objects are evolved beyond the age where a luminous 6.7\,GHz maser can be sustained.
Finally, segregation by association with secondary maser species identifies evolutionary differences within the population of 6.7GHz sources. 
\end{abstract}




\section{Introduction}
\label{sec:intro}
The formation of massive stars within our galaxy is currently a poorly constrained process relative to our understanding of the mechanisms which produce stars of a similar mass to the Sun\,\citep{Krumholz2014TheFunction}.
For stars of $>$8\Msun, the onset of fusion in a protostellar core before it has finished accreting material from its surroundings leads to a complicated interplay between the strong outwards feedback and the infalling material\,\citep{Kudritzki2002LinedrivenStars,Zinnecker2007TowardFormation}. 
Further complexity is added by the tendency for parsec-scale clumps within molecular clouds to form clusters of stars with a range of masses, rather than isolated massive stars\,\citep{Lada2003EmbeddedClouds}.
The evolution of a single, rare, massive protostar is therefore difficult to follow as the accretion of mass and subsequent evolution may be influenced by other cluster members.
The feedback from multiple high-mass protostars also rapidly processes their natal cloud, both kinematically and chemically, to quickly erase any of the initial conditions of the environments in which they form.

Although a framework comparable to the Class 0/I/II/III classification of low-mass protostars is currently lacking in the high-mass regime, several distinct evolutionary phases of a young high-mass protostar have been identified \citep[e.g.][]{Zinnecker2007TowardFormation, Svoboda2016THECLUMPS}. 
This includes the formation of cold starless (or very young protostellar) clumps \citep[e.g.][]{alessio2017quietclumpsI}, hot cores \citep[e.g.][]{Cesaroni2005HotCores} and finally H{\sc ii} regions\,\citep[e.g.][]{Kurtz2005HypercompactRegions}. 
To fully understand the mechanisms governing the evolution of massive young stellar objects (MYSOs), snapshots along the entire evolutionary path are required. 
Due to the relatively short timescales between the onset of fusion and the dispersal of the parent cloud, the point at which an MYSO attains its final mass is not well characterised.
So far, the deeply embedded nature of MYSOs within the star forming clumps makes it challenging to identify a population of sources at this point in evolution, often requiring interferometric resolutions to probe cores of $\sim$0.1\,pc size forming individual stars\,\citep{Tan2014MassiveFormation}.

Aside from directly observing the emission from a protostar itself, other features of high-mass star forming regions have been tied to the evolutionary status of protostellar objects.
Examples of this include extended green emission (EGOs), which previous authors have found to be associated with MYSOs in the earliest stages of evolution with ongoing outflow activity \citep[e.g. ][]{cyganowski2008catalog,Chen2013NEWLYCATALOG}, and various maser species\,\citep{Forster1989TheRelation,Ellingsen2007InvestigatingSurveys,Breen201012.2-GHzSchemes}.
Found only in the hot, dusty inner circumstellar environments of high-mass protostars, class II methanol masers allow us to isolate individual sources within a clump and have been proposed as a marker of a particular, although poorly constrained, stage of massive star formation that is expected to last between 2.5$\times10^4$ and 4.5$\times10^4$ years \citep{vanderWalt2005OnMasers}.

Class II methanol masers are found in MYSOs prior to the destruction of methanol within the immediate environment of the protostar and emit strongly at a main line frequency of 6.7\,GHz, \citep{Minier2003TheMasers, ellingsen2006methanol}.
Unlike class I methanol masers which are collisionally pumped, class II methanol masers are radiatively pumped by strong infrared radiation at $\sim$70\microns\,\citep{Cragg1992PumpingMasers}.
Sufficient flux at this wavelength is provided by the thermal re-emission of the strong UV emission of a massive protostar by surrounding dust\,\citep{shari2013exclusivemethanol}.
In addition to the main 6.7\,GHz line, masing of a second methanol line at 12.2\,GHz also occurs under similar physical conditions\,\citep{Cragg2001MultitransitionMasers}.
The focus of this paper is to comprehensively characterise the Galactic population of MYSOs hosting class II methanol masers, so that the evolutionary status of these sources can be constrained.

Other maser species in a clump may also trace various evolutionary stages of a host protostar.
Collisional masers such as water and class I methanol masers are often associated with the shocks from molecular outflows\,\citep{Slysh1994TheGHz,Walsh2011TheData,Cyganowski2009ASURVEY}, both in low and high-mass star formation, but are also found in other environments such as evolved stars\,\citep{Deacon2007HSources}. 
Hydroxyl masers are found outside H{\sc ii} regions including those created as the protostar begins to ionize its surrounding in the late stages of its evolution.
Observations of excited-state hydroxyl masers can also be used to measure the magnetic field in star forming regions through their hyperfine splitting\,\citep[][Avison et al. in prep.]{Caswell1995Excited-stateGHz,Caswell2003SpectraMHz}.
Unlike the 6.7\,GHz methanol masers, each of these maser species may be diagnostic of a range of astrophysical scenarios that share common physical conditions, and can trace multiple stages of evolution even when associated with an MYSO.
Class II methanol masers are therefore the most appropriate for isolating a population of protostars in a given evolutionary state, and additional masers can offer further insight into the characteristics of a massive protostar.

Catalogues from large surveys of Galactic masers have recently been released, with the Methanol MultiBeam Survey (MMB) providing a complete sample of luminous 6.7\,GHz class II methanol masers for the southern Galactic plane\,\citep{green20096}.
Only observed as a secondary line to 6.7\,GHz, the targeted follow-up study of all MMB sources to detect the second class II methanol line at 12.2\,GHz has also been completed\,\citep{Breen201612.2-GHz2060}.
An untargeted survey of hydroxyl masers for the inner Galactic plane is provided by Southern Parkes Large-Area Survey in Hydroxyl \citep[SPLASH,][]{Dawson2014SPLASH:Region}, and the rarer excited-state hydroxyl (ex-OH) masers were co-observed during the MMB.
\citet{adam2016exohI} released a catalogue of the ex-OH masers for the entire MMB survey range, with their magnetic field properties studied further in Avison et al. (in prep.). 
An untargeted survey of the 22\,GHz maser line of water has also been observed by the H$_2$O Southern Galactic Plane Survey \citep[HOPS,][]{Walsh2011TheData} for the inner Galactic Plane, giving a large sample of water maser detections to supplement the higher sensitivity follow-up towards a small number of MMB sources by \citet{Titmarsh2014A620,Titmarsh2016A6}.
\citet{Breen2018TheHOPS} recently investigated the Galactic populations of these masers and the associations between them, finding further evidence to support an evolutionary sequence for the common maser species found in the vicinity of young stars.

To complement the large coverage of the maser studies, the \textit{Herschel} Infrared Galactic Plane Survey (Hi-GAL) has mapped the entire Galactic plane at high-resolution in the far-infrared\,\citep{molinari2010clouds,molinari2010hi}.
Peaking in this wavelength regime and visible over several of the 5 wavelengths observed, the thermal dust emission from MYSOs visible as compact objects with Hi-GAL can be used to reconstruct the spectral energy distribution (SED) and derive properties for each source such as temperature and mass surface density, in addition to luminosity and mass if the distance to a source is known.
Counterparts at wavelengths $\leq70$\microns may also be used as indicators that a clump is no longer starless but hosts at least one protostar \citep{Dunham2008IdentifyingCores}.
In addition to the 70\microns coverage of Hi-GAL, compact source catalogues for surveys in the mid-infrared (MIR), such as the GLIMPSE \citep{benjamin2003glimpse} and MIPSGAL \citep{carey2009mipsgal} surveys with the \textit{Spitzer} \citep{churchwell2009spitzer} satellite, are now available for the inner Galactic plane.
Previous studies have already been performed to determine the properties of clumps associated with known star formation sites, such as infrared dark clouds (IRDCs) \citep{traficante2015initial} and MALT90 clumps \citep{Guzman2015FAR-INFRAREDSAMPLE}, as well as generally over the inner Galactic plane \citep{Elia2017The67_.circ0}.

As the most appropriate masers to exclusively select high-mass protostars in a constrained range of evolutionary states, we identify a large sample of protostellar clumps in a similar stage of evolution through association with a class II methanol maser.
The availability and Galactic scale of the catalogues from both Hi-GAL and the MMB survey allows us to select a sample of sufficient size to fully characterise the properties of clumps hosting 6.7\,GHz methanol masers.
As this is a large sample of objects (several hundred), we adopt analysis techniques to compare populations of objects and identify statistically significant differences in the underlying distributions of properties, rather than compare individual sources.
The first portion of this work (Sections~\ref{sec:counterparts}-\ref{sec:FIRresults}) details the identification and infrared characterisation of this population of objects.
We then address the question of whether the presence of a methanol maser in a protostellar clump has any significance, making use of mid-infrared data principal component analysis to identify a sample of objects with identical properties to a methanol maser host clump but lacking a strong 6.7\,GHz maser (Sections \ref{sec:requirements}-\ref{sec:MIRresults}). 

Section \ref{sec:maserassoc} includes additional star formation masers to identify differences in populations of sources within the sample of 6.7\,GHz host clumps.
Previous work has found correlation between the appearance of and properties of masers during the progression of the star formation process in a clump \citep[e.g.][]{Breen201012.2-GHzSchemes,Titmarsh2014A620,Titmarsh2016A6}.
This work does not aim to characterise the Galactic populations of star formation masers themselves, but presents an investigation into evolutionary differences between clumps with a variety of other maser species secondary to the 6.7\,GHz line.
The relationship between clump and maser properties is also discussed.

\section{Data sets}
\label{sec:datasets}

\subsection{Methanol MultiBeam survey}
\label{sec:MMBdata}

The MMB survey completely searched Galactic longitudes from -174$^{\circ}$ to 60$^{\circ}$ through the Galactic centre, and latitudes $|b|$ $\leq$ 2$^{\circ}$ for 6.7GHz methanol masers \citep{green20096}. 
A total of 972 6.7\,GHz methanol masers were detected above the 3-$\sigma$ detection limit of 0.51\,Jy and are presented in a series of five catalogue papers \citep{Caswell2010The6,Green2010The20,Caswell2011The345,Green2012TheRegion,Breen2015The2060}. The catalogue papers provide precise maser positions derived from targeted interferometric observations accurate to 0.4\,arcsec, as well as peak flux density information. The integrated flux densities are provided for the full catalogue in \citet{Breen2015The2060}.

In this paper, we adopt the distances for MMB sources published in \citet{Green2011DistancesSelf-absorption} and \citet{Green2017TheMasers}.
These are primarily kinematic distances, with near-far ambiguities resolved through H{\sc i} self absorption, and utilise the kinematic parameters of \citet{Reid2016ASOURCES}. 
As reported in \citet{Green2017TheMasers}, the near-far kinematic distance ambiguity is resolved for 778 of the MMB sources, and we exclude the remaining sources with near-far ambiguities from any distance-dependent analysis.

\subsection{Additional maser studies}
\label{sec:extramaserdata}

A targeted follow-up towards each MMB source to search for a second class II methanol line at 12.2\,GHz has been completed, with the catalogues published in \citet{breen201212,Breen201212.2-GHz186-330,Breen201412.2-GHz20,Breen201612.2-GHz2060}. 
A total of 438 12.2\,GHz masers were detected (45.3 per cent of sources).
The MMB survey also covered the 6035\,MHz excited-state maser transition of OH, with 127 ex-OH masers detected over the southern Galactic plane\,\citep{adam2016exohI}.
The most recent follow-up to the MMB survey is the search for water masers associated with the 6.7\,GHz masers by \citet[][hereafter the Titmarsh et al. sample]{Titmarsh2014A620, Titmarsh2016A6}.
This survey targeted all 217 MMB masers between Galactic longitudes -19$^{\circ}$ and 20$^{\circ}$ with the Australia Telescope Compact Array (ATCA), finding 110 to be associated with a 22\,GHz water maser. 

Additional water masers have been detected by HOPS \citep{Walsh2011TheData}, which surveyed 100\,deg$^2$ of Galactic plane between $-70^{\circ}<l<30^{\circ}$ with the Mopra Radio Telescope between 19.5 and 27.5\,GHz at lower sensitivity than the Titmarsh et al. studies.
As water masers are less localised than methanol masers, precise positions were obtained from follow-up observations with ATCA\,\citep{Walsh2014AccurateHOPS}.
These were used to group detected masers into sites, 435 of which are associated with star formation.
For this work, we use the association of these masers with MMB masers given in \citet{Breen2018TheHOPS}.

The final maser survey that we make use of is the SPLASH survey \citep{Dawson2014SPLASH:Region} of ground-state hydroxyl masers, with the initial survey with the Parkes telescope between $-28^{\circ}<l<10^{\circ}$ through the Galactic centre and Galactic latitudes $-2^{\circ}$ and $+2^{\circ}$ detecting $\sim$600 OH maser sites. 
\citet{Qiao2016ACCURATEREGION,Qiao2018AccurateRegion} obtained accurate interferometric positions with the ATCA for OH masers in two regions of $334^{\circ}<l<344^{\circ}$ and $-5^{\circ}<l<5^{\circ}$, and also identify associations between the OH masers and MMB masers in these ranges.
Only the main line transitions of 1665 and 1667\,MHz are associated with high-mass star forming regions.

\subsection{Hi-GAL compact source catalogues}
\label{sec:firdata}

The Hi-GAL survey mapped the entire Galactic plane in a 2$^{\circ}$ strip following the Galactic warp in five wavelengths in the far-infrared\,\citep{molinari2010clouds,molinari2010hi}.
Observations at 70, 160, 250, 350 and 500\microns were carried out simultaneously, with angular resolutions of 10.0, 13.6, 18.0, 24.0 and 34.5\,arcsec in each band respectively \citep{alessio2011romagal}.
Following map creation, photometric catalogues of compact sources extracted using the CuTEx algorithm \citep{molinari2011source} have been published by \citet{molinari2016dr1} for each wavelength for the inner Galactic plane.
These catalogues contain 123\,210, 308\,509, 280\,685, 160\,972 and 85\,460 sources at 70, 160 250, 350 and 500\microns respectively between Galactic longitudes $-71^{\circ}$ and $67^{\circ}$ through the Galactic centre.
The Hi-GAL compact source catalogues made use of in this work are similarly extracted with CuTEx but cover the entire Galactic plane (S. Molinari et al., priv. comm.).
For the full Galactic plane, the numbers of objects in the compact source catalogues for each wavelength increases to 158\,092 at 70\microns, 580\,295 at 160\microns, 468\,394 at 250\microns, 251\,679 at 350\microns and 129\,489 at 500\microns.

In this work, we carry out a multi-wavelength analysis using the CuTEx catalogues to derive the physical properties of massive clumps hosting methanol masers.
\citet{Elia2017The67_.circ0} have derived the properties of the population of massive clumps visible with Hi-GAL and we use their results to provide a sample of objects against which to compare maser-hosting clumps.

\subsection{Mid-infrared data}
\label{sec:mir_data}
The FIR observations are complemented by data sets in the MIR from the \textit{Wide-field Infrared Survey Explorer} (\textit{WISE}) and \textit{Spitzer} satellites.
The GLIMPSE and MIPSGAL surveys carried out with the \textit{Spitzer} satellite provide coverage of the inner Galactic plane at high resolution\,\citep{churchwell2009spitzer,benjamin2003glimpse,carey2009mipsgal}, with the two surveys covering $295^{\circ}<l<65^{\circ}$ and $298^{\circ}<l<63^{\circ}$ respectively for at least $|b|<1^{\circ}$ over the survey range.
For this study, we make use of the GLIMPSE I, II and 3D catalgoues at 8\microns and MIPSGAL 24\microns point source catalogues produced by \citet{Gutermuth2015A/MIPSGAL}.

At a lower resolution, the \textit{WISE} satellite has surveyed the entire Galactic plane, with the most recent data set AllWISE combining both the initial \textit{WISE} and \textit{NEOWISE} data sets\,\citep{wright2010wide,mainzer2011preliminary}.
To supplement the inner Galactic plane coverage of the MIPSGAL catalogue, the AllWISE point source catalogue is also used to identify counterparts at 22\microns. 

In addition to the point-like counterparts identified in the mid-infrared, we also make use of the catalogue of Extended Green Objects (EGOs) of \citet{cyganowski2008catalog}.
These are defined as regions of excess extended emission in the 4.5\microns band of the original \textit{Spitzer} GLIMPSE survey. 
As only the GLIMPSE I data was used for the \citep{cyganowski2008catalog} catalogue, the Galactic plane coverage is limited to the regions of $295^{\circ}<l<350^{\circ}$ and $10^{\circ}<l<65^{\circ}$ for $|b|<1^{\circ}$.
\citet{Cyganowski2009ASURVEY} and \citet{Cyganowski2011BIPOLAREGOs} find class I methanol masers, commonly associated with outflow activity, to spatially coincide with EGO emission towards high-mass star forming regions. 
Predominantly tracing emission from species shocked through outflow interactions, EGOs provide another possible signpost of the evolutionary state of a 6.7\,GHz host clump alongside the secondary masers\,\citep{Reach2006AGalaxy,DeBuizer2010DIRECTREGIONS}.

\section{Identification of counterparts}
\label{sec:counterparts}
\subsection{Hi-GAL compact sources}

The thermal dust emission associated with the clump hosting each maser may be identified in the catalogues of compact sources visible in each of the wavelengths in the Hi-GAL survey.
Association with a methanol maser is determined through on-sky position.
Due to the degradation of resolution with increasing wavelength present in the Hi-GAL data, the maximum allowed distance between the centroid of a clump and the maser position is varied with wavelength when identifying maser counterparts.

This reduces the number of false associations at short wavelengths and avoids missed counterparts at long wavelengths due to the positional uncertainty.
For each of the Hi-GAL catalogues, a counterpart is assigned to a maser if the source centroid is separated by less than half of the corresponding Hi-GAL beam widths from the maser position.
The maximum angular separations are 5.1, 6.8, 9.0, 12.0 and 17.25\,arcsec at 70, 160, 250, 350 and 500\microns respectively.
In the instance that several counterparts are identified, the source with the smallest angular separation from the maser coordinates is selected.
Table \ref{tab:counterparts} shows the number of masers assigned a counterpart in each wavelength, with $>84$\% of masers detected with a counterpart in any single band.
Overall, nearly all sources (95.7\%) are associated with a Hi-GAL compact source in at least one band.

For the longest wavelength Hi-GAL catalogues, a greater number of false associations between masers and sources in this catalogue will be returned due to the increased beam size.
The overall percentage associated at 500\microns is however less than at other wavelengths, most likely due to the comparatively poor resolution resulting in missed detections. 
For clumps hosting evolved protostellar sources, such as those capable of sustaining a class II methanol maser, the 160\microns emission is expected to be close to the peak of the thermal SED. 
In line with this, we find the number of sources with a counterpart recovered to be greatest for 160\microns.

\begin{table}
\centering
\caption{Infrared counterparts to the 972 MMB masers found in each of the single wavelength Hi-GAL source extractions. The number of sources with a counterpart identified within the maximum counterpart angular distance $\theta$ in Column 2 is reported in Column 3 for each band. The number is also given as a percentage of the maser sample in Column 4. The bottom row gives the number of MMB sources with a counterpart identified in $\geq1$ Hi-GAL band.}
	\begin{tabular}[width=\columnwidth]{lccc}
		\hline
		$\lambda$ [$\mu$m] & $\theta$ [$^{\prime\prime}$] & No. with counterpart & \%\\ \hline
		70 & 5.1 & 859 & 88.4\\ 
		160 & 6.8 & 873 & 89.8\\
		250 & 9.0 & 823 & 84.7\\ 
		350 & 12.0 & 826 & 85.0\\ 
		500 & 17.3 & 822 & 84.6\\ 
		Any & - & 930 & 95.7\\\hline
\end{tabular}
\label{tab:counterparts}
\end{table}

\subsection{Multi-wavelength sample selection}

The sample of masers studied further in this work is selected based on the visibility of compact emission over several Hi-GAL wavelengths.
Since Class II methanol masers are pumped by radiation at $\sim$70\microns \citep{Sobolev2005MethanolFormation}, their host clumps are expected to be `protostellar' in nature, visible as both compact emission from the cool dust envelope at wavelengths $\geq$160\microns and at 70\microns from a warmer inner component\,\citep{motte2010initial}.
In order to fit to the spectral energy distribution (SED) of the dust envelope of a protostellar clump, a detection in at least 3 wavelengths $\geq$160\microns is required, as the emission at 70\microns is neither optically thin nor tracing the same cold material.

Although a high percentage of sources are associated with a 500\microns counterpart, this is artificially increased due to the larger beam size including false associations with sources that may not truly host the methanol maser.
In these cases, the 500\microns counterparts may belong to cold dust condensations that do not show any evidence of emission at $\leq$160\microns associated with an embedded protostar.
Therefore, requiring a catalogued 500\microns detection severely limits our multi-wavelength sample size as all associations that are less likely to be true associations are removed through shorter wavelength requirements, and the percentage of maser sources with a 500\microns counterpart falls below that of the other bands.
So, the sample analysed further are defined as masers associated with a counterpart in 70, 160, 250 and 350\microns only.

Subject to these constraints, 731 6.7\,GHz methanol masers (72.5 per cent) are identified with infrared emission in all bands between 70 and 350\microns
In addition to these sources, we also note that 22 masers (2.2 per cent) are identified with emission at wavelengths between 160, 250 and 350\microns but lack a detected 70\microns counterpart. 
\citet{Elia2017The67_.circ0} show that a deeper targeted extraction at 70\microns towards such sources is likely to reveal a counterpart or provide an upper limit on the flux from any warm component present.
By visual inspection of the Hi-GAL maps, we confirm that these maser sources do all have a counterpart at 70\microns but are not recovered by CuTEx.

Methanol maser sources detected in the FIR in previous studies may also be missing from the sample in this work due to the saturation of the Hi-GAL maps at 250 and/or 350\microns. 
An example of this are the two maser-hosting protostellar clumps in the infrared dark cloud SDC335, with \citet{adam2015sdc335} making use of the 70 and 160\microns Hi-GAL data only.
In total, 80 sources are removed due to unreliable fluxes reported in the Hi-GAL catalogues.

The sample is further refined to exclude sources displaying irregular FIR SEDs as it will not be possible to fit these with the simple model chosen to describe the objects.
As the CuTEx algorithm used to produce the Hi-GAL catalogues performs source deblending, such sources are likely to be false associations giving rise to the unphysical shape rather than blending within a Hi-GAL beam.
For prestellar and protostellar cores, the dust envelope of a source is expected to have a SED peaking at $\leq 250$\microns.
Defining the [250-350]\microns colour of a source as the logarithmic ratio of 250 to 350\microns flux, we use this property to assess the validity of the infrared fluxes assigned to maser.

Prior to the calculation of the [250-350]\microns colour of a source, the fluxes at long wavelengths require scaling to ensure that the flux from the same volume of material is considered at all wavelengths, as excess gas is included if a source is unresolved. 
The size at 250\microns is taken to most reliably trace the size of the cold clump envelope, the fluxes at 350\microns are scaled according to the relation
\begin{equation}
\label{eq:scale}
F_{\nu} ^{SED} = F_{\nu} \times \frac{\mathrm{FWHM}^{dec}_{250}}{\mathrm{FWHM}^{dec}_{\nu}}
\end{equation} 
where the $\mathrm{FWHM}^{dec}$ values are the deconvolved circularised FWHM source sizes \citep{NguyenLuong2011TheMini-starburst} given in the Hi-GAL catalogues.
The sources are assumed to be resolved at 70 and 160\microns and no scaling is applied.
On average, the scaled flux at 350\microns is 73 per cent of the original counterpart flux, with a standard deviation of 13 per cent.
The overall effect is to steepen the SED of each source between 250\microns and 350\microns, i.e. shifting the [250-350]\microns colour towards more positive values, with an example shown in Figure \ref{fig:greybody}.

We exclude sources of negative [250-350]\microns colour (more emission at 350\microns than 250\microns), as these are irregular SEDs given the expected peak emission wavelength.
Imposing further constraints based on the source colours is not possible, as the temperature dependence of the peak position may lead to either increasing or decreasing SEDs in all other consecutive wavelength bands.
Initially 24 of the 731 sources fall below the colour limit at 0, reducing to 4 following flux scaling.
In conjunction with the removal of sources with unreliable fluxes, a total of 84 sources are removed from the sample, giving a sample size of 647 for further analysis.

\subsection{Mid-infrared counterparts}

To extend the spectral energy distribution, positional association was similarly used to search the mid-infrared point source catalogues described in Section \ref{sec:mir_data} for all 647 maser sources appearing in the four required Hi-GAL bands with reliable fluxes.
The 24\microns MIPSGAL catalogue was initially searched for counterparts to the sample of masers identified above using a half-beam association radius of 3.0\,arcsec \citep{Gutermuth2015A/MIPSGAL}.
Of the 627 sources falling within the MIPSGAL survey range, only 40.8 per cent of the sample (304 sources) were found to have an associated MIPSGAL 24\microns point source. 
As for the 70\microns catalogues, \citet{Elia2017The67_.circ0} also performed a deeper targeted extraction in the MIPSGAL data towards their protostellar objects to recover faint sources to either identify missing counterparts, or provide an upper limit on the 24\microns flux.

However, the masers lacking 24\microns counterparts show a trend towards higher FIR luminosities (see Section \ref{sec:luminosity_results}).
As the FIR luminosity derived from the Hi-GAL data correlates with the luminosity at 24\microns, this implies that saturation of the MIPSGAL images is a plausible cause for the small fraction of sources associated with a counterpart, rather than the failing to detect faint 24\microns sources \citep{Dunham2008IdentifyingCores}. 
Visual inspection of the MIPSGAL images confirms that this is the case towards a majority of the sources.

The AllWISE point source catalogue at 22\microns was also searched for counterparts as a substitute for those lacking 24\microns sources.
This data is at much lower resolution, with a beam size of 24.0\,arcsec, but does not suffer from the same saturation effects.
Further to this, AllWISE covers the full range of coordinates of the sample.
Although the lower resolution AllWISE data will not detect the small and faint counterparts detectable in the MIPSGAL data, the detection of bright sources and extended survey range recovers additional counterparts.
A total of 451 counterpart sources are found in the 22\microns AllWISE catalogue, 280 of which were not assigned a MIPSGAL counterpart previously.

For the purpose of determining the existence of a $\sim$20\microns counterpart and estimating the flux by tabulated integration, the 22\microns WISE and 24\microns MIPSGAL data points are used interchangeably.
Combining the WISE and MIPSGAL detections finds a majority (518, 80.1 per cent) of maser host sources to be associated with a counterpart at $\sim$20\microns.
For the masers with a counterpart identified in both the MIPSGAL and AllWISE catalogues, the MIPSGAL counterpart is used preferentially.
For sources with both a MIPSGAL and AllWISE counterpart found, a comparison of the two flux values finds no systematic differences.

The \textit{Spitzer} GLIMPSE I, II and 3D point source catalogues were similarly used to assign 304 counterparts at 8\microns within 2.0\,arcsec of each maser coordinate.
The association of 6.7\,GHz methanol masers with 8\microns counterparts has previously been investigated by \citet{gallaway2013mid}.
However, these associations are only performed for the subset of MMB masers with interferometric positions available at the time of publication.
We do not use these associations but match to the GLIMPSE catalogues to cover the full MMB range in addition to the Hi-GAL protostellar objects, and ensure that the matching criteria are consistent for the two samples.
The numbers of sources with an MIR counterpart in each of these surveys is shown in Table \ref{tab:MIR}, alongside the number of the methanol maser hosts within each survey region.

Approximately half (52.1 per cent) of the maser sources bright at $\sim$20\microns also have an 8\microns GLIMPSE object assigned to them and therefore the presence of an 24\microns source does not necessarily indicate that emission is expected at 8\microns.
Conversely, most sources (88.8 percent) visible at 8\microns also have a MIPSGAL or AllWISE counterpart.

\citet{gallaway2013mid} previously reported that 83 per cent of methanol masers are associated with emission in at least one of the 4 GLIMPSE bands (67 per cent in all bands).
When performing the same matching against the GLIMPSE point source catalogues, the authors recovered a counterpart in any GLIMPSE band for only 55 per cent of the MMB masers.
This difference is attributed to the lack of extended or slightly extended emission included in the point source catalogues, as is common towards maser sources, and report that the number of MIR counterparts to the MMB masers increases by a factor of approximately 2 if such sources are included alongside point-like counterparts.
Performing targeted source extraction and inspection towards both the maser and protostellar samples to recover extended emission is beyond the scope of this paper.
Our result that 48.3 per cent of methanol masers are associated with an 8\microns counterpart is therefore an underestimate of the true number of sources bright at 8\microns.

\begin{table*}
	\centering
	\caption{The single wavelength searches for maser counterparts in GLIMPSE 8\microns, MIPSGAL 24\microns and AllWISE 22\microns point source catalogues. The intersection of each survey coverage with the MMB survey is given in the second column. The number of FIR detected masers falling within the range of each survey is stated, alongside the number with a counterpart assigned. Counterparts were assigned using a maximum distance of one half-beam size for each survey. The last two rows indicate the total number of the maser sources with a visible counterpart in either of the $\sim$20\microns catalogues and the subset of these also associated with 8\microns emission.}
	\begin{tabular}{l c c c }
		\hline
		Survey (wavelength [\microns]) & Common coverage & Masers in range & No. with counterpart (\%)\\ \hline
		GLIMPSE I, II, 3D (8) & 295$^{\circ}<l<60^{\circ}$, $|b|\leq 1$& 629  & 304 (48.3)\\
		MIPSGAL (24) & 298$^{\circ}<l<60^{\circ}$, $|b|\leq 1$ & 627 & 238 (40.0) \\
		AllWISE (22) & 186$^{\circ}<l<60^{\circ}$, $|b|\leq 2$  & 647 & 451 (69.7) \\
		AllWISE (22) or MIPSGAL (24) & 186$^{\circ}<l<60^{\circ}$, $|b|\leq 2$ & 647 & 518 (80.1) \\
		AllWISE (22)/MIPSGAL (24) \& GLIMPSE (8) & 295$^{\circ}<l<60^{\circ}$, $|b|\leq 1$ & 629 & 270 (42.9)\\ 			\hline	 
	\end{tabular}
\label{tab:MIR}
\end{table*}

\section{Infrared clumps analysis}

\subsection{SED fitting}

Within the far infrared regime, each methanol maser host clump is taken to emit thermally at the dust temperature as a single temperature modified blackbody, or greybody. 
This approximation describes the emission well for wavelengths $\geq$160\microns, but the 70\microns flux is found to trace the warm embedded protostellar component and not the emission from the envelope\,\citep{motte2010initial,Dunham2008IdentifyingCores}.
Additionally, the emission at 70\microns is not necessarily optically thin and cannot be reliably fitted without also modelling optical depth effects.

\begin{figure}
	\includegraphics[width=\columnwidth]{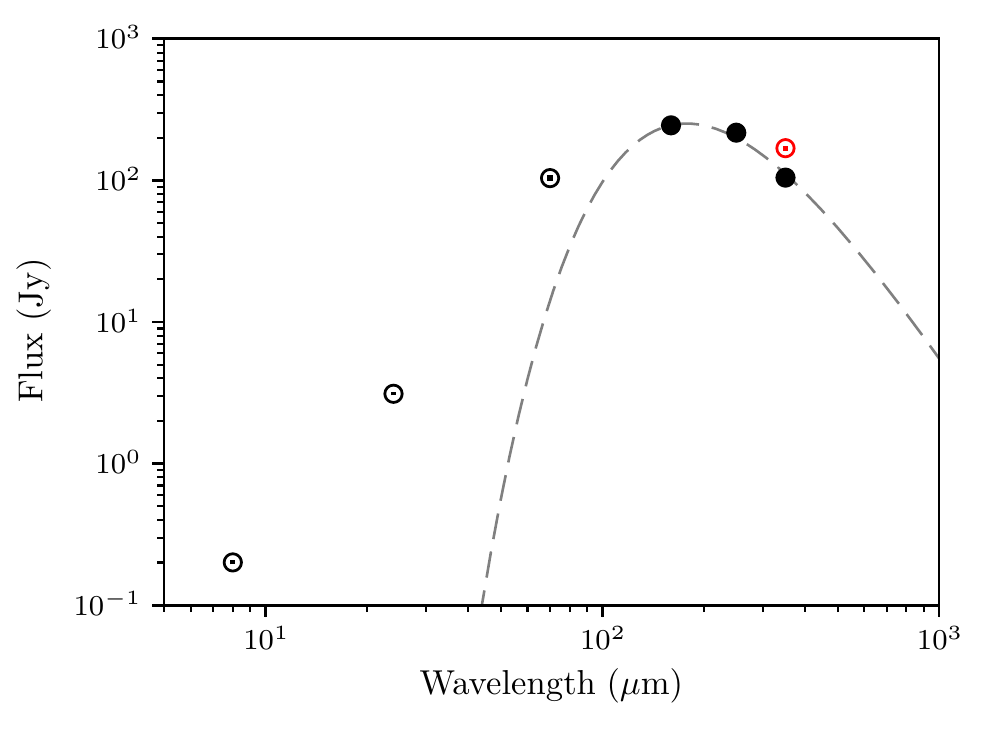}
    \caption{The fluxes integrated over the source size in the Hi-GAL maps, for a typical maser hosting clump in the sample. The filled circles are the data points used for fitting to the SED at 160, 250 and 350\microns, with the unscaled flux at 350\microns shown in red. The data points likely originating from a warmer inner component at 8, 24 and 70\microns are marked as open circles and are not used for fitting. The grey dashed line is the fitted greybody for this source, returning a temperature of 17.0\,K and a clump envelope mass of 806\Msun, given that the distance to the host maser is 2.7\,kpc.}
    \label{fig:greybody}
\end{figure}

At wavelengths $\geq$160\microns, the emission is optically thin and the observed intensity of emission from a clump is given by 
\begin{equation}
\label{eq:greybody}
S_{\nu} =  \frac{M\kappa_{\nu}}{d^2} B_{\nu}(T).
\end{equation}
where $B_{\nu}(T)$ is the Planck function evaluated at the observation frequency $\nu$ for clump temperature $T$, $M$ is the total gas+dust mass of the envelope and $d$ is the distance to the clump.
The dust mass opacity coefficient $\kappa_{\nu}$ given in units of cm$^2$g$^{-1}$ is frequency dependent and assumed to follow the scaling relation
\begin{equation}
\label{eq:scaling}
\kappa = \kappa_0\left(\frac{\nu}{\nu_0}\right)^{\beta}
\end{equation}
where the values of $\kappa_0$ and $\nu_0$ are set to 0.005\,cm$^2$g$^{-1}$ and 230\,GHz respectively for a gas-to-dust ratio of 100\,\citep{preibisch1993influence}.
The spectral index $\beta$ is fixed rather than fitted as the limited number of data points for each source does not allow reliable determination and the degeneracy between $\beta$ and $T$ cannot be accounted for.
A value of 1.8 is chosen as this was found to best characterise the shape of the observed SED of the clumps, consistent with previous work\,\citep[e.g.][]{Juvela2015GalacticCores,Juvela2018iHerschel/isup/sup,Guzman2015FAR-INFRAREDSAMPLE}.
A greybody is fitted to each constructed SED by least-squares minimisation using the Levenberg-Marquardt algorithm to return best fit values of $T$ and $\frac{M}{d^2}$.
For masers with a known distance from the data described in Section \ref{sec:MMBdata}, values of physical parameters such as total mass and radius are also obtained for a clump.
An example of a the FIR SED of a typical maser host source and the fitted greybody is shown in Figure \ref{fig:greybody}.

\subsection{Calculating luminosity}
\label{sec:luminosity_calc}

For this analysis, the FIR luminosity of a source is defined as the luminosity in the range 70-500\microns.
We do not include the mid-infrared fluxes in this luminosity calculation to ensure that the integral is evaluated over the same interval for all sources in the sample.
Although distance-dependent quantities themselves, parameters such as the ratio of $L$ to $M$ may still be evaluated for sources lacking a distance value.

Taking the total FIR luminosity as the area under the fitted greybody SED returns an estimate of the luminosity originating from the dust envelope of the clump.
This does not reliably describe the total far-infrared luminosity of a source as an excess of energy is likely to be emitted by a warmer inner component due to the embedded protostar at shorter wavelengths, including 70\microns\,\citep{Dunham2008IdentifyingCores}.
Despite this, 26 sources show a deficiency in 70\microns flux relative to the greybody estimate, which can be attributed to unreliable recovery of the true flux at 70\microns due to complex environments and source blending\,\citep[e.g.][]{Elia2013THE225.5,molinari2016dr1,Persi2016The15507-5359}.
We therefore define the FIR luminosity of a source $L_{\mathrm{FIR}}$ as the integral under the tabulated data points in this range calculated, supplementing the four required Hi-GAL fluxes with a 500\microns flux returned by extrapolating the best-fit greybody, as the envelope described by the single temperature greybody is responsible for emission at this wavelength.
We perform tabulated integration using a 5-point Newton-Cotes formula to allow for deviation from ideal greybody behaviour at 70\microns and better characterise the total energy emitted in the FIR by the clump (see Section \ref{sec:luminosity_results}).

Throughout the following sections the properties of this sample of maser host clumps are shown alongside the same distributions for the general Galactic protostellar clump population visible with Hi-GAL, defined by \citet{Elia2017The67_.circ0}.
Prior to comparison with the protostellar sample, it should be noted that there is a small difference in how the luminosity of the two samples (masers and protostars) is calculated.
For this work, the maser clump luminosity is strictly defined as the integral given by the tabulated Hi-GAL data between 70 and 500\microns, and so is a measure of the \textit{far-infrared} luminosity. On the other hand, 
\citet{Elia2017The67_.circ0} calculate the \textit{bolometric} luminosity of each source by including the flux at wavelengths outside the Hi-GAL wavelength range. 
For the sources in the protostellar sample with a counterpart detected in either the MIPSGAL, 22\microns WISE or 21\microns \textit{MSX} \citep{Egan2003The2.3} data, the range of fluxes used to calculate source luminosity have been extended to have a lower limit of $\sim$20\microns.
For all sources, \citet{Elia2017The67_.circ0} calculate the luminosity at wavelengths $\geq160$\microns from the greybody fit and use the tabulated fluxes to calculate the luminosity contribution at shorter wavelengths.
For the maser-bearing sources associated with 24\microns emission, we find an average increase in luminosity by a factor of $2.0\pm0.78$ compared to the \textit{far-infrared} luminosity used here.
As \citet{Elia2017The67_.circ0} find only 10\% of sources to be MIR dark (i.e. lacking a counterpart at $\sim 20$\microns), this correction factor is likely to apply to most sources.
Given that the infrared emission is responsible for the pumping of the class II methanol maser line, this luminosity is the most appropriate for this work, although typically evolutionary diagnostics such as $L/M$ are designed for use with the bolometric luminosity.

\section{FIR Results}
\label{sec:FIRresults}

\begin{figure}
	\includegraphics[width=\columnwidth]{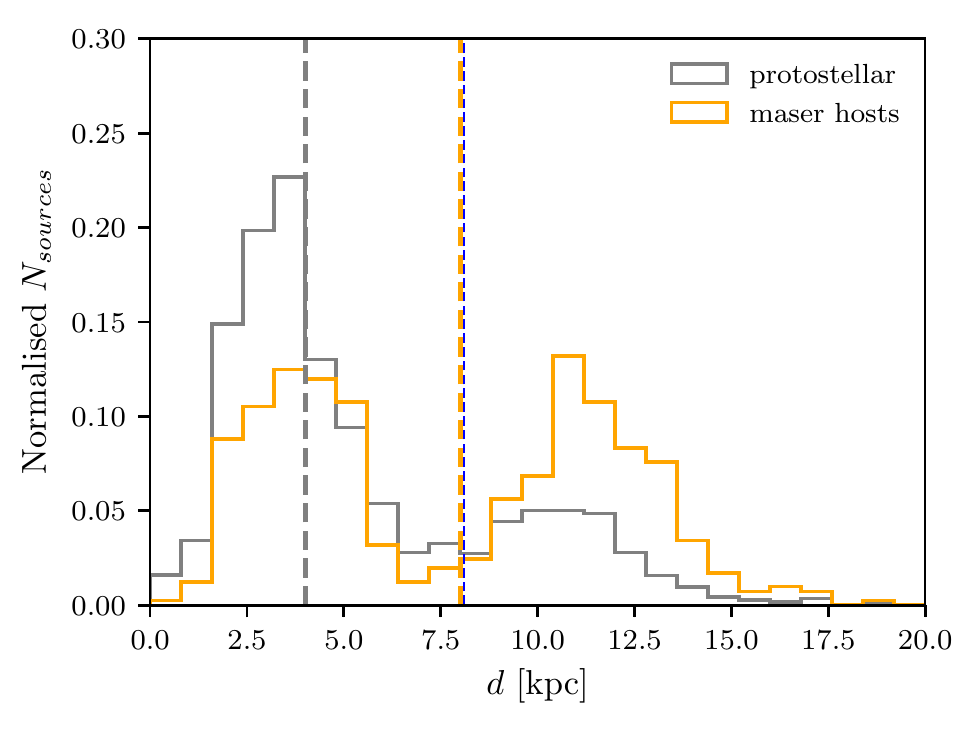}
    \caption{Heliocentric distance distributions of the protostellar (grey) and maser (orange) samples, with the median distances of 4.0\,kpc and 8.0\,kpc marked in respective colours. The assumed distance to Galactic centre at 8\,kpc is marked in blue.}
    \label{fig:distance}
\end{figure}

The derivation of clump properties for the 6.7\,GHz hosts from their FIR fluxes allows us to compare this sample against existing catalogues of high-mass star forming clumps.
Within the 22,426 protostellar sources reported by \citet{Elia2017The67_.circ0}, we select the 22,293 sources satisfying the same multi-wavelength (70 to 350\microns) visibility criteria as the maser sample, and restrict to the Galactic longitude and latitude range of the MMB survey.
Additionally, we only consider the 25 per cent of the above sources with `good' distances (i.e. resolved between near and far through the use of other tracers) as the assumption of far kinematic distance applied to the remaining sources may artificially increase in the number of sources at high luminosities and masses.
The distance distributions of the two samples are shown in Figure \ref{fig:distance}.
Each displays bimodality around the Galactic centre, which is unsurprising as IR observations and kinematic distance resolution is challenging in this direction, resulting in a dip at the assumed distance to Galactic centre of 8.0\,kpc.
The symmetry in the distances of the MMB sources between the near and far side of the Galaxy supports that the MMB survey has recovered the majority of Galactic methanol masers.
However, a significant shift towards distances on the near side of the Galactic plane is seen for the protostellar sample relative to the maser hosts.
This reflects the ability of Hi-GAL to detect sources of a given angular size, corresponding to a larger physical radius at further distances, and the primary methods used to confirm distances or resolve kinematic ambiguities, such as association with absorption features\,\citep{Elia2017The67_.circ0}.
The effects of the bimodality in distance are discussed in Section \ref{sec:l-m}.

We select this sample of 5497 protostellar objects for comparison, hereafter the `protostellar sample', with the aim to determine whether the subset of protostellar sources hosting a visible methanol maser occupies a constrained space within the total population of protostellar objects in the inner Galactic plane.

The following sections detail the results for the individual properties calculated for the maser host clumps, compared against the protostellar sample in each case.
Table \ref{tab:ir_results_all} presents the distribution medians and standard deviations for all properties.
Approximate thresholds on the clump properties required to host a methanol maser are given in each section, and also included in this table.
In each case, these represent requirements of a protostellar environment on the clump scale, rather than the conditions in the circumstellar region of an individual massive protostar, to sustain maser activity.

\subsection{Temperature and mass surface density}
\label{sec:t_and_sigma}

\begin{figure}
	\includegraphics[width=\columnwidth]{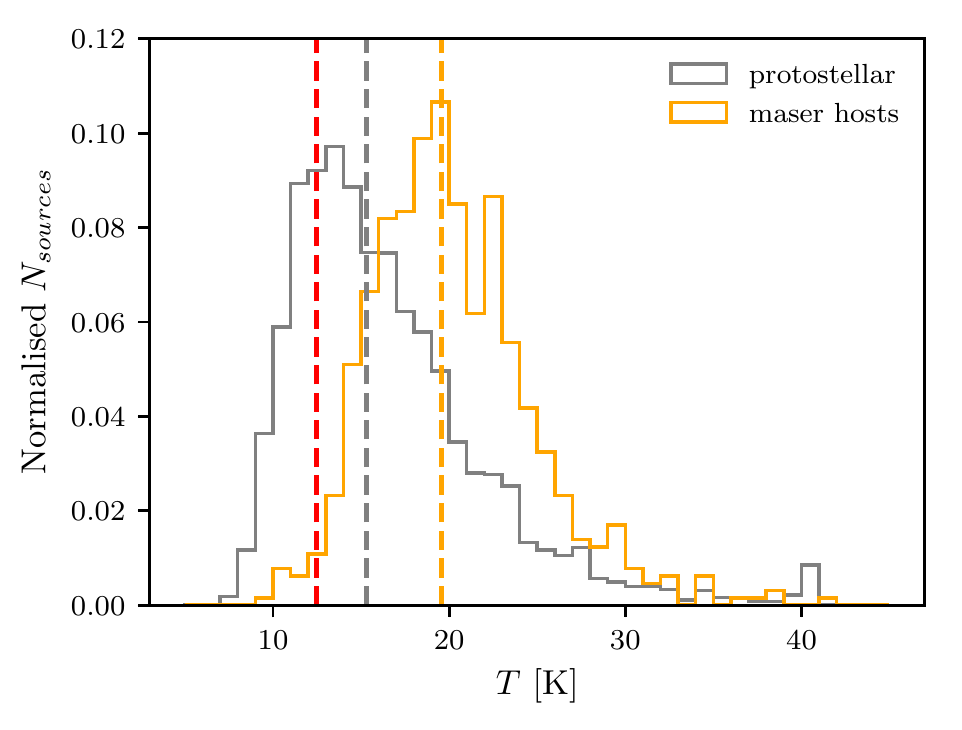}
    \caption{Distribution of temperatures determined by greybody fitting to the SED of clumps for the protostellar (grey) and maser (orange) samples. 
    The sample medians of 19.5\,K and 15.3\,K are also marked for the maser and protostellar sample respectively.
    The threshold clump temperature taken to be required to sustain a class II methanol maser is marked in red at 12.5\,K.}
    \label{fig:T}
\end{figure}

For clumps in the sample, the distance independent quantities of temperature and mass surface density $\Sigma$ are obtained for all sources following SED fitting.
The methods adopted by \citet{Elia2017The67_.circ0} to characterise their protostellar sources are similar to our own, both using the 250\microns angular clump size to define physical size and modelling each SED as a modified blackbody.
A key difference in our methods is the choice of spectral index $\beta$ and reference dust mass opacity $\kappa$.

To correct for differences in $\beta$ when considering FIR SEDs fitted with a modified blackbody covering the peak of emission, \citet{Guzman2015FAR-INFRAREDSAMPLE} give a relationship between spectral index and temperature of
\begin{equation}
    \frac{h\nu_{\mathrm{peak}}}{k_{\mathrm{B}}T_{\mathrm{d}}} \simeq \beta + 3
\end{equation}
where $T_{\mathrm{d}}$ is the dust temperature, $\nu_{\mathrm{peak}}$ is the wavelength at which the SED peaks, and $k_{\mathrm{B}}$ and $h$ are the Boltzmann and Planck constants respectively, with this relationship accurate to 10 per cent for values of $\beta>1$.
Following the method used by \citet{Guzman2015FAR-INFRAREDSAMPLE} to compare temperatures derived from the far-infrared with different values of $\beta$, we obtain a scaling of $T_{\beta={2.0}} = 0.96T_{\beta={1.8}}$.
The median temperature of the maser sample decreases to 18.7\,K under this correction, and does not significantly affect the conclusion that the maser sources are significantly warmer.
Assuming a worst-case of 10 per cent uncertainty on this scaling, the largest possible decrease in median temperature from the use of a different $\beta$ is 2.7\,K which, although not completely negligible, is also significantly less than the 4.2\,K offset in temperature seen between the protostellar and maser clumps.
We therefore do not consider the choice of $\beta$ to significantly impact the results obtained throughout this work.

When comparing median $\Sigma$, masses or thresholds between different works, it is important to be aware of the offset factor which the adoption of different values and forms for the dust mass opacity can introduce. 
For the dust mass opacity, \citet{Elia2017The67_.circ0} use a reference value of 0.1\,cm$^2$g$^{-1}$ at 300\microns.
Rescaling our reference value to 300\microns with Equation~\ref{eq:scaling} gives a value of 0.07\,cm$^2$g$^{-1}$ and we find that the $\Sigma$ (and mass) presented for a clump in this work 
offset to higher $\Sigma$ and mass by a factor of 1.4 relative to the protostellar sample due to this difference.

Since all the objects with irregular SEDs and unreliable fluxes in the Hi-GAL compact source catalogues have already been excluded from the sample, the remaining maser sources are well fitted with temperatures up to a maximum of $\sim42\,$K.
The distribution of fitted $T$ values is shown in Figure \ref{fig:T} which has a median of 19.5\,K for the maser hosts.
The protostellar sample has a lower median temperature of 15.3\,K.
To set thresholds on the properties required for a source to host a methanol maser, we adopt 2-standard deviation lower limits on the distributions of properties of the maser sources, unless otherwise stated. 
As the presence of a high temperature tail in the temperature distribution causes significant asymmetry, we only use points below the median to calculate the standard deviation and obtain a $2\sigma$ lower threshold.
This gives a minimum temperature for a clump to host a methanol maser of 12.5\,K, and is marked in red in Figure \ref{fig:T}.

Also making use of the Hi-GAL data, \citet{Guzman2015FAR-INFRAREDSAMPLE} find an average temperature of 18.6\,K for protostellar objects, and the study of \citet{Breen2018TheHOPS} also makes use of these temperature maps to derive a temperature of $\sim$24\,K towards 6.7\,GHz methanol masers.
Both of these are significantly warmer than the respective temperatures of 15.3\,K and 19.5\,K for the protostellar and maser objects considered in this work. 
This offset arises from the difference in aperture that is integrated over to obtain the flux of a source in each Hi-GAL wavelength.
\citet{Guzman2015FAR-INFRAREDSAMPLE} assume a fixed aperture at all wavelengths, similar to the clump size at 250\microns, whereas this work and \citet{Elia2017The67_.circ0} allow this size to vary according to the appearance in each map.
The size reported in the Hi-GAL compact source catalogues is often significantly smaller at 160\microns relative to the size at 250\microns for a source.
This is therefore a difference in definition rather than a true inconsistency between our two studies.
As this work is primarily focused on identifying differences between objects with methanol masers and the protostellar population of \citet{Elia2017The67_.circ0}, both of which have been characterised using the same flux extraction method, this does not affect the conclusions we draw from this analysis.

\begin{figure}
	\includegraphics[width=\columnwidth]{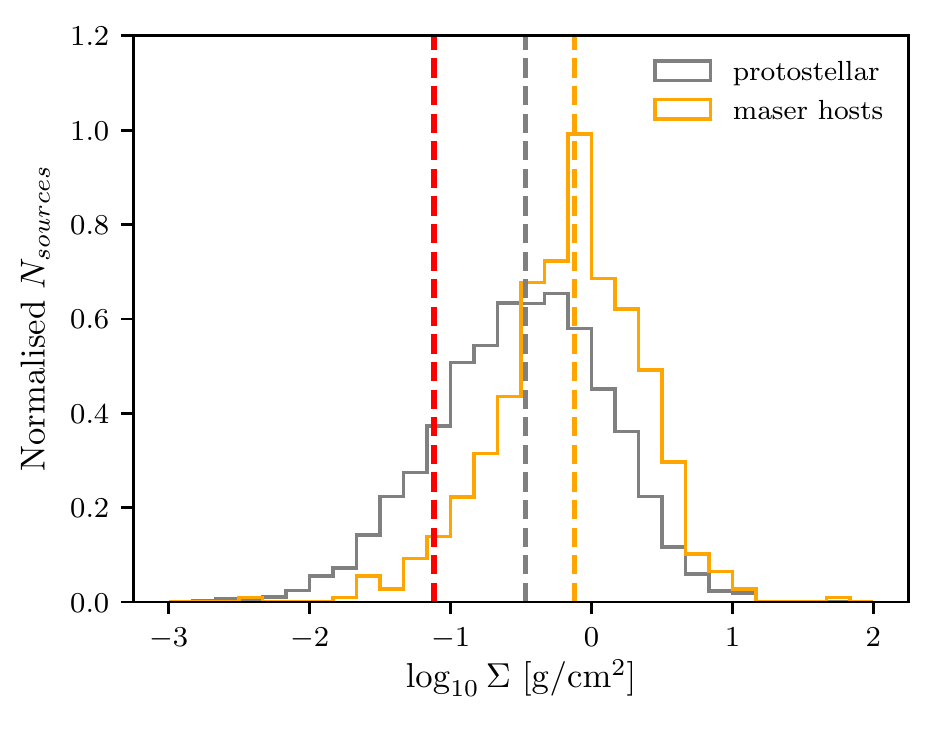}
    \caption{Distribution of mass surface densities calculated from the prefactor returned following greybody fitting and the clump size at 250\,$\mu$m. The protostellar sample (grey) has a median of 0.34\gpercmsq and the maser sample (orange) has a median of 0.76\gpercmsq, with a lower limit on the mass surface density required to host a methanol maser (red) set at 0.08\gpercmsq.}
    \label{fig:surfdens}
\end{figure}

Figure \ref{fig:surfdens} shows the distributions of mass surface density $\Sigma$ in logarithmic bins, where $\Sigma=\frac{M}{d^2}\frac{1}{\Omega}$ for solid angle  $\Omega$ subtended by the clump (calculated from the source FWHM angular size at 250\microns), $M$ is the clump mass and $d$ the distance.
The median $\Sigma$ for the maser host sample is 0.76\gpercmsq with 248 (38 per cent) sources above the approximate threshold of $\Sigma \geq 1$\gpercmsq for massive star formation proposed by \citet{Krumholz2008AFormation}.
Only 21.0 per cent of the protostellar population fall above the same threshold, with a sample median of 0.34\gpercmsq.
Including a correction factor of 1.4 to account for the systematic offset due to the choice of dust mass opacity reduces the maser median value to 0.54\gpercmsq, and this cannot account for the difference between the two samples.

This threshold is assumed for non-magnetised regions whereas other studies support a lower threshold \citep[e.g.][]{Butler2012MID-INFRAREDCLUMPS,Urquhart2015TheDust,traficante2015initial,alessio2018darkclumpsII}, with \citet{Tan2014MassiveFormation} suggesting a threshold of $\Sigma \sim 0.1$\gpercmsq below which massive stars cannot form in the presence of magnetic fields.
For the maser and protostellar objects, the percentages of sources above this threshold are 94 and 80 per cent respectively. 
The maser hosting sources are therefore found to be at a higher median mass surface density than the protostellar sample and a lower limit of 0.08\gpercmsq is taken as the requirement for a clump to host a methanol maser.

\subsection{Mass and radius}

\begin{figure}
	\includegraphics[width=\columnwidth]{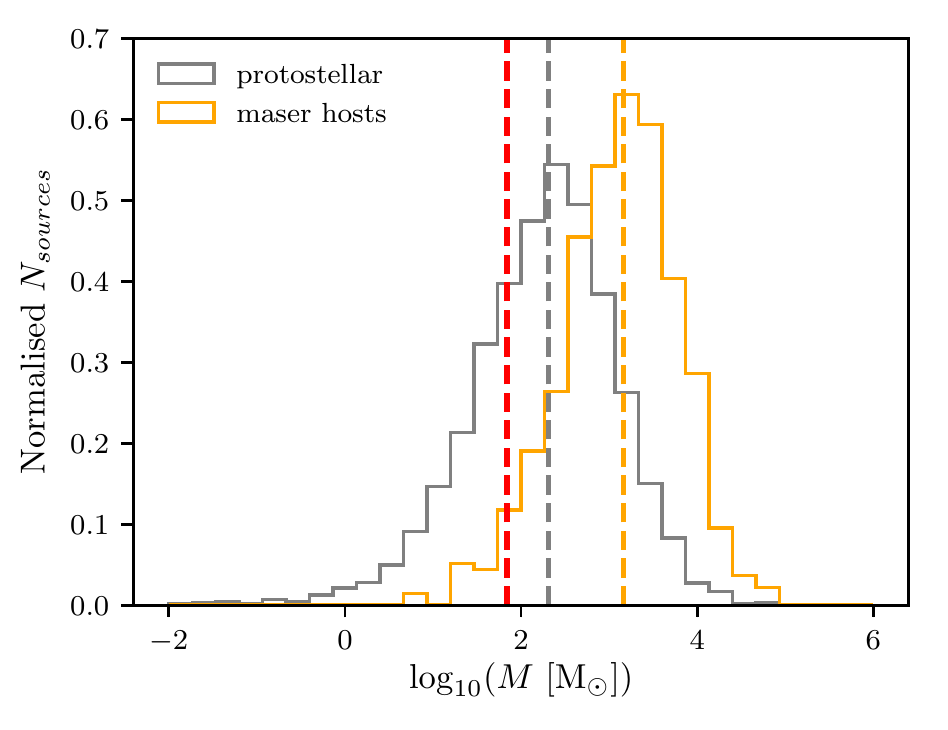}
    \caption{Mass distributions for the protostellar (grey) and maser hosting (orange) samples, with medians of 206\Msun and 1457\Msun respectively and a minimum threshold (red) to host a methanol maser set at 69\Msun. }
    \label{fig:mass}
\end{figure}

\begin{figure}
	\includegraphics[width=\columnwidth]{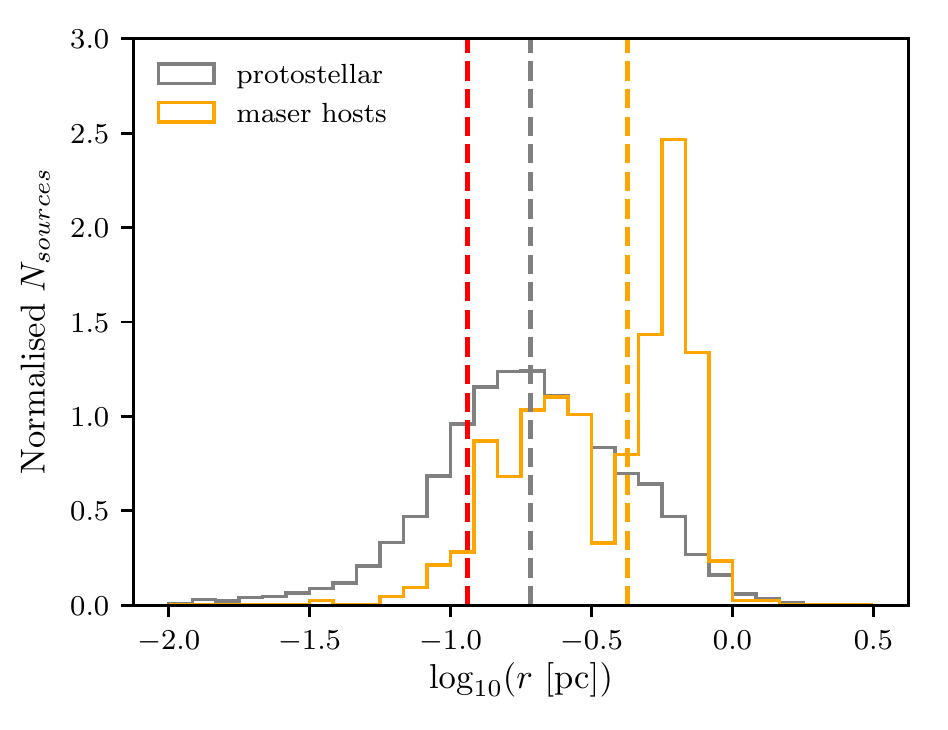}
    \caption{Distributions of physical radii derived from the source size at 250\microns for the maser host clumps (orange) and the Galactic protostellar sample (grey), with medians of 0.42\,pc and 0.19\,pc respectively.
    The bimodality in radius of the maser sources is due to a bimodality in the distance distribution.
    The minimum size required for a methanol maser to be present is taken to be 0.11\,pc (red).}
    \label{fig:radius}
\end{figure}

\begin{figure}
	\includegraphics[width=\columnwidth]{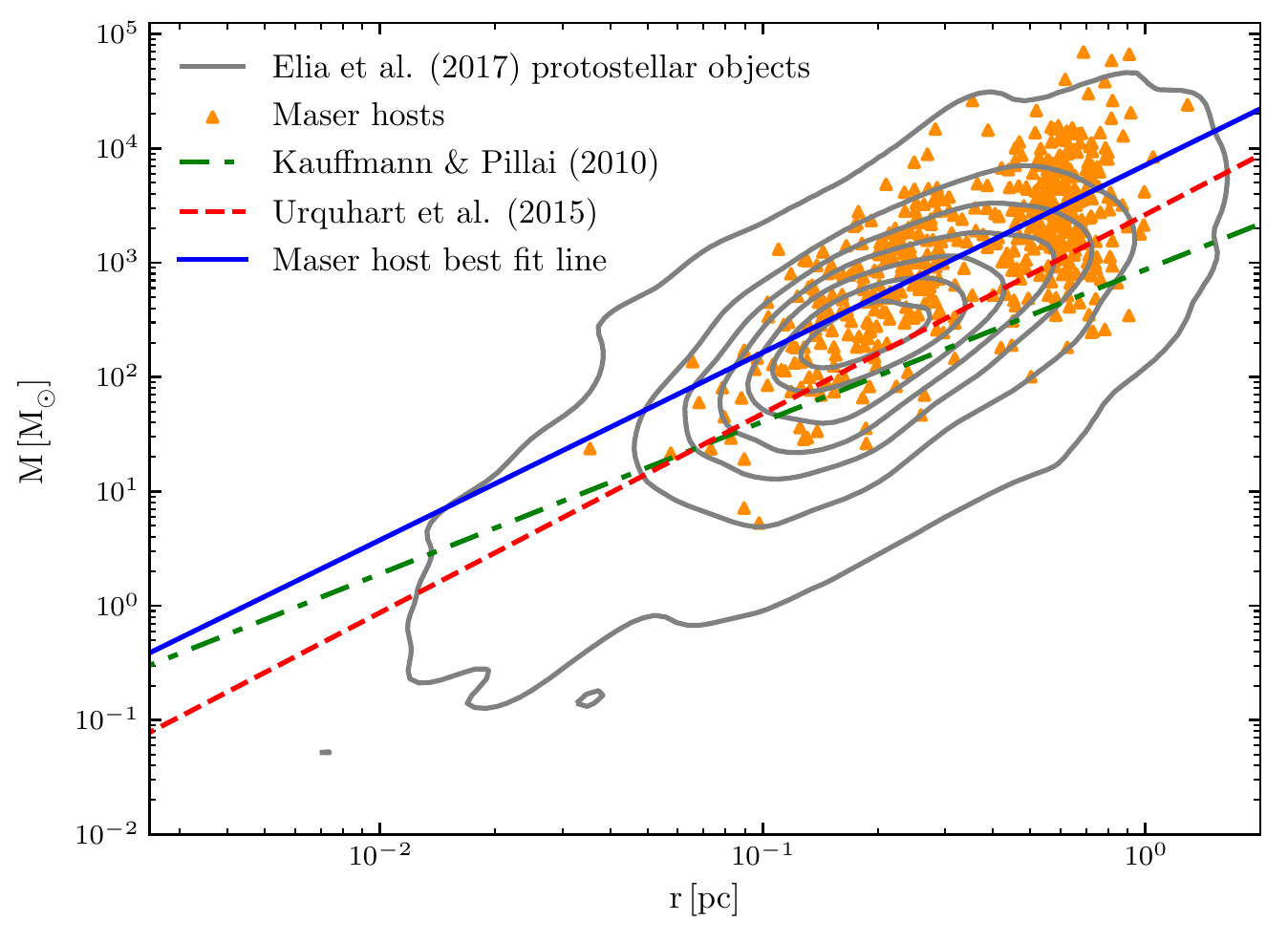}
    \caption{Mass-radius diagram of the maser sources (orange) with the protostellar sources shown as grey contours, with the levels at intervals of 15\% between 1\% and 90\% of the peak density in $M$-$r$ space. The approximate threshold for star formation to occur in a clump of \citet{Kauffmann2010HOWCLUSTERS} is shown as the dash-dotted green line, given by $M(r) \leq 870$\Msun$\left(r/\mathrm{pc}\right)^{1.33}$. 
    The red dashed line is the power law fit of observed properties towards 6.7\,GHz methanol maser clumps by \citet{Urquhart2015TheDust}, given by $\log \left( \frac{M}{1\mathrm{M}_{\odot}} \right) = 1.74\left( \frac{r}{1\mathrm{\,pc}} \right) + 3.42$. 
    For comparison, the best fit line to our data is shown as a blue solid line of $\log \left( \frac{M}{1\mathrm{M}_{\odot}} \right) = (1.64\pm0.07)\left( \frac{r}{1\mathrm{\,pc}}\right) + (3.85\pm0.04)$.}
    \label{fig:mass_radius}
\end{figure}

\begin{figure}
	\includegraphics[width=\columnwidth]{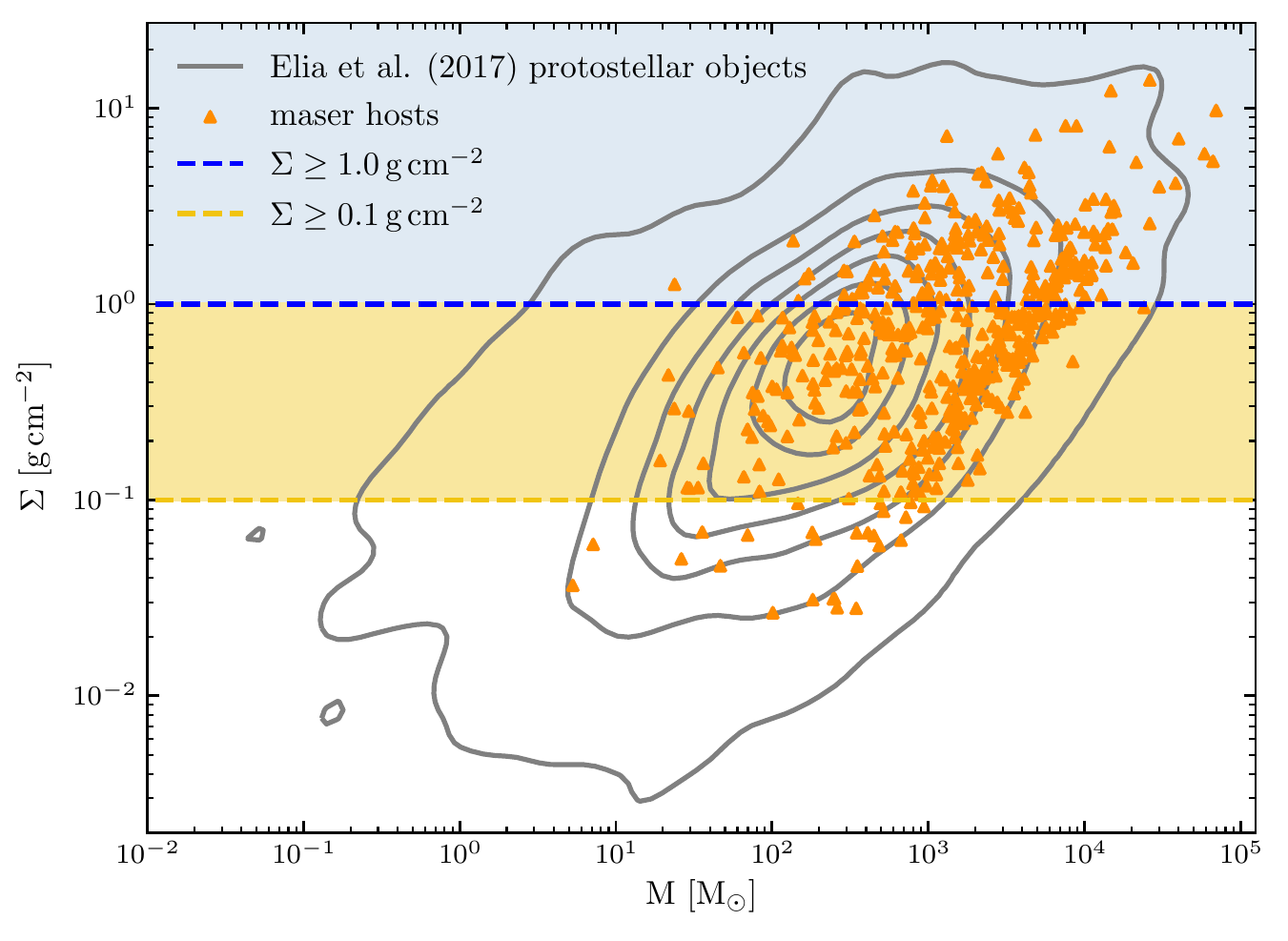}
    \caption{$\Sigma$-M diagram for the protostellar (grey contours, levels at 1\%, 15\%, 30\%, 45\%, 60\%, 75\% and 90\% of peak density in $\Sigma$-M space) and maser hosting (orange) sources. The approximate surface density criterion for star formation to proceed of $\Sigma\geq 1.0$\gpercmsq is marked in blue.}
    \label{fig:mass_sigma}
\end{figure}

For the 511 sources with a reliable distance value and a luminosity $>100$\Lsun (see Section \ref{sec:luminosity_results}), the mass of a clump may be calculated from the $M/d^2$ prefactor previously used to calculate mass surface density, as well as conversion of the angular source size at 250\,$\mu$m to a physical clump radius.
The distributions of the mass and radius are plotted in Figures \ref{fig:mass} and \ref{fig:radius} respectively.
The median mass of the maser hosting clumps is 1457\Msun with a median radius of 0.42\,pc, whilst the protostellar sample is less massive and smaller on average, with median values of 206\Msun and 0.19\,pc.
Again, we note that the difference in choice of dust mass opacity may bias our results towards higher masses by a factor of 1.4, although this is a small correction relative to the difference in median mass.
The lower limit on mass and radius to host a methanol maser are taken to be 69\Msun and 0.11\,pc.
We note here that the bimodality in radius seen in Figure \ref{fig:radius}, and later separation into two clusters in Figures \ref{fig:mass_radius} and \ref{fig:mass_sigma}, is caused by the underlying bimodal distance distribution shown in Figure \ref{fig:distance} and the limited range of angular sizes \citep[1-3 times the PSF in each band,][]{molinari2016dr1} for objects to be included in the Hi-GAL compact source catalogues. 

Figure \ref{fig:mass_radius} shows the mass of the clumps as a function of the radius, with the \citet{Kauffmann2010HOWCLUSTERS} scaling relation shown in green for comparison, with only 6 per cent of the maser hosts falling below this limit.
The best fit line of $\log \left( \frac{M}{1\mathrm{M}_{\odot}} \right) = 1.74\left( \frac{r}{1\mathrm{\,pc}} \right) + 3.42$ obtained by \citet{Urquhart2015TheDust} for objects hosting methanol masers is shown in red.
For comparison, the line of best fit through our maser hosts is shown in blue and is given by $\log \left( \frac{M}{1\mathrm{M}_{\odot}} \right) = (1.64\pm0.07)\left( \frac{r}{1\mathrm{\,pc}}\right) + (3.85\pm0.04)$.
The differences between the results of \citet{Urquhart2015TheDust} and this work are discussed further in Section~\ref{sec:IR_summary}.

Mass and radius are both derived using the distance to a source, and this may introduce artificial correlation in $M-R$ space.
As mass is proportional to $d^2$ and radius to $d$, we would expect to find a relationship of $M\propto r^2$ if the correlation in Figure \ref{fig:mass_radius} was purely due to the underlying distribution of source distances.
We obtain a power-law index of 1.6 from our line of best fit in $M-R$ space, indicating that this is not the case.

The derivation of mass for each source also allows the maser sources to be placed in $M-\Sigma$ space.
The locations of protostellar sample and the maser sources are plotted in Figure \ref{fig:mass_sigma}, and the thresholds of $\Sigma=0.1$ and 1.0\gpercmsq are also shown.
As stated in Section \ref{sec:t_and_sigma}, nearly all maser sources (94 per cent) and the majority of protostellar sources (80 per cent) fall above the revised threshold of 0.1\gpercmsq for magnetised regions to form massive stars proposed by \citet{Tan2014MassiveFormation}.
\citet{alessio2018darkclumpsII} find an approximate threshold of 0.12\gpercmsq from studies of MYSOs dark at 70\microns, and therefore mostly starless.
Our approximate lower limit of 0.08\gpercmsq for a clump to host a class II methanol maser, and a massive protostar to sustain it, is remarkably similar to these estimates despite vast differences in our methods. 

\subsection{Luminosity}
\label{sec:luminosity_results}

\begin{figure}
	\includegraphics[width=\columnwidth]{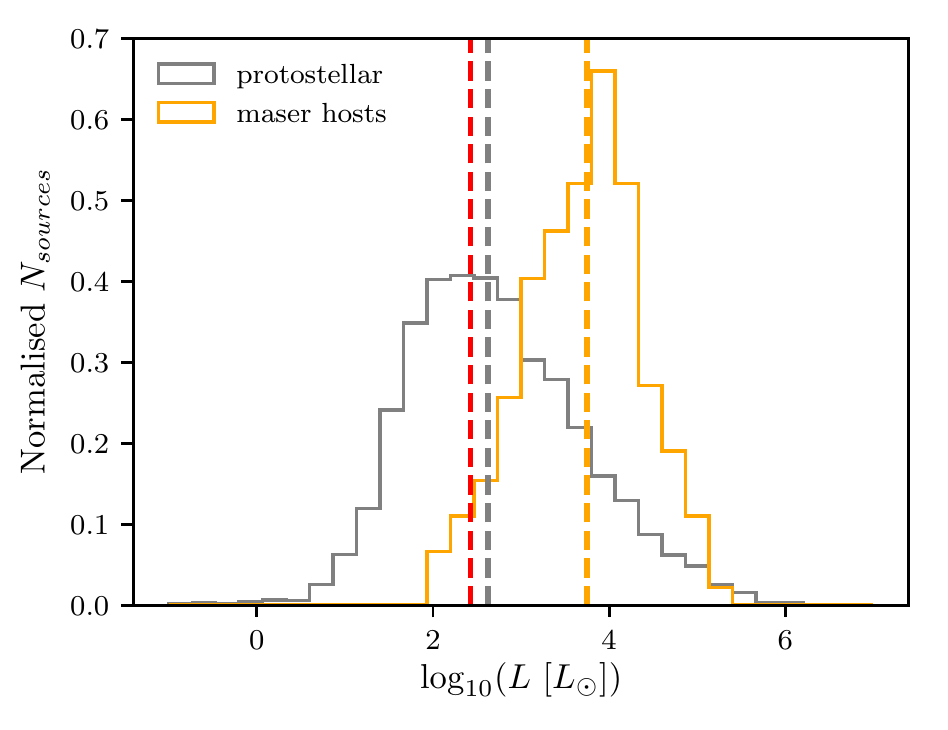}
    \caption{{The far-infrared luminosity distributions of sources that host a methanol maser (orange) and the total Galactic population of clumps with high-mass protostars (grey). The maser sources have a median luminosity of 5617\Lsun, with a 2$\sigma$ lower bound of 264\Lsun (red). The median luminosity of the protostellar sources is 422\Lsun. It should be noted that for the maser sources, the lowest luminosity sources have large distance uncertainties.}}
    \label{fig:luminosity}
\end{figure}

\begin{figure}
	\includegraphics[width=\columnwidth]{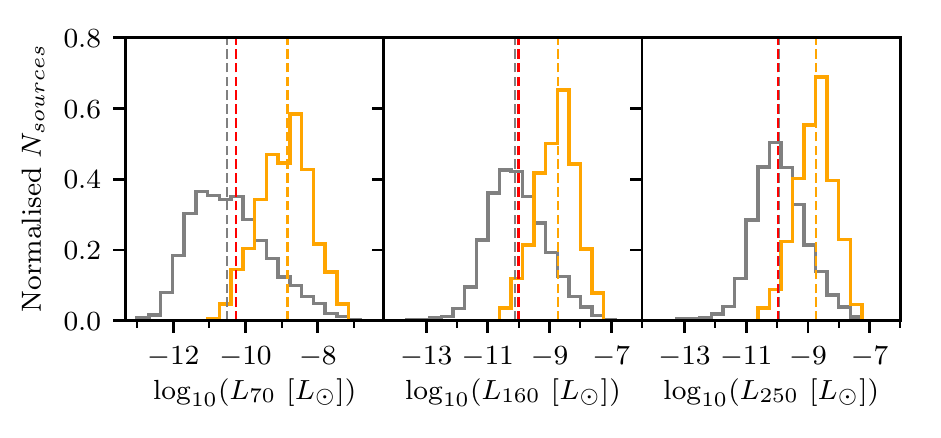}
    \caption{{The luminosities at 70\microns, 160\microns and 250\microns for the maser (orange) and general population of protostellar MYSOs in the Galaxy (grey), calculated from the flux in a given Hi-GAL band multiplied by $4\pi d^2$ where $d$ is the distance to the source. From left to right: the maser sources have medians of $L_{70}$, $L_{160}$ and $L_{250}$ of $1.42\times10^{-9}$\Lsun, $1.87\times10^{-9}$\Lsun and $1.83\times10^{-9}$\Lsun respectively. The corresponding medians for the protostellar sources are $3.05\times10^{-11}$\Lsun, $7.50\times10^{-11}$\Lsun and $1.11\times10^{-10}$\Lsun, and the lower thresholds on each luminosity for a source to host a methanol maser (red) are $5.39\times10^{-11}$\Lsun, $9.64\times10^{-11}$\Lsun and $1.07\times10^{-10}$\Lsun.}}
    \label{fig:spec_luminosity}
\end{figure}

Calculated from the tabulated fluxes from each source rather than fitting (Section \ref{sec:luminosity_calc}), the FIR luminosity distributions are shown in Figure \ref{fig:luminosity}, in addition to the specific luminosities at 70, 160 and 250\microns in Figure \ref{fig:spec_luminosity}.
These luminosities are given by $L_{\lambda}=4\pi f_{\lambda}d^2$ for sources with distance, with no correction for bandwidth as we are only making a comparison between Hi-GAL fluxes.
As described in Section \ref{sec:MMBdata}, the distances assigned to each MMB source are compiled from HI self-absorption studies and supplemented with values from previous literature using techniques such as astrometry\,\citep{Green2011DistancesSelf-absorption,Green2017TheMasers,Reid2016ASOURCES}.

As high-mass protostellar objects are not expected to have far-infrared luminosities $\lesssim$100\Lsun, all 16 maser-hosting clumps with a derived luminosity below this threshold have been checked for reliability.
Sources occupying the lowest luminosity region of the sample tend to have small heliocentric distances, and therefore a small component of their observed velocity relative to the local standard of rest is due to Galactic rotation.  
This results in a large percentage uncertainty on the inferred kinematic distance for these sources relative to the majority of the sample, in addition to as yet untested accuracy of the applied \citet{Reid2016ASOURCES} method for calculating distances in the Southern Galactic plane due to the lack of parallax measurements for comparison.
The luminosities of these sources are therefore poorly constrained, with the distance distribution of low-luminosity objects in the sample discussed further in Section \ref{sec:lmatched}.

Twelve of the 16 sources are found at near distances ($\sim$1-2\,kpc).
The uncertainty in the distance to the maser is sufficient to account for the low luminosity, with each source consistent with a luminosity of $>$100\Lsun within $3\sigma$.
These 12 sources are therefore excluded from further distance-dependent analysis.
The remaining 4 sources have very low derived luminosities of $<$20\Lsun, and in conjunction with other unusual derived properties.
These sources are also excluded as the SEDs are unrealistic given the expected approximate range of source properties (false associations, unreliable flux measurement etc.) despite being physically valid.

The median value of $L_{\mathrm{FIR}}$ is found to be 5617\Lsun for the maser sources, with a 2$\sigma$ lower bound of 264\Lsun used to define the minimum maser-hosting clump luminosity.
The corresponding medians for $L_{70}$, $L_{160}$ and $L_{250}$ are $1.42\times10^{-9}$\Lsun, $1.87\times10^{-9}$\Lsun and $1.83\times10^{-9}$\Lsun, with thresholds set at $5.39\times10^{-11}$\Lsun, $9.64\times10^{-11}$\Lsun and $1.07\times10^{-10}$\Lsun respectively.

The protostellar sample has a median bolometric luminosity of 422\Lsun with only 58 per cent of the sample falling above the 264\Lsun threshold set from the FIR estimate for the maser sample.
Although the factor of 2 over-estimation in luminosity for the protostellar sources relative to the maser sources is small when compared to the offset in the median values, this further enhances the difference in average luminosity between the two populations.
In order of increasing wavelength, the matching median values of specific luminosity are $3.05\times10^{-11}$, $7.50\times10^{-11}$ and $1.11\times10^{-10}$\Lsun and the thresholds set on specific luminosities are similarly strong, with 41, 46 and 51 per cent of the protostellar sample meeting the assumed requirement to host a methanol maser in $L_{70}$, $L_{160}$ and $L_{250}$ respectively.

\subsection{Distance dependence and luminosity-to-mass ratio}
\label{sec:l-m}

\begin{figure*}
	\includegraphics[width=\textwidth]{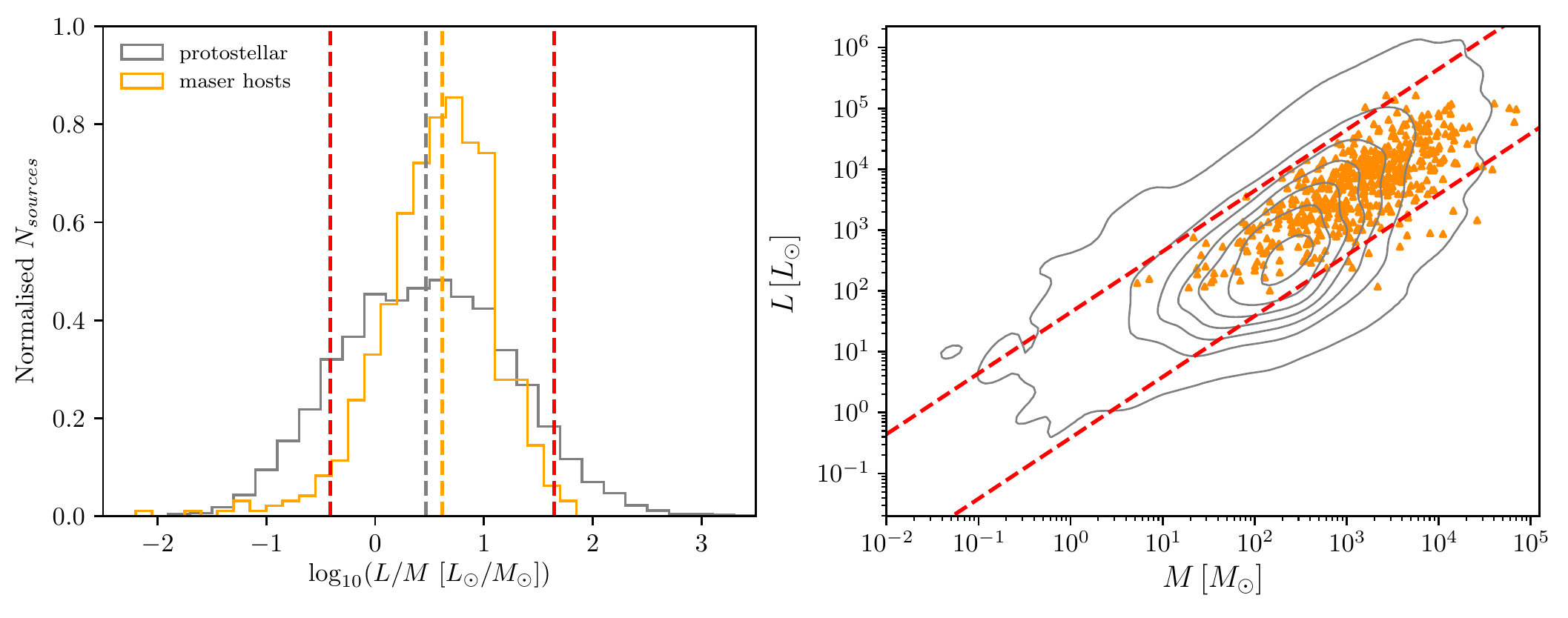}
    \caption{The distribution of luminosity-to-mass ratio as derived from the far-infrared bolometric luminosity and the mass from SED fitting for each source in the maser (orange, $L_{FIR}$) and protostellar (grey, $L_{bol}$) sample. Left: The distributions of the two samples, with the medians of 4.1\Lsun/\Msun (maser) and 2.9\Lsun/\Msun (protostellar) marked and thresholds taken to be the minimum and maximum $L/M$ to host a methanol maser marked in red at 0.39\Lsun/\Msun and 44.32\Lsun/\Msun respectively. Right: The position of the sources in each sample in $L$-$M$ space, with the same assumed thresholds to host a maser marked as red lines. The maser hosts are marked in orange, with the grey contours showing the distribution of the protostellar sources. The contour levels are at intervals of 15\% of the peak density in L-M space, with the lowest contour at 1\% and the highest contour at 90\%.}
    \label{fig:mass_luminosity}
\end{figure*}

As shown in Figure \ref{fig:distance}, the separation between sources at the near and far heliocentric distances about Galactic centre is relatively distinct, and significant bimodality is observed in subsequent parameters such as radius.
The catalogued compact sources visible with Hi-GAL are selected on their angular scale, which corresponds to different selection criteria on physical radius for the populations of objects at the near and far sides of the Galaxy.
We are therefore selecting two different populations of objects as the sources at greater distances are selected to be larger.
Comparable objects at nearby distances may be resolved into several objects, and similarly the population of small clumps hosting methanol masers on the far side of Galaxy will also not be detected.
Although nearly all masers have a counterpart detected in at least one wavelength, we do not detect the full population of 6.7\,GHz as multiwavelength sources with Hi-GAL as 270 lack a detection all required bands.

Separating the sources into `near' and `far' distances about the Galactic centre at 8.0\,kpc, we select 256 and 255 maser sources at the near and far distances respectively.
For the protostellar objects, the split is skewed towards near distances, finding 4229 at $<8$\,kpc and 1268 at $>8$\,kpc.
Table\,\ref{tab:nearfar} reports the typical distance-dependent properties for the `near' and `far' objects for the maser and protostellar objects.
As expected, sources at the far distances are found to have significantly larger radii than nearby sources, and are a factor of $\sim$3 larger.
Similarly, an increase in median mass and luminosity is also found.
In many cases, the uncertainty on the distance to a source is also significant and corresponds to an increased uncertainty in parameters calculated from the measured distance relative to parameters that are not.

\begin{table}
\centering
\caption{Median properties of the maser hosts and protostellar clumps at heliocentric distances $<8$\,kpc (near) and $>8$\,kpc (far). The values in brackets are the standard Gaussian errors.}
	\begin{tabular}[width=\columnwidth]{l c c c c }
		\hline
		Property & \multicolumn{2}{l}{Maser}  & \multicolumn{2}{l}{Protostellar}  \\
		& Near & Far & Near & Far \\ \hline
		$d$ [kpc] & 3.8(1.5) & 11.4(1.8) & 3.5(1.5) & 10.7(2.1)\\ 
		$r$ [pc] & 0.2(0.1) & 0.6(0.1) & 0.2(0.1) & 0.5(0.2)\\
		$\log_{10}$($M$ [\Msun]) & 2.8(0.6) & 3.5(0.5) & 2.1(0.8) & 3.0(0.6)\\ 
		$\log_{10}(L_{\mathrm{FIR}}$ [\Lsun]) & 3.4(0.6) & 4.0(0.5) & 2.3(0.9) & 3.6(0.8)\\ 
		$\log_{10}$($L/M$ [\Lsun/\Msun]) & 0.7(0.5) & 0.6(0.5) & 0.4(0.8) & 0.7(0.7)\\ \hline
\end{tabular}
\label{tab:nearfar}
\end{table}

The primary focus of this work is to characterise the evolutionary status of MYSOs hosting class II methanol masers, and we are therefore concerned with the evolutionary bias between the near and far sources.
Mass-luminosity diagrams are a common diagnostic used to place an MYSO along its evolutionary sequence, with the overall trend in the space for a sample characterised by the luminosity-to-mass ratio, $L/M$.
Again, we emphasise that this work defines this as the ratio between FIR luminosity and mass, whereas discussions in other work including \citet{Elia2017The67_.circ0} discuss the luminosity-to-mass ratio using bolometric luminosity.
Although luminosity and mass are both calculated from measured distance, this parameter is not (as both are proportional to $d^2$), giving an estimate of the evolutionary status of a clump without introducing significant uncertainties, and associated correlations, through the use of a distance value.
Shown in the last row of Table\,\ref{tab:nearfar}, the median $L/M$ for sources at the near and far distances are very similar for the maser sources.
We therefore conclude that although we are sampling different sized clumps with heliocentric distances either side of Galactic centre, the two populations maser hosts are of similar evolutionary status and we do not separate them when discussing parameters that are not dependent on the scale of a clump.

Shown in the right panel Figure \ref{fig:mass_luminosity}, the methanol maser hosts occupy a band within the protostellar regime, towards the upper end, in a mass-luminosity diagram.
Although the median $L/M$ of the maser sample of 4.1 is close to the protostellar sample median of 2.9, the resulting distribution is significantly narrower for the maser host clumps, as shown in the left panel of Figure \ref{fig:mass_luminosity}.
Both lower and upper limits are therefore set on the $L/M$ value of a clump that can host a methanol maser, at 0.39\Lsun/\Msun and 44.32\Lsun/\Msun respectively, marked in red on both panels.
This threshold is low, given that \citet{Molinari2016CALIBRATIONFORMATION} find the expected range of $L/M$ for protostellar sources to be $1\lesssim L/M \lesssim 10$.
However, our estimate of $L/M$ is not directly comparable as we use the far-infrared luminosity only, rather than the bolometric luminosity over all frequencies.
Combining the factors of 2 (due to the definition of luminosity, Section~\ref{sec:luminosity_calc}) and 1.4 (choice of reference dust mass opacity, Section~\ref{sec:t_and_sigma}) results in a factor of $\sim$3 lower $L/M$ for the maser sources compared with the protostellar sample from \citet{Elia2017The67_.circ0}. 
Accounting for this factor enhances the offset in median between the two samples, strengthening the conclusion that the maser sources are, on average, found at greater $L/M$.

\subsection{Colours}
\label{sec:colours}
\begin{figure}
	\includegraphics[width=\columnwidth]{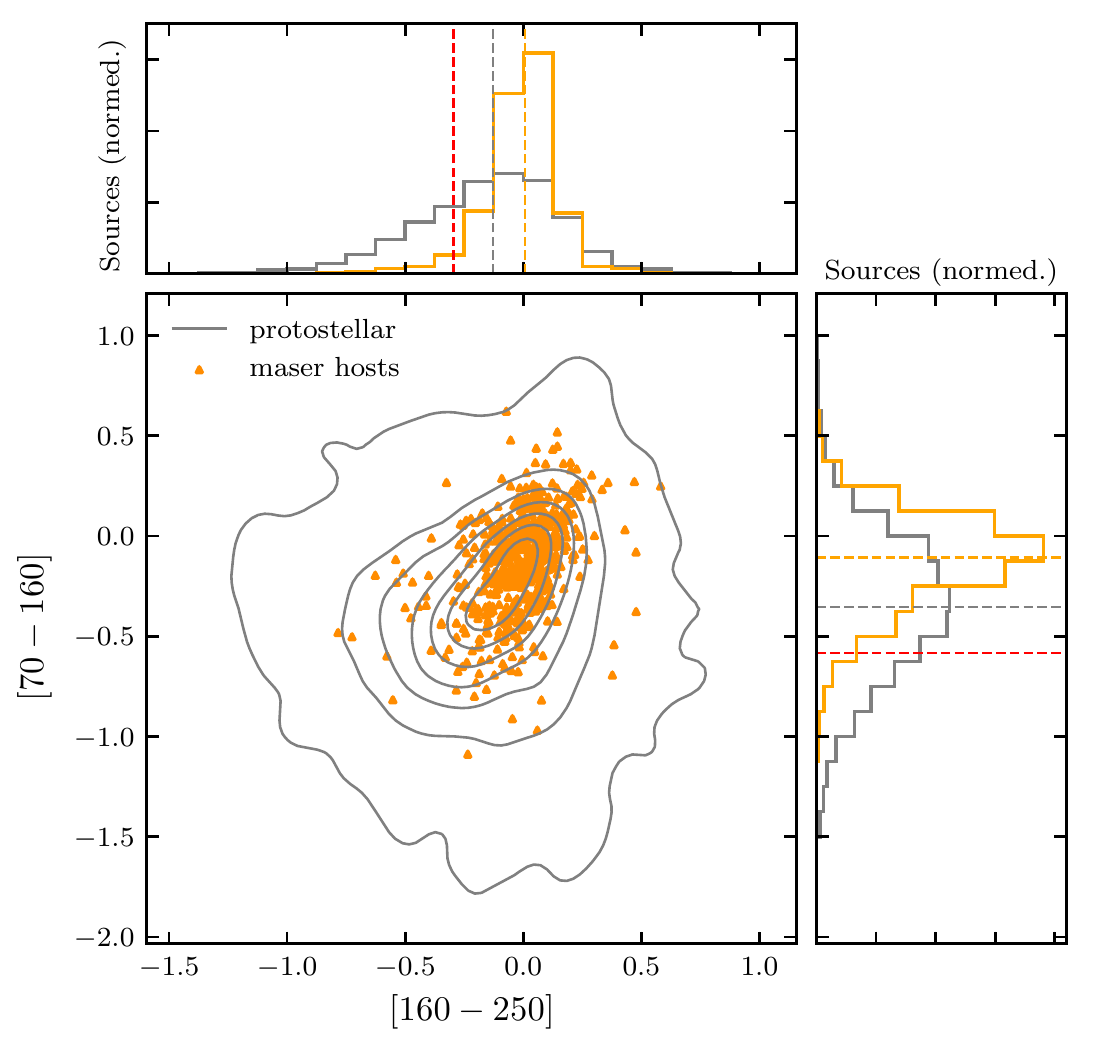}
    \caption{Colour-colour diagram showing the [70-160]\microns colour against the [160-250]\microns colour for the protostellar (grey contours, from lowest to highest density in colour-colour space: 1\%, 15\%, 30\%, 45\% 60\%, 75\%, 90\%) and maser host (orange) sources respectively. The normalised distributions of each colour are shown to the right of and above the scatter plot respectively, with the medians marked in addition to the lower limit on each colour taken as a requirement to host a methanol maser.}
    \label{fig:colour_colour}
\end{figure}

The temperature derived from fitting to the far-infrared SED will primarily trace the cold protostellar envelope of a clump rather than the temperature of the warm embedded protostar responsible for the excess of flux over the greybody value at 70\microns.
The colour of a source, defined as the logarithmic ratio of two fluxes, can characterise the slope of an SED and therefore the temperature of an object without full knowledge of the form of the spectral energy distribution.
Figure \ref{fig:colour_colour} displays the colour-colour diagram for the clumps from the [160-250]\microns colour and the [70-160]\microns colour, with the axis histograms showing the distribution of each. 
The 70\microns flux of the maser clumps is on average stronger relative to the 160\microns flux than the protostellar objects, resulting in a more positive [70-160]\microns colour, as is the ratio of 160\microns to 250\microns flux.
This implies that the warm component responsible for the emission at the shortest wavelengths is more significant relative to the total thermal emission from all components seen at longer wavelengths.
A similar result is obtained when considering the [70-250]\microns colour, with the medians and thresholds of the source colours for each sample summarised in Table \ref{tab:ir_results_all}.

\subsection{Completeness}
\label{sec:completeness}
We find the majority of methanol masers to be associated with compact dust emission with over 95\% of the maser sources detected in at least one wavelength of Hi-GAL. 
However, to associate a maser with a Hi-GAL source we require a detection in four Hi-GAL bands (70, 160, 250 and 350\,\microns) and 27\% do not satisfy this requirement. 
Also, 18\% of the maser sources are excluded from the analysis as they do not have a distance determined. 

In the following analysis of the maser sources (Section~\ref{sec:IRvsMaser}), we derive a relation between the FIR clump luminosity and the integrated luminosity of its 6.7\,GHz maser.
The line of best fit to the medians $L_{6.7}$\,$[\text{Jy km\,s$^{-1}$kpc}^2]$ of $\log_{10}(L_{6.7}) = (0.8\pm0.1)\log_{10}(L_{FIR}) + (0.4\pm0.2)$.
For a given $L_{FIR}$ in \Lsun, the average scatter in 6.7\,GHz luminosity is 10$^{0.6}$.
For a clump of 100\Lsun, the expected maser luminosity from this relation is 100\,Jy\,km\,s$^{-1}$\,kpc$^2$, decreasing to 10$^{1.4}$\,Jy\,km\,s$^{-1}$\,kpc$^2$ when considering the quoted scatter.
At 2\,kpc, these masers would have integrated fluxes of 2.0\,Jy\,km\,s$^{-1}$ and 0.5\,Jy\,km\,s$^{-1}$ respectively. 

\citet{Green2017TheMasers} give the luminosity sensitivity of the MMB as 0.15\,Jy\,km\,s$^{-1}$, assuming a maser line to span 3 spectral channels each with a sensitivity of 0.51\,Jy.
Both the average and lower bound on expected luminosity of a methanol maser in a 100\Lsun clump are therefore detectable to greater than 3$\sigma$ with the MMB sensitivity out to a distance of 2.7\,kpc for a maser of average luminosity in a clump of 100\Lsun, so we consider the MMB to be complete out to this limit. 

\begin{figure}
	\includegraphics[width=\columnwidth]{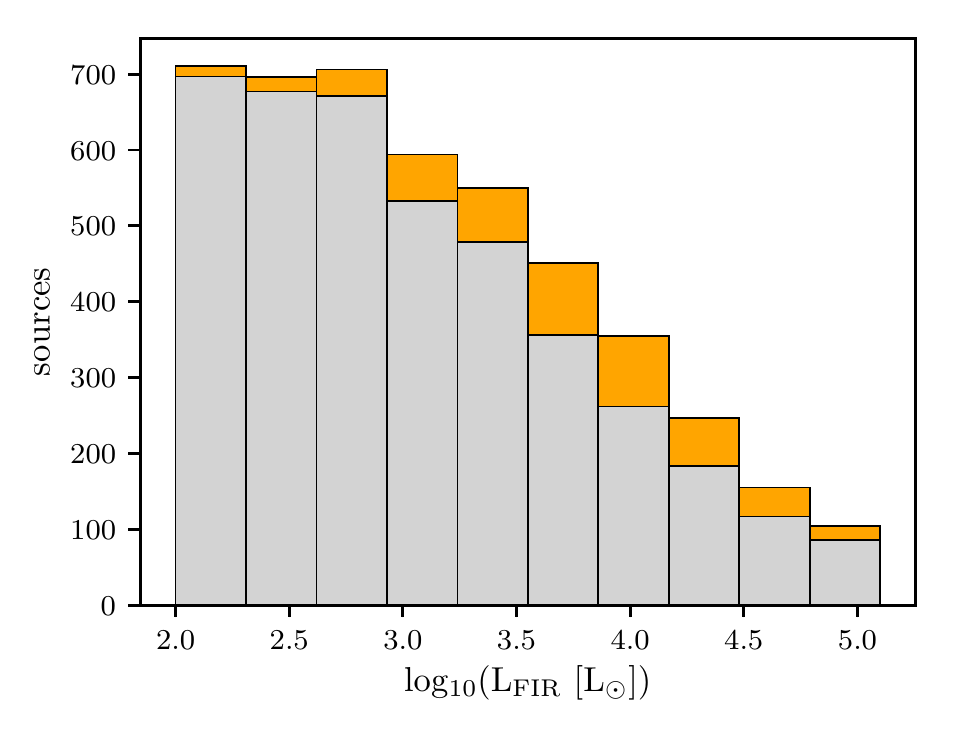}
    \caption{Stacked histogram of the protostellar sources (grey) and maser sources (orange) as a function of far-infrared luminosity.}
    \label{fig:lfrac}
\end{figure}

The completeness of the Hi-GAL catalogue is harder to assess. \citet{Elia2017The67_.circ0} find the 90\% mass completeness limit to be better than $10^{1.45}$\Msun for all objects closer than 2.7\,kpc.
Adopting a minimum luminosity to mass ratio of -0.1 (Fig.~\ref{fig:mass_luminosity}), this mass limit corresponds to a luminosity limit of 22\,\Lsun, significantly below the lowest luminosity of one of the maser sources, 100\,\Lsun, suggesting that the completeness of the Hi-GAL catalogue is not an issue for our study. 
The dominant issue limiting the completeness of the protostellar sources is the large fraction missing a reliable distance, with only one in four sources assigned a reliable distance. 

Figure~\ref{fig:lfrac} displays the number of maser sources relative to the number of protostellar sources as a function of luminosity for the entire sample.
This shows a distinct lack of maser sources relative to the number of protostellar sources at low luminosities.
Comparing the number of sources detected within 2.7\,kpc only (where the MMB is complete for low luminosity sources of $\sim100$\,\Lsun) we find 3 maser sources with luminosity between 80 and 120\,\Lsun and 92 protostellar sources, a ratio of 1:31. 
Within the same distance, there are 24 maser sources with luminosities $>10^3$\Lsun and 64 protostellar sources, a ratio 1:3. 
This is considerably larger than for the lower luminosity sources and demonstrates that the paucity of low luminosity maser sources is not a result of incompleteness. 
Note that these values have not been corrected for the completeness of the Hi-GAL sample. 
However, this should not affect the conclusion as the Hi-GAL completeness is dominated by sources with poorly defined distances. 

A final potential consideration for the completeness of maser studies is the effect of beaming, as the amplitude of emission from a maser `spot' can be a strong function of the direction to the line of sight. 
Strong class II methanol masers have several spectral components or spots so that statistically the beaming can be assumed to be homogeneous.
However, for lower luminosity maser sources with fewer components, beaming may have a more significant, but difficult to quantify, effect on the source counts.

Recent VLBI studies \citep[e.g.][]{Bartkiewicz2009TheObservations,Bartkiewicz2014EuropeanObjects,Bartkiewicz2016EuropeanMasers,Pandian2011THEMORPHOLOGIES,Fujisawa2014ObservationsObservations} have found 6.7\,GHz masers with a range of morphologies consistent with several different physical origins of the maser emission within a protostellar clump. 
One of these maser morphologies is a ring structure which has been used to suggest that some methanol masers arise in the disks associated with young high-mass stars. 
Naively, it might be thought that masers associated with some of these structures, such as disks, for example, might have a limited range of preferred beaming angles. 
However, VLBI observations have shown ring-like structures at a range of inclination angles \citep[e.g.][]{Moscadelli2010RevisingW3OH,Bartkiewicz2016EuropeanMasers,Sanna2017PlanarCepheusAHW2}. 
Therefore it is unlikely that there is a consistent preferred beaming direction within most maser sites and we consequently assume isotropic emission. 

\subsection{Summary of results}
\label{sec:IR_summary}
The above analysis finds, on average, that protostellar clumps hosting class II methanol masers are hotter, more luminous in both bolometric and single wavelength luminosities, more massive and at greater mass surface densities than the general Galactic population of protostars, although they are at a comparable luminosity-to-mass ratio with a narrower distribution of values.
The median properties and standard deviations for all properties are listed in Table \ref{tab:ir_results_all} for the maser hosting and protostellar objects, alongside the threshold set on each property as the clump requirements to host a methanol maser.

For comparison, the values obtained by \citet{Urquhart2015TheDust} in their study of the ATLASGAL 870\microns dust clumps associated with class II methanol masers have also been included in this table. 
For their derivation of clump properties, a single-temperature greybody with fixed $T=20$\,K was assumed.
As for the protostellar objects, \citet{Urquhart2015TheDust} use a difference dust mass opacity which results in systematically increasing our masses by a factor of 1.8 ($\sim10^{0.3}$) relative to their results.
The average mass derived for maser host clumps here is slightly smaller than the value derived by \citet{Urquhart2015TheDust}.

Converting their average H$_2$ column density of $N_{\text{H}_2}=10^{22.74}$\,cm$^{-2}$ to a mass surface density using
\begin{equation}
\label{eq:coldens_to_surfdens}
\Sigma = \frac{71}{\pi}\text{\Msun}\text{\,pc}^{-2}\left[ \frac{N_{\text{H}_2}}{10^{21}\text{\gpercmsq}}  \right]
\end{equation}
and a conversion factor of 4800\Msun\,pc$^{-2}$ = 1\gpercmsq \citep{Kauffmann2010THECLOUDS}, we obtain a mass surface density of $\log_{10}(\Sigma\,[\text{\gpercmsq}])$ of -0.59$\pm$0.51. 
Although both mass and mass surface density are consistent within errors, a direct comparison of the two results is not particularly meaningful as the significantly lower value can be mostly attributed to the relatively larger radius averaged over each clump \citep[average size 0.93\,pc in][compared with 0.42\,pc in the current work]{Urquhart2015TheDust}.
The assumption of $T=20$\,K will also modify the results for individual sources, but as this is comparable to the median temperature for maser hosts of 19.5\,K, it is unlikely the cause of a systematic offset in mass surface density between the two studies.

\begin{table*}
    \centering
    \begin{tabular}{l | @{\extracolsep{1cm}}  c  @{\extracolsep{12pt}}  c  @{\extracolsep{1cm}}   c @{\extracolsep{12pt}} c @{\extracolsep{1cm}}  c c @{\extracolsep{12pt}} c}
        \hline
        Property & \multicolumn{2}{c}{Maser} & \multicolumn{2}{c}{Protostellar} & \multicolumn{1}{c}{Threshold} & \multicolumn{2}{c}{Urquhart et al. (2015)} \\
        & Median & Std. dev. & Median & Std. dev. & & Median & Std. dev. \\
        \hline
        $T$ [K]                       & 19.5  & 4.6    & 15.3   & 5.6     & 12.5      & (20)  & -  \\ 
        $\log_{10}$($\Sigma$ [\gpercmsq])   & -0.12  & 0.50   & -0.47   & 0.60      & -1.12     & -0.59  & 0.51  \\
        $r$ [pc]$^{\dagger}$                      & 0.42  & 0.23   & 0.19   & 0.21    & 0.11      & 0.93  & 0.96 \\
        $\log_{10}$($M$ [\Msun])$^{\dagger}$      & 3.16  & 0.66   & 2.31   & 0.82    & 1.84      & 3.28  & 0.67 \\
        $\log_{10}(L_{\mathrm{FIR}}$ [\Lsun])$^{\dagger}$      & 3.75  & 0.66   & 2.63   & 0.99    & 2.42      & -     & -    \\ 
        $\log_{10}$($L_{70}$ \Lsun])        & -8.85  & 0.71   & -10.52  & 1.08      & -10.27    & -  & -  \\ 
        $\log_{10}$($L_{160}$ \Lsun])       & -8.72  & 0.64   & -10.12  & 0.95      & -10.02    & -  & -  \\ 
        $\log_{10}$($L_{250}$ \Lsun])       & -8.74  & 0.62   & -9.95   & 0.86      & -9.97     & -  & -  \\ 
        $\log_{10}$($L/M$ [\Lsun/\Msun])$^{\dagger}$      &  0.62  & 0.52   & 0.47    & 0.77      & -0.41, 1.65   & -  & -  \\
        {[70-160]} colour           & -0.11  & 0.24  & -0.35  & 0.37    & -0.58     & -  & -  \\
        {[160-250]} colour          &  0.01  & 0.15  & -0.13  & 0.30    &  -0.30    & -  & -  \\
        {[70-250]} colour           & -0.11  & 0.33  & -0.52  & 0.52    & -0.77     & -  & -  \\
        \hline
    \end{tabular}
    \caption{Summary of all properties derived from the Hi-GAL fluxes for the clumps hosting 6.7\,GHz class II methanol masers, and the corresponding properties for the general Galactic protostellar population. Following the median and standard deviation of a property for each set of objects, the lower threshold set as the clump requirement to host a methanol maser given by the 2-$\sigma$ width of the maser host distribution are given. For $L/M$, an upper threshold is also given. The final two columns give the values derived from dust emission towards 6.7\,GHz host clumps from \citet{Urquhart2015TheDust}, where a dust temperature of 20\,K was assumed for all clumps and we have converted the reported average column density to a mass surface density by Equation \ref{eq:coldens_to_surfdens}.\newline$^{\dagger}$ These properties are given for sources at near ($<8$\,kpc) and far ($>8$\,kpc) distances separately in Table \ref{tab:nearfar}. }
    \label{tab:ir_results_all}
\end{table*}


\section{Requirements to host a methanol maser}
\label{sec:requirements}

From the results presented in Section \ref{sec:FIRresults}, the higher temperature and stronger flux at 70\microns that we find for maser-bearing clumps particularly highlights that the differences in properties may be due to evolutionary differences.
The distinct band occupied by the methanol maser hosts within the total $L/M$ range of the protostellar sample may imply that methanol masers are associated with a shorter-lived phase of protostellar evolution, appearing once an embedded protostar is sufficiently evolved to sustain the maser pumping but dispersing before the protostar reaches its final mass, possibly a result of the destruction of methanol in the circumstellar environment.

The thresholds set throughout Section \ref{sec:FIRresults} on each property are used to identify sources capable of hosting a class II methanol maser based on the FIR host clump properties. 
Applying each of the thresholds and removing clumps associated with methanol masers from the protostellar sample, which are subject to the selection criteria stated at the start of Section \ref{sec:FIRresults}, 896 (out of 5497, 16 per cent) of the protostellar clumps are found to satisfy all requirements yet have no detected 6.7GHz maser within them.
With 731 of methanol maser host clumps detected as protostellar sources with Hi-GAL, and 647 with SEDs appropriate for analysis, this implies that class II methanol masers are detected towards only approximately half of the clumps which, based on the properties discussed here, could host a maser. 

We investigate the differences between maser hosting and this subset of non-maser clumps further by first searching for segregation between the types of objects in any combination of FIR properties, followed by a comparison of these two types of objects for sources of similar luminosities.
The findings of this analysis and possible underlying physical causes for sources to lack a detected 6.7\,GHz methanol maser are then discussed in Section \ref{sec:nonmaser_interp}.

\subsection{Principal component analysis}

\begin{figure}
	\includegraphics[width=\columnwidth]{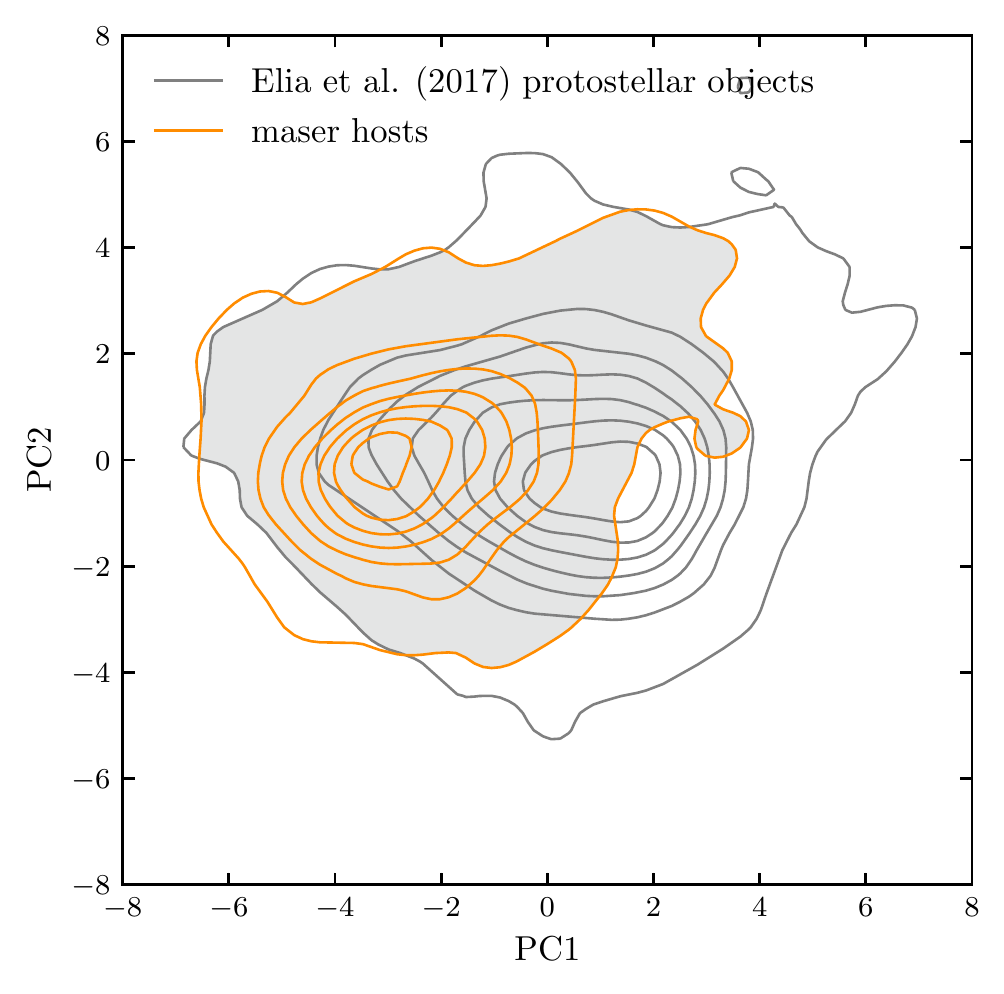}
    \caption{The distributions of maser (orange contours) and non-maser (grey contours) sources in the plane of the first and second principal component. The shaded overlap indicates the 896 sources maser and non-maser sources with identical properties in the 8-dimensional infrared parameter space defined by the input to the PCA. For each set of sources, the contour levels are at intervals of 15\% between 1\% and 90\% of the peak density in PC1-PC2 space.}
    \label{fig:PCAspace}
\end{figure}

When studying objects defined by a large number of properties (high-dimensionality), identifying trends between every possible combination of parameters is an incredibly lengthy task. 
Although we have presented the individual properties of the methanol maser host clumps and a select few combinations of these in Section \ref{sec:FIRresults}, this is not a full investigation of mutual trends between the properties.
Failing to consider all combinations may result in missed significant differences between maser and non-maser protostellar objects, such as finding maser objects to be at greater mass and luminosity individually but finding them to be restricted to a narrow middle band in luminosity-to-mass ratio.

Principal component analysis (PCA) provides a method with which to reduce the dimensionality of the data by defining new variables along mutual trends in parameters, so that we only need to consider a few variables to fully define the clumps.
In addition to identifying trends in several parameters within a set of data, PCA may also be used to identify differences in two sets of objects if they are sufficiently offset in the original parameter space.
Due to the large overlap in each property seen in Section \ref{sec:FIRresults}, we do not expect to find clear separation in the new parameter space but instead define a region of overlap between the two samples to identify a sample of objects matched in all properties to the methanol maser hosts, but lacking a maser.

Both the maser and non-maser protostellar clumps are taken to be defined by eight properties: $L_{70}$, $L_{250}$, $[70-160]$ colour, $[70-250]$ colour, $L_{\mathrm{FIR}}$, $M$, $r$ and $T$.
These are used as the original variables, chosen as they are both intrinsic properties and not necessarily related through combinations of each other.
PCA is a linear technique and to allow for power-law dependencies between parameters, logarithms of $L_{70}$, $L_{250}$, $L_{\mathrm{FIR}}$, $M$ and $r$ were taken prior to input to the PCA.
We find only the first two principal components, PC1 and PC2, to be significant trends.
PC1 represents a mutual increase in  $L_{\mathrm{FIR}}$, $L_{\mathrm{70}}$, $L_{\mathrm{250}}$ $M$ and $r$ and therefore represents the trend in observed parameters with the scale of a clump.
An independent trend in increasing $[70-160]$ and $[70-250]$ colours and temperature with decreasing mass and radius forms PC2, and we interpret this as a likely evolutionary trend.
Further discussion of the individual components and techniques is included in Appendix~\ref{append:pca}.

Figure \ref{fig:PCAspace} displays the first and second principal components plotted against each other with points categorised by sample, showing clear separation in the peak location of the maser and non-maser objects, although there is significant overlap.
The non-maser objects falling within the contours occupied by the maser sources are the 896 satisfying the conditions to host a methanol maser in all infrared parameters.
This large overlap (an approximately equal numbers of maser and non-maser sources) confirms that whilst there are significant statistical differences between the populations of maser and non-maser protostellar objects, we cannot easily identify whether a MYSO will host a methanol maser through any combination of the derived infrared parameter.

\subsection{Luminosity matched comparison}
\label{sec:lmatched}

Although 896 of the non-maser sources meet all individual requirements on each IR-derived property to host a methanol maser, a comparison of these sources against maser sources of the same luminosity excludes the scatter introduced by considering objects of vastly different size, evolutionary state or energy output.
This allows us to examine any subtle differences between protostellar clumps that are matched in their IR properties to methanol maser hosts, but lack a detection with the MMB survey.

\begin{figure}
    \centering
    \includegraphics[width=\columnwidth]{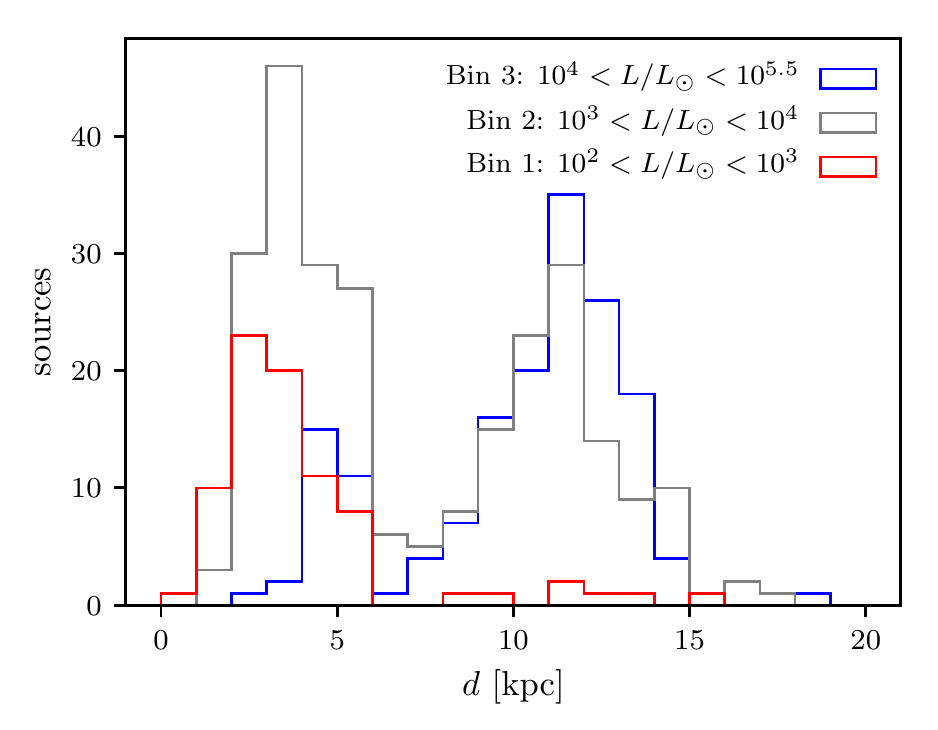}
    \caption{Heliocentric distance distributions of the MMB sources with far-infrared clump luminosities between $10^2 - 10^3$\Lsun, $10^3 - 10^4$\Lsun and $10^4 - 10^{5.5}$\Lsun for bins 1 (red), 2 (grey) and 3 (blue) respectively.}
    \label{fig:lbin_dist}
\end{figure}

The sources are split into three bins in $L_{FIR}$: $10^2 - 10^3$\Lsun, $10^3 - 10^4$\Lsun and $10^4 - 10^{5.5}$\Lsun.
These bins were chosen to include a sufficient number of sources in each bin to determine differences with significance and to include the majority of the high-luminosity tail.
The distributions of luminosity within each bin are found to be identical between the two types of object, therefore excluding any luminosity biases in each bin as a possible cause for any other property differences.
However, the distance distribution is shown in Figure \ref{fig:lbin_dist} for the MMB sources in each luminosity bin and the majority of sources in the low-luminosity bin are found at relatively small distances.
As discussed in Section \ref{sec:luminosity_results}, small distances carry a much greater uncertainty and the distance-dependent properties of these sources, such as luminosity, are less reliable compared to sources at greater distances.
The average properties derived here for the sources in the lowest luminosity bin are therefore unlikely to give an accurate characterisation of methanol maser sources with luminosities $<1000$\Lsun.

The restricted non-maser sample is compared against the maser sources in each other IR property with the median values shown in Table \ref{tab:lmatched}.
Unlike the previous comparisons shown, the differences in the median value of each property between the non-maser and maser sources are less distinct in comparison to the offsets reported in Section \ref{sec:FIRresults}.
The KS test p-values between the two categories of objects are also given to determine whether a difference in median likely corresponds to a difference in the underlying distribution.


\begin{table*}
	\centering
	\caption{The median far-infrared derived properties of non-maser and maser sources when binned by luminosity to allow comparison of sources with an approximately equivalent energy output. For each bin, the median for each category is shown alongside the KS p-value for the two distributions.}
	\begin{tabular}{l | l l l | l l l | l l l }
		\hline
		Property & \multicolumn{3}{l}{Bin 1 ($10^2-10^3$\Lsun)}  & \multicolumn{3}{l}{Bin 2 ($10^3-10^4$\Lsun)} & \multicolumn{3}{l}{Bin 3 ($10^4-10^{5.5}$\Lsun)} \\ 
		& Non-maser & Maser & KS p-value & Non-maser & Maser & KS p-value & Non-maser & Maser & KS p-value \\ \hline
		No. sources                         & 96 & 80 & -                 & 567 & 258 & -           & 217 & 165 & - \\
		T [K] 		                        & 15.4  & 17.8 & $10^{-7}$  & 17.0 & 19.2 & $10^{-15}$  & 19.8 & 21.6 & $10^{-5}$ \\
		$\Sigma$ [\gpercmsq]                & 0.24  & 0.44 & 0.003      & 0.34 & 0.70 & $10^{-13}$  & 0.66 & 1.08 & $10^{-4}$ \\
		M [\Msun] 	                        & 232   & 185  & $10^{-4}$  & 465  & 991  & $10^{-17}$  & 1629 & 3873 & $10^{-12}$\\
		L$_{70}$ [$10^{-10}$\Lsun]  	    & 0.74  & 1.17 & $10^{-9}$  & 2.84 & 9.11 & $10^{-34}$  &21.72 &56.26 & $10^{-21}$\\ 
		L$_{160}$ [$10^{-10}$\Lsun]  	    & 1.43  & 1.61 & 0.002      & 4.48 & 12.71& $10^{-39}$  &25.43 &63.73 & $10^{-32}$\\ 
		L$_{250}$ [$10^{-10}$\Lsun]  	    & 1.74  & 2.00 & 0.020      & 4.55 & 13.74& $10^{-39}$  &25.13 & 58.34& $10^{-23}$\\ 
		{[70-160]}\microns            		& -0.28 & -0.18& $10^{-4}$  & -0.19& -0.12 & $10^{-5}$  &-0.01 &-0.06 & 0.061 \\ 
		{[70-250]}\microns            		& -0.38 & -0.24& $10^{-4}$  & -0.19& -0.13 & 0.009      & 0.02 & -0.01& 0.198\\ 
		{[160-250]}\microns            		& -0.11 & -0.06& 0.076      & -0.02& 0.00  & 0.074      & 0.03 & 0.05 & 0.065\\ \hline
	\end{tabular}
\label{tab:lmatched}
\end{table*}

For the reasons above, we consider the statistics for the second and third luminosity bins to best reflect the physical conditions of each sample.
The most notable difference between the maser and protostellar objects, both in difference in the median and statistical significance, is seen in 70\microns luminosity with a factor of 3.2 increase in the median for maser hosts over the protostellar objects in the mid luminosity bin between $10^3$ and $10^4$\Lsun.
We also see this for the other specific luminosities at 160 and 250\microns, although the contrast between the two median values is slightly smaller. 
As noted in Section~\ref{sec:luminosity_calc}, this may be due to the inclusion of additional wavelengths during the luminosity calculation for the protostellar sample.
However, the factor of two due to this is not sufficient to account for the offset in the specific luminosities, and therefore this implies that the maser sources are stronger in these shorter wavelengths than the protostellar objects relative to their total luminosity. 
This trait is also reflected in the [70-160] and [70-250] colours which trace the inner colour temperature of a clump, and the overall envelope temperature although these results are less statistically significant.

\subsection{Physical interpretation}
\label{sec:nonmaser_interp}

A number of physical scenarios may lead to some clumps hosting class II methanol masers and others lacking this feature.
Firstly, we discount the possibility that only the most extreme (e.g. most massive, most luminous, etc) sources may host a methanol maser as all sources fall above the thresholds on each property to host a methanol maser. 
Notably, we find that sources down to a threshold of $\sim0.08$\gpercmsq are capable of hosting a methanol maser and therefore a massive protostar, comparable to the threshold of approximately 0.05\gpercmsq for massive star formation found by \citet{urquhart2018atlasgal}.
As 6.7\,GHz methanol masers exclusively trace high-mass protostars, this suggests that this all sources above this threshold on mass surface density may form high-mass stars.
We would also not expect to recover differences in typically evolutionary properties such as temperature and luminosity-to-mass ratio if this were the case.
Additionally, if the two types of object are distinct and only those above some threshold in any combination of properties host methanol masers, separation of the two categories in principal component space would be seen.

There are three scenarios that our analysis does not consider, as each would require follow-up spectral observations towards a large number of sources to confirm.
A possible case is that the two types of objects are not distinct and that the non-maser sources do have masers but they are not present in the MMB catalogue due to low strengths or a beaming angle of the maser away from the line of sight.
As 6.7\,GHz maser strength shows correlation with the 70\microns luminosity (Section \ref{sec:IRvsMaser}), it is possible that the non-maser sources host weak methanol masers due to their comparatively weak 70\microns emission.
The survey of \citet{Pandian2007TheData} has a flux completeness limit a factor of 1.9 lower than the MMB and detected 12 additional weak maser sources which are not in the MMB survey \citep{Breen2015The2060}. 
In the same longitude range the MMB survey detected 93 sources, so the weaker sources correspond to only 13\% of the MMB population. 
This suggests that if the non-maser sources (which are similar in total number to the MMB sources) do in fact have weak masers, these masers are considerably below the MMB sensitivity, or indeed that of \citet{Pandian2007TheData}. 
The deeper `piggyback' survey carried out with the MMB receiver will provide further insight on the population of weak masers (Ellingsen et al., in prep.). 

The greater variability of weak 6.7\,GHz masers compared with strong masers may also contribute to weaker sources being missed by the MMB survey.  
However, even for weak maser sources, the typical variability is not extreme, with a measured mean variability of 36 per cent for sources of peak maser flux of $<1.5$\,Jy \citep{Green2017TheMasers}. 

Finally, due to either evolutionary effects or the properties of the natal environment of a clump, chemical differences in the circumstellar region may prevent a protostellar source from masing at 6.7\,GHz.
Although plausible, these scenarios cannot fully account for the differences in the distributions of IR properties that we find between the two categories.

\begin{figure*}
	\includegraphics[width=0.9\textwidth]{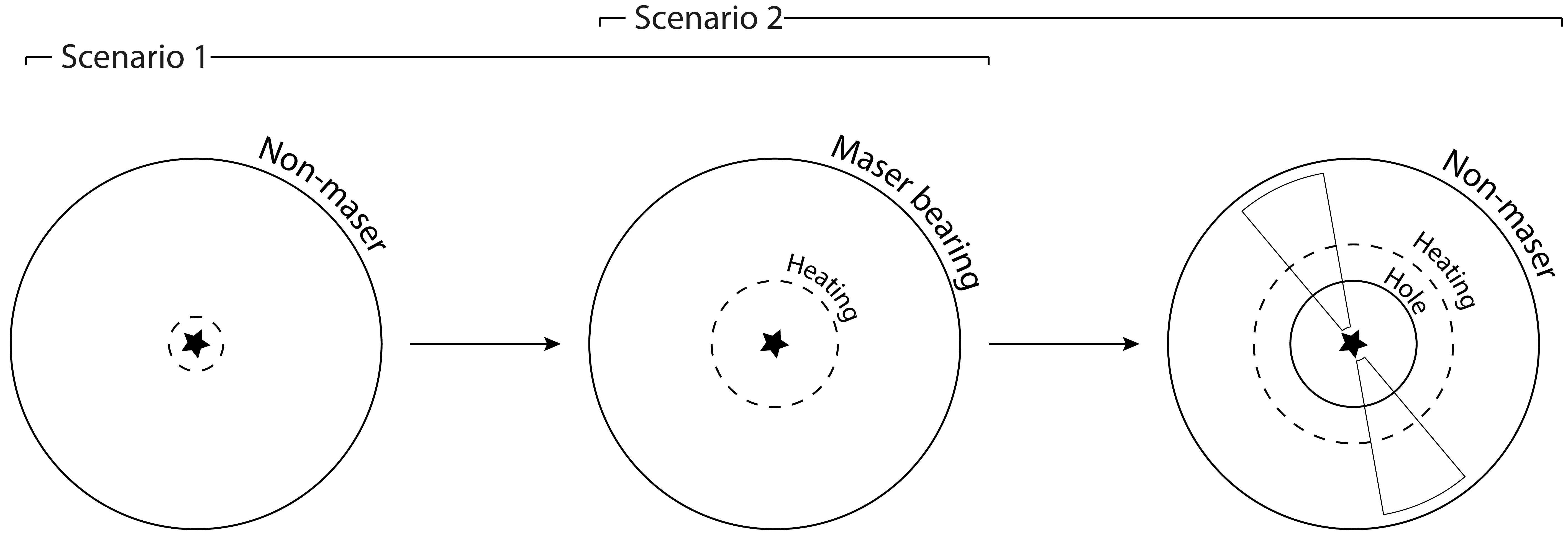}
    \caption{The two possible evolutionary scenarios considered in this work to account for the lack of detection of a methanol maser towards protostellar sources of identical infrared properties to clumps hosting methanol masers.
    Evolution increases from left to right, with the solid lines showing the location of the dusty envelope material and the dotted lines indicating heated regions around a central protostar (filled star).
    With a methanol maser source represented by the middle evolutionary state, scenario 1 considers the non-maser sources to fall at an earlier stage of evolution, and scenario 2 considers them to be at a later stage of evolution, in which a central hole in the dusty material has been created and outflows may be present (cones).}
    \label{fig:scenarios}
\end{figure*}

Shown in Figure \ref{fig:scenarios}, we consider two evolutionary scenarios in this work that may cause the observed differences in the infrared properties of maser and non-maser sources, with evolution advancing from left to right:
\begin{enumerate}[label=\arabic*)]
    \item Maser sources (middle) are at a more advanced stage of evolution and have had longer to irradiate and heat their envelopes than the protostellar objects (left).
    \item The protostars in non-maser sources have cleared their immediate surroundings of dust and possibly methanol (right) and are more evolved than the maser sources (middle).
\end{enumerate}
Scenario 1 is supported by our findings that maser sources have warmer envelope temperatures for a given luminosity and more prevalent emission from a warm inner region at 70\microns, as indicated by greater $L_{70}$, [70-160]\microns colour and [70-250]\microns colour.
Scenario 2 instead implies that the non-maser sources are at a more advanced stage of evolution than maser hosts, and have sufficiently cleared either the methanol or hot dust from their surroundings required for methanol maser emission. 
As the 70\microns emission from a MYSO originates from hot dust in the immediate protostellar environment, it is therefore also feasible that a decrease in $L_{70}$ indicates that a source is sufficiently evolved to remove this dust, rather than failing to strongly heat this dust.
Removing the contribution of the warm inner component to the FIR SED of a source may also result in a lower clump-averaged temperatures recovered from fitting, and is a potential origin of the difference found in temperature between the non-maser and maser sources.

The MIR data can be used to probe which scenario is more likely correct.
The emission from the central regions of a protostellar object is severely attenuated by any dusty envelope material along the line of sight.
Initially, such as in the first scenario we are considering, very little of this emission is expected to escape.
In the final evolutionary state shown in Figure \ref{fig:scenarios}, the cleared central region results in a reduced column density of material that the MIR emission must pass through to escape, resulting in brighter emission from the object.
In addition to this, lines of sight through the envelope into the central regions may be cleared by outflows, also making brighter emission more likely towards objects in this evolutionary state.
However, this may also result in the 8\microns emission being extended, and so the source may not be included in the GLIMPSE point source catalogue.
The bias towards nearer distances for the non-maser sources (Section~\ref{sec:FIRresults}) may also contribute to their 8\microns counterparts being too extended to be included in the GLIMPSE point source catalogue.

\begin{figure}
	\includegraphics[width=\columnwidth]{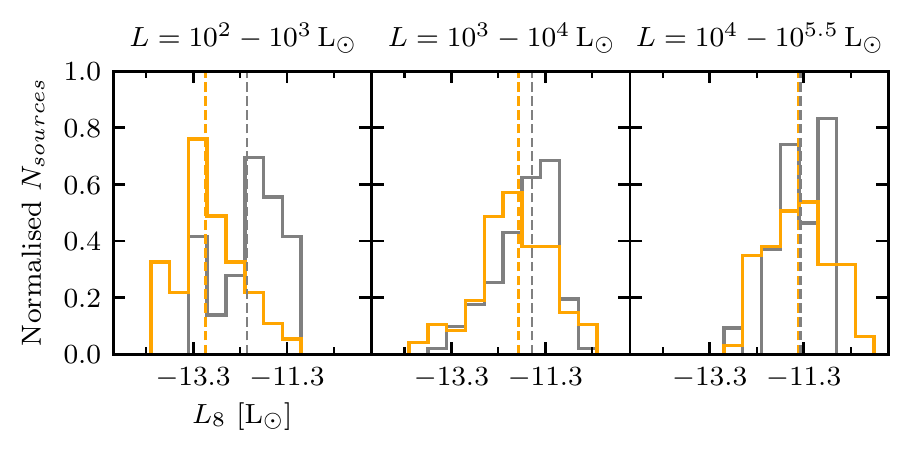}
    \caption{The distribution of 8\microns spectral luminosities for maser (orange) and non-maser (grey) sources with a counterpart in the GLIMPSE point source catalogues, binned by FIR luminosity (increases left to right). The number of sources in each distribution and the values of the marked medians are given in Table \ref{tab:mircat}.}
    \label{fig:8mu}
\end{figure}

\begin{figure}
	\includegraphics[width=\columnwidth]{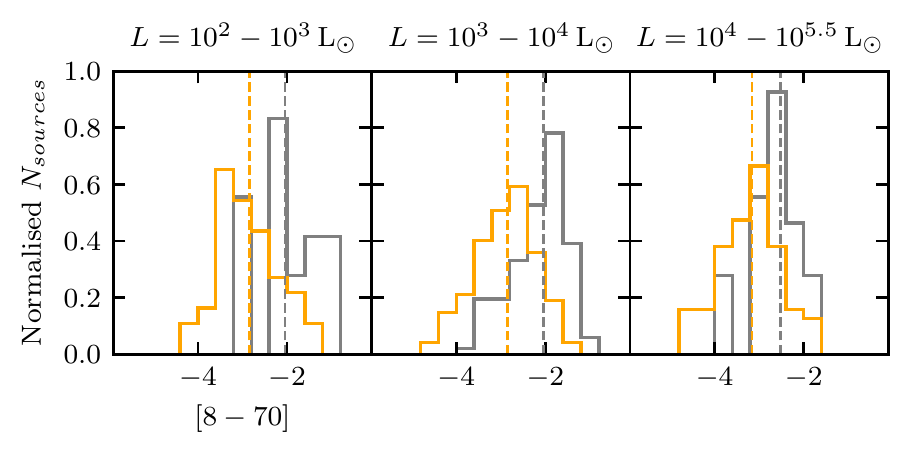}
    \caption{The distributions of [8-70]\microns colour for maser (orange) and non-maser (grey) sources binned by FIR luminosity with a counterpart in the GLIMPSE point source. The distribution medians are given in Table \ref{tab:mircat}.}
    \label{fig:8-70}
\end{figure}

Associating the non-maser sources with objects in the GLIMPSE 8\microns catalogue, as done for the maser sources, 173 of the 896 (19.3\%) non-maser sample have an 8\microns counterpart.
This is notably smaller than the 48.3\% of maser sources found with 8\microns GLIMPSE point sources counterparts.
The distributions of 8\microns luminosities for the maser and non-maser objects are shown in Figure~\ref{fig:8mu} and summarised in Table~\ref{tab:mircat}. 
On average the 8\microns luminosities of the non-maser sources are found to be greater than the maser sources, a result evident as a significant difference through the use of KS tests.
There is a more noticeable difference between the categories in the [8-70]\microns colour, shown in Figure~\ref{fig:8-70}.

\begin{table*}
	\centering
	\caption{Median values for the mid-infrared properties derived from 8\microns data for the maser and non-maser objects with a counterpart in the GLIMPSE catalogues to investigate the likelihood of each scenario shown in Figure \ref{fig:scenarios}. The sources are binned in luminosity, with the sample median in each bin given for the two types of object, with the KS p-value reported in the bottom row. The corresponding distributions are shown in Figure \ref{fig:8mu} and \ref{fig:8-70}.}
	\begin{tabular}{l | l l l | l l l | l l l }
		\hline
		& \multicolumn{3}{l}{Bin 1 ($10^2-10^3$\Lsun)}  & \multicolumn{3}{l}{Bin 2 ($10^3-10^4$\Lsun)} & \multicolumn{3}{l}{Bin 3 ($10^4-10^{5.5}$\Lsun)} \\ 
		& No. & $L_8$ [$10^{-12}$\Lsun] & [8-70]\microns & No.& $L_8$ [$10^{-12}$\Lsun] & [8-70]\microns & No. & $L_8$ [$10^{-12}$\Lsun] & [8-70]\microns \\ \hline
		Non-maser 		    & 18    & 0.70 & -2.04      & 128  & 2.59 & -2.05       & 27   & 4.24 & -2.52    \\
		Maser               & 46    & 0.09 & -2.85      & 118  & 1.34 & -2.84       & 79   & 3.75 & -3.16    \\
		KS p-value	        & - & $10^{-8}$ & $10^{-8}$ & - & $10^{-9}$ & $10^{-14}$ & - & $10^{-11}$ & $10^{-11}$\\ \hline
	\end{tabular}
\label{tab:mircat}
\end{table*}

Finding the non-maser sources to have brighter 8\microns counterparts is unlikely if the underlying evolutionary scenario is that these sources are less evolved.
This therefore supports the suggestion that the matched non-maser sources are in a more evolved state.
However, it should be noted that a significant number of objects in each sample are not associated with a compact 8\microns counterpart.

As mentioned previously, there is a large overlap in all properties for the two types of clump and they are therefore not distinct types objects as seen in the infrared.
A detailed investigation of all differences responsible for this behaviour requires follow-up observations.
Observations to determine the chemistry of the two types of region would be beneficial, and detection of class II masers that are too weak to detect with the MMB or angled away from the line of sight would also result in re-categorisation of some non-maser objects.

To assess the plausibility of this evolutionary scenario, we consider the relative lifetimes of the two scenarios implied by the number of sources in each, taking the maser sources to occupy the mid evolutionary state in Figure~\ref{fig:scenarios} and the matched protostellar sources to occupy the final scenario. 
A comparison of the total number of sources in each category requires quantification of the sources missing from the total count for both the maser-bearing sources and the catalogue of \citet{Elia2017The67_.circ0}.
In Section~\ref{sec:completeness} we determined that the MMB is complete in methanol masers hosted by clumps of $>100$\Lsun within 2.7\,kpc, and that we do not miss a significant number of nearby methanol maser sources due to sensitivity. 
However, the number of clumps hosting methanol maser sources requires correction for by the percentage not detected over multiple wavelengths in Hi-GAL as well as the percentage of the sample that lacks a reliable distance value. 
As we do not detect 33\% of maser host clumps over all required Hi-GAL bands and lack a distance for 18\% of the sources in the sample, the total correction factor to obtain the true number of methanol maser hosts is 1.8.
A factor of 4 correction is required for the protostars to account for those objects without reliable distance determinations. 

Within 2.7\,kpc we find 50 maser sources and 9 matched non-maser sources, corrected to 90 and 36 sources respectively following the above. 
This implies the lifetimes to be within a factor of approximately 2.5, which we take to be reasonable for the proposed evolutionary scenario.

In the case that a significant population of 6.7\,GHz masers have a preferred beaming direction (such as a circumstellar disks with a restricted beaming angle centred on the disk planes), we would expect a population of matched maserless sources identical to the host clumps.
However, the sources without a detected maser are found to be weaker at 70\microns, less likely to be associated with a point-like 8\microns counterpart and more luminous at 8\microns, and so are not an identical population to the maser host clumps.
This result suggests that it is unlikely that a significant fraction of the Galactic class II methanol maser population have a preferred beaming direction, consistent with recent VLBI observations (Sec.~\ref{sec:completeness}). 


\section{Mid-infrared results}
\label{sec:MIRresults}

\begin{figure}
	\includegraphics[width=\columnwidth]{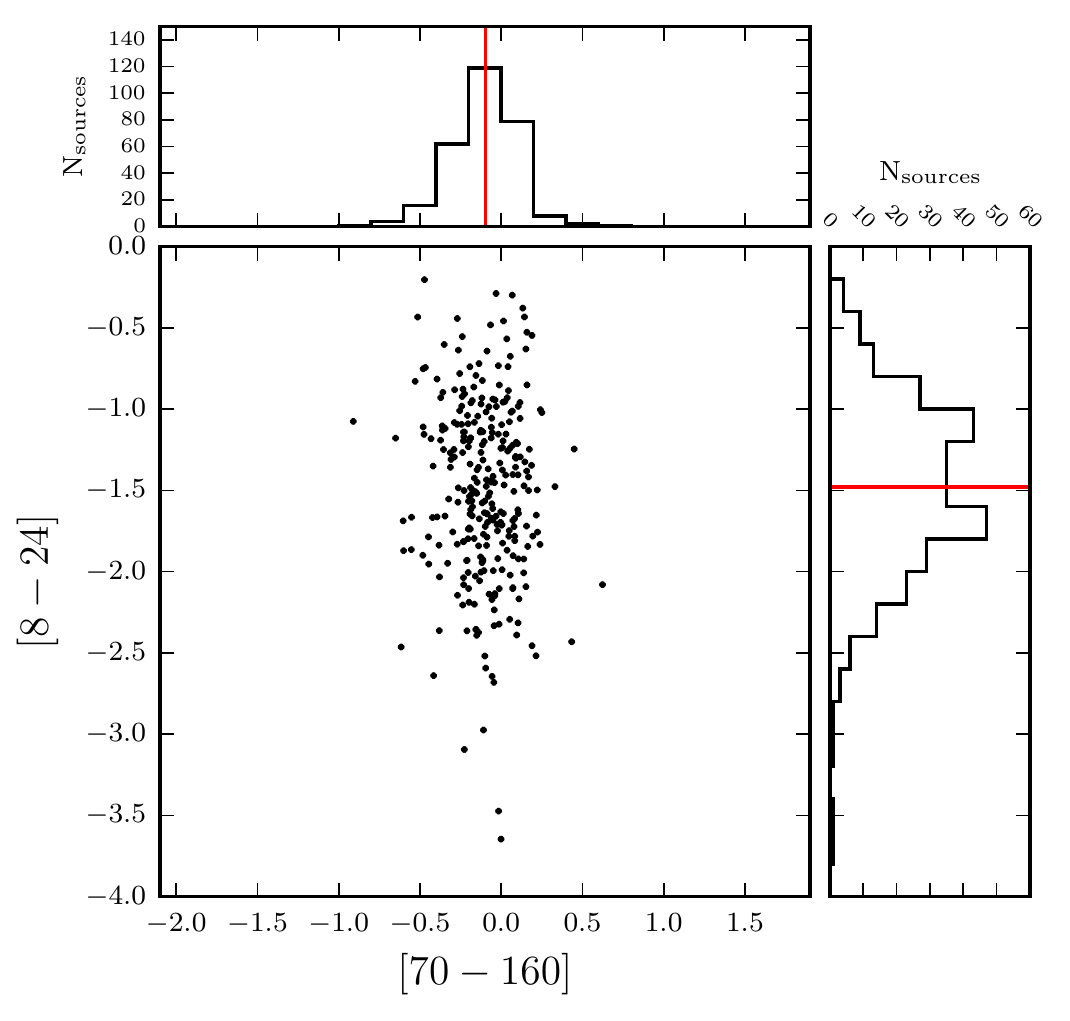}
    \caption{Colour-colour diagram of mid-infrared [8-24]\microns colour against [70-160]\microns colour shown in Section \ref{sec:colours} with the axis scale set to be equal. The respective marginal distributions are shown for each property, with the medians marked in red at -1.48 and -0.1 for [8-24]\microns and [70-160]\microns colour respectively, with corresponding 1-$\sigma$ widths of 0.55 and 0.20.}
    \label{fig:colour_colour_mir}
\end{figure}

\begin{figure}
	\includegraphics[width=\columnwidth]{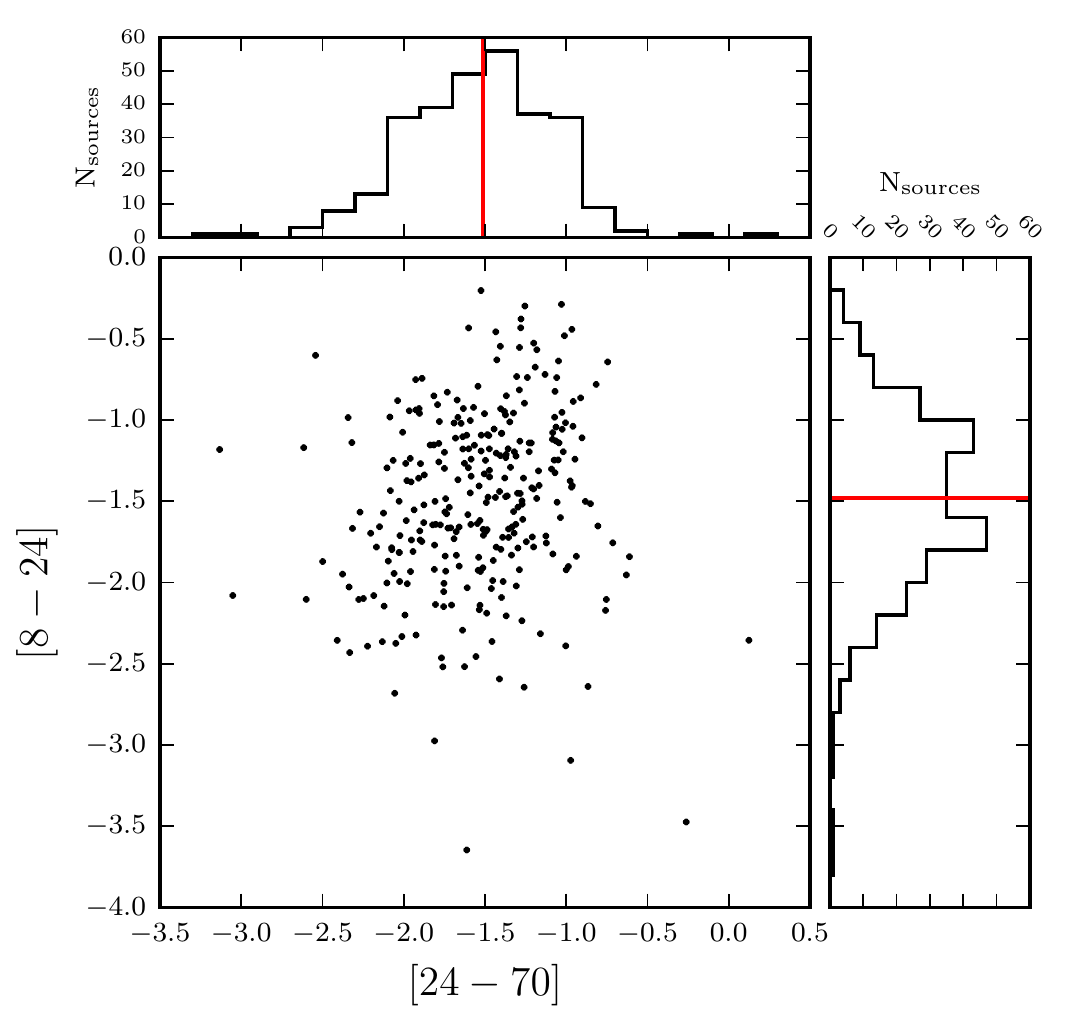}
    \caption{The [8-24]\microns colour plotted against [24-70]\microns colour with the axis scale set to be equal. As shown in the marginal distribution on the y-axis, the [8-24]\microns colour has a median of -1.48 (red) and standard deviation of 0.55. Similarly, the [24-70]/microns colour has a median of -1.51 and a standard deviation of 0.43.}
    \label{fig:colour_colour_mir2}
\end{figure}

For the 292 maser host clumps with a counterpart identified at both 8\microns and 22/24\microns, the colour-colour diagram between the [8-24]\microns and [70-160]\microns colour is shown in Figure \ref{fig:colour_colour_mir}, displaying no evidence of correlation.
For the maser sources with 8\microns and 24\microns counterparts identified, the [70-160]\microns colour is narrowly distributed with a 1-$\sigma$ of 0.20 about a median of -0.10.
The [8-24]\microns colour shows a larger scatter, with a median of -1.48 and standard deviation of 0.55.
A similar result is obtained for the [8-24]\microns and [24-70]\microns colour-colour diagram shown in Figure \ref{fig:colour_colour_mir2}, where the [24-70]\microns colour has a median of -1.51 and standard deviation of 0.43.
A likely cause of the increasing scatter at 8 and 24\microns is the dependence of flux on the viewing angle of the clump as the increase in opacity towards shorter wavelengths causes a high dependence on the amount of obscuring envelope material along the line of sight.
For example, a line of sight cleared of dust through outflow activity results in a greater flux than if the same object was observed perpendicular to the outflow axis.
Often observed to be associated with water maser emission, an outflow system such as this is a likely scenario for the clumps hosting methanol masers.

\begin{figure}
	\includegraphics[width=0.9\columnwidth]{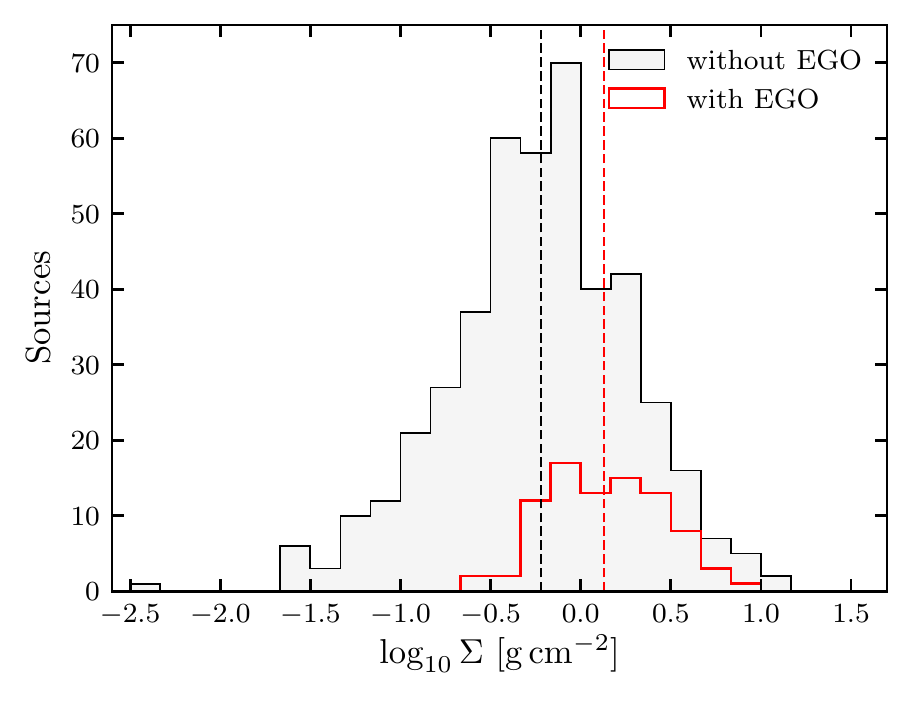}
    \caption{Distribution of surface densities for maser host clumps with (red) and without (grey) an associated EGO, with respective medians of 1.61\gpercmsq and 0.60\gpercmsq marked by the vertical lines.}
    \label{fig:egohist}
\end{figure}

The final mid-infrared data associated with the methanol masers in this work is the catalogue of EGOs produced by \citet{cyganowski2008catalog}, with counterparts again identified through positional association with a maximum separation of 5\,arcsec.
The survey limits of GLIMPSE, from which this catalogue is derived, contain 528 of the 647 methanol masers in the sample.
Of these, 86 are found to display EGO emission, yielding an association percentage of 16.3 per cent.
The 6.7\,GHz masers with an associated EGO are found to be indistinct in every derived property of the host clump aside from the mass surface density.
As shown in Figure \ref{fig:egohist}, the clumps with an EGO are found at a median mass surface density of 1.61\gpercmsq, which is significantly greater than that of those without an EGO (KS p-value $\sim 10^{-9}$) with a median value of 0.60\gpercmsq.

\section{Evolutionary differences in maser associations}
\label{sec:maserassoc}
\subsection{Identification of counterpart masers}
\label{sec:maserID}
All maser catalogues detailed in Section \ref{sec:extramaserdata} were used to find other maser emission arising from similar on-sky coordinates, and therefore the same protostellar host clump, for each 6.7\,GHz maser in the sample selected for this work.
The survey range for both the 12.2\,GHz and excited hydroxyl masers matches that of the MMB survey as the 12.2\,GHz study was a targeted follow-up to all MMB sources by \citet{breen201212,Breen201212.2-GHz186-330,Breen201412.2-GHz20,Breen201612.2-GHz2060} and the ex-OH masers reported by \citet{adam2016exohI} which were co-observed during the MMB survey.
An ex-OH maser is taken to be associated with a 6.7\,GHz source if the on-sky separation is less than 2.0\,\arcsec.
\citet{Breen2018TheHOPS} also identified associations between the 6.7\,GHz MMB masers and the water masers from the HOPS data, with positional association a more complex task due to the less localised nature of water maser emission.
The authors take this into account, and 6.7\,GHz methanol maser is associated with a water maser if a HOPS maser spot falls within 2.0\,arcsec.

A full analysis of maser associations and their number statistics over the full Galactic Plane is presented in \citet{Breen2018TheHOPS}.
For this work, we are only concerned with investigating population differences in FIR-derived properties of protostellar clumps hosting 6.7\,GHz class II methanol masers based on the presence of additional maser emission, i.e. to allow determination of whether certain masers define a particular evolutionary stage or subset of class II methanol maser hosts.
Within our sample of 647 Hi-GAL bright masers, 291 are listed to have observed 12.2\,GHz emission and from the sample of 127 ex-OH masers, 32 are associated with a clump in the Hi-GAL bright 6.7\,GHz host sample.
Over a limited range of $20^{\circ}$ in Galactic longitude, 31 of 143 clumps are also associated with ground state hydroxyl maser sites detected with the SPLASH survey.
From the 501 host clumps within the HOPS survey region, 122 are associated with water maser emission.
The percentage of methanol masers also associated with a water maser improves with the addition of the more sensitive \citet{Titmarsh2014A620} and \citet{Titmarsh2016A6} studies towards a subset of 217 of these clumps, giving 110 associations.

\begin{table*}
\centering
\caption{The median properties of 6.7\,GHz host clumps associated with a secondary maser of a given type, and for clumps lacking the same maser. The KS p-values are also given to assess whether the presence (or lack) of the maser holds any significance for each clump property.}
	\begin{tabular}{ll c c c c c c}
		\hline
		Maser & & Number & $T$ [K] & $\Sigma$ [\gpercmsq] & $M$ [\Msun] & $L_{\mathrm{FIR}}$ [\Lsun] & $L/M$ [\Lsun/\Msun] \\ \hline
		12.2\,GHz CH$_3$OH & with & 291 & 20.2 & 0.94 & 1440 & 6410 & 3.95 \\
		& without & 356 & 19.2 & 0.60 & 1270 & 4420 & 4.50 \\
		& p-value & - & 0.072 & $\sim10^{-5}$ & 0.12 & 0.004 & 0.15 \\\rule{0pt}{4ex}
        {OH} & with & 31 & 22.0 & 1.53 & 1806 & 14330 & 7.02 \\
        & without & 112 & 18.3 & 0.87 & 1538 & 4315 & 2.60 \\
        & p-value & - & $\sim10^{-4}$ & 0.003 & 0.14 & 0.025 & $\sim10^{-4}$ \\\rule{0pt}{4ex}
		{Ex-OH} & with & 32 & 22.2 & 1.61 & 1420 & 9040 & 7.82 \\
		& without & 615 & 19.4 & 0.74 & 1340 & 5220 & 4.04 \\
		& p-value & - & 0.012 & $\sim10^{-4}$ & 0.97 & 0.12 & 0.005 \\\rule{0pt}{4ex}
		{HOPS H$_2$O} & with & 122 & 21.1 & 1.36 & 1570 & 9100 & 5.69\\
		& without & 379 & 19.3 & 0.67 & 1290 & 4800 & 3.96\\
		& p-value & - & 0.002 & $\sim10^{-8}$ & 0.26 & 0.006 & 0.004\\\rule{0pt}{4ex}
		{T14/T16 H$_2$O} & with & 110 & 20.2 & 0.95 & 1884 & 8854 & 4.72\\
		& without & 107 & 17.9 & 0.72 & 2133 & 7649 & 2.79\\
		& p-value & - & 0.009 & 0.059 & 0.94 & 0.16 & 0.005\\
		 \hline
\end{tabular}
\label{tab:maser_assoc}
\end{table*}

\subsubsection{12.2\,GHz class II methanol masers}

As for the 6.7\,GHz maser line, the 12.2\,GHz class II methanol masers are radiatively pumped by the re-emission of the strong UV radiation from the embedded protostar as infrared by nearby dust, and therefore are also co-located with the central protostellar object in a MYSO.
This is the most common additional maser line observed towards the 6.7\,GHz host clumps, with approximately half (291 of 647) of the sample exhibiting both lines.
We find no difference between the 12.2\,GHz maser properties in the sample in this paper and the complete sample of 431 12.2\,GHz masers detected in the follow-up study \citep[][Section \ref{sec:extramaserdata}]{breen201212,Breen201212.2-GHz186-330,Breen201412.2-GHz20,Breen201612.2-GHz2060}.
Similarly, the percentage of 6.7\,GHz masers with an additional 12.2\,GHz detection is not found to be a function of distance and is approximately constant at 45 per cent. 
Both of the above indicate that we have not biased the sample of 6.7\,GHz masers also with 12.2\,GHz detections by selecting only those with visible Hi-GAL counterparts, and our sample is representative of the entire population.

\begin{figure*}
	\includegraphics[width=\textwidth]{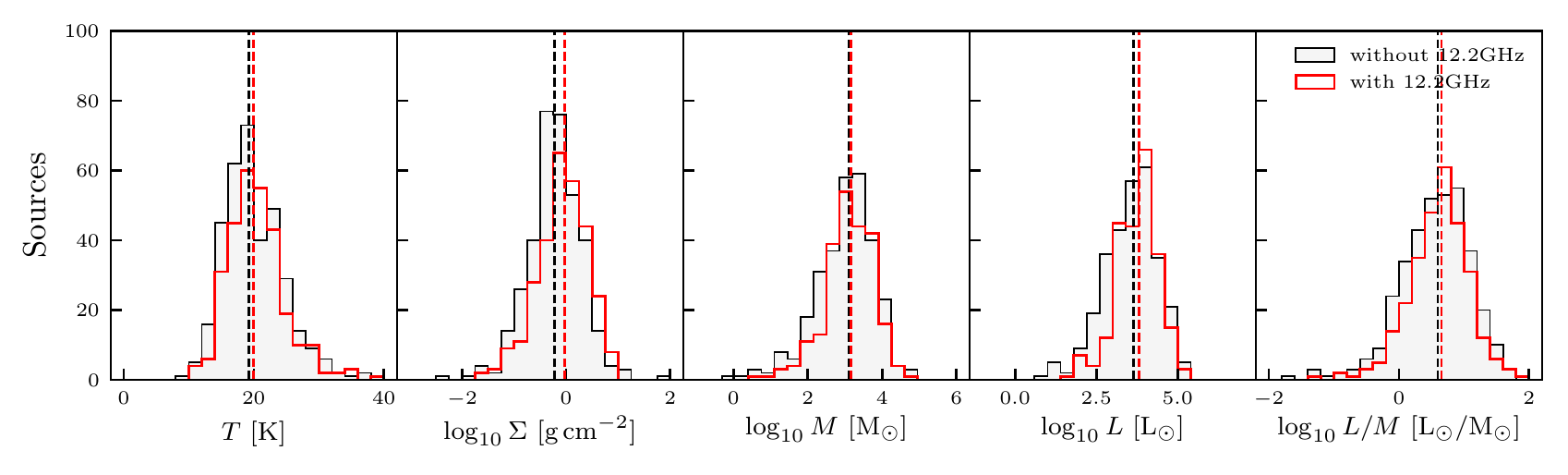}
    \caption{Distributions of the host clump properties derived in Section \ref{sec:FIRresults} categorised into clumps also associated with a 12.2\,GHz class II methanol maser detected (red) and those without (black). The median values of each distribution are given in Table \ref{tab:maser_assoc}.}
    \label{fig:12ghz}
\end{figure*}

Figure \ref{fig:12ghz} shows the distributions of the FIR-derived properties for the host clumps with and without 12.2\,GHz maser emission detected.
The medians and KS p-values shown in the top row of Table \ref{tab:maser_assoc} are used to determine which properties display a significant difference based on this categorisation.
A significant difference is found in the mass surface density (Figure \ref{fig:12ghz}, second panel from left).
Clumps emitting at 12.2\,GHz have at a median surface density of 0.94\gpercmsq whereas the clumps with no 12.2\,GHz emission have a much lower sample median of 0.60\gpercmsq, although this is still significantly higher than that of the general protostellar population.
The median luminosities are 6410 and 4220\Lsun respectively for clumps with and without 12.2\,GHz emission, and the host clumps with 12.2\,GHz are found to be more luminous (p-value of 0.004).
Although there is a small increase in median temperature in sources which host a 12.2\,GHz maser, we do not find this to be statistically significant, similar to the result found by \citet{Breen2018TheHOPS} for similar samples. 

\subsubsection{Hydroxyl masers}

Ground-state and excited-state OH maser lines are pumped by a combination of collisional and radiative processes \citep[predominantly the latter,][]{Field1988TheCase,Cragg2002ModellingRegions}. The ground state are found offset from the central protostar, typically associated with the edge of the enveloping ultra-compact H{\sc ii} region, whilst the excited-state are more closely bound to the protostar \citep[e.g.][]{Fish2007Expanded1, Caswell2009Maser300.969+1.147, Chanapote2019TracingArray}.
Due to the association with the formation of an H{\sc ii} region, we expect to find IR properties typical of the more advanced stages of protostellar evolution towards objects with a hydroxyl maser.

\begin{figure*}
	\includegraphics[width=\textwidth]{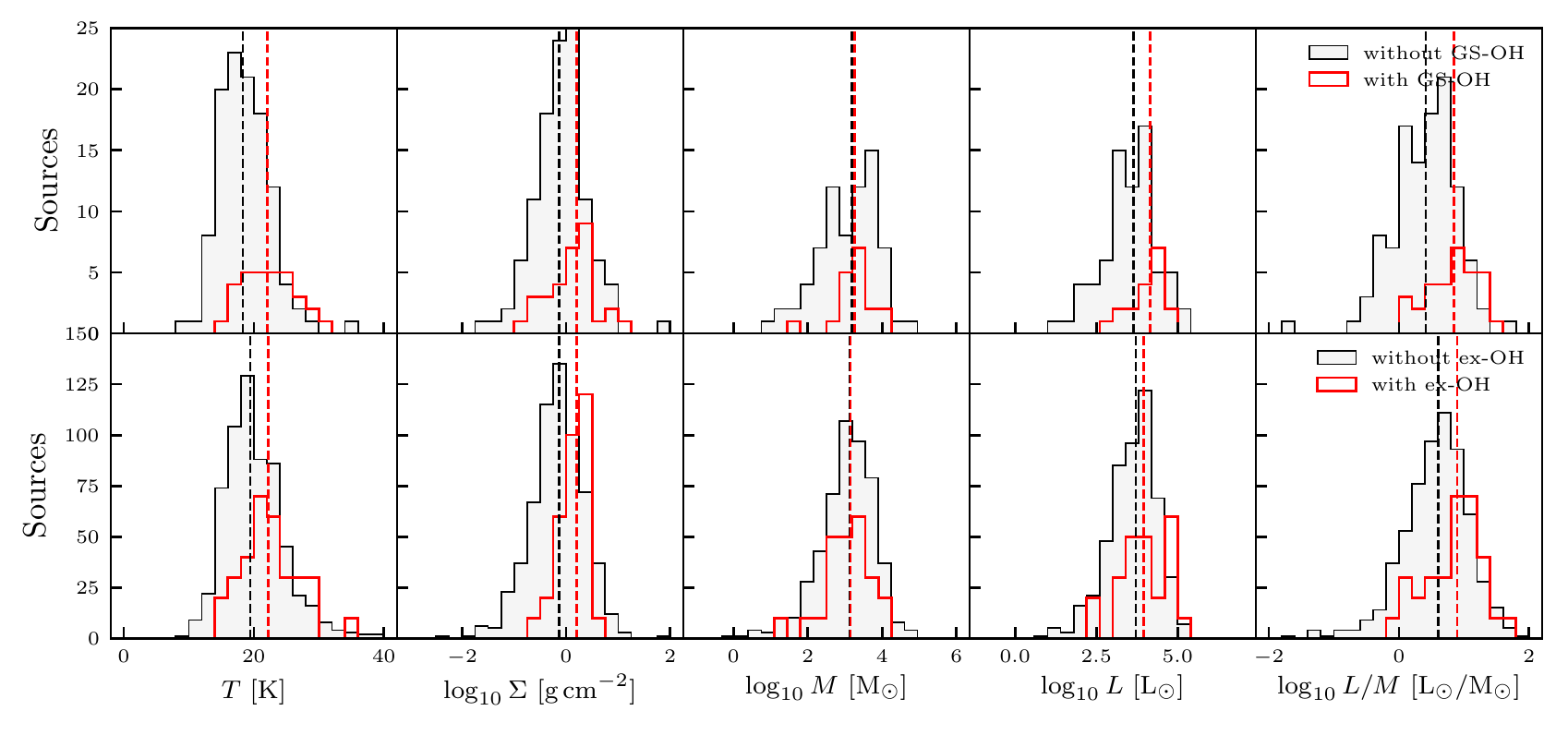}
    \caption{For clumps with (red) and without (black) a secondary hydroxyl maser detection, the distributions of host clump properties are shown with the marked median values given in Table \ref{tab:maser_assoc}.
    The top row shows the distribution of secondary ground state hydroxyl masers detected for clumps within the SPLASH survey region. The bottom row displays the distribution of properties for clumps with excited state hydroxyl masers, covering all MMB sources. The number of masers in each bin with an ex-OH maser has been scaled up by a factor of 10 for visibility.}
    \label{fig:exoh}
\end{figure*}

As described in Section \ref{sec:extramaserdata}, the excited state hydroxyl transition at 6035\,MHz was co-observed during the main MMB survey, therefore providing full coverage of the sample of Hi-GAL bright MMB masers for associations.
The clumps associated with both a 6.7\,GHz methanol maser and an ex-OH maser are the smallest sample considered in this work with only 32 members, and statistically significant results are therefore more challenging to obtain.
The distributions and median values of the IR properties are displayed in the bottom row of Figure \ref{fig:exoh} and Table \ref{tab:maser_assoc} respectively for methanol masers associated with an ex-OH maser and those without. 
Despite the small sample size, we find statistically significant increases in T, $\Sigma$ and L/M for clumps also associated with an ex-OH maser, with large increases in median value for each.
A median temperature of 22.1\,K and mass surface density of 1.61\gpercmsq are consistent with the increase in temperature and surface density expected in the later stages of collapse, further supported by an increase in median L/M to 7.82\Lsun/\Msun.
There is no statistical evidence to suggest a difference in mass or luminosity between clumps with and without an ex-OH maser, and this IR analysis implies that the presence of an ex-OH maser is an evolutionary trait as opposed to occupying only the most luminous or most massive objects.

In addition to the ex-OH masers, the associations between SPLASH ground state OH maser sites and MMB masers performed by \citet{Qiao2016ACCURATEREGION} and \citet{Qiao2018AccurateRegion} are also used to separate the methanol maser sample into those with and without an association.
Only the 143 clumps within the two SPLASH regions of $334^{\circ}<l<344^{\circ}$ and $-5^{\circ}<l<5^{\circ}$ through the Galactic centre are considered, giving a 31 masers associated with a ground-state OH transition and 112 without. 
As for the ex-OH masers, we find a comparable increase in median clump temperature, mass surface density and luminosity to mass ratio for 6.7\,GHz clumps associated with a ground state OH maser, with no distinction in clump mass.
A significant increase in luminosity is also found towards methanol maser clumps with a secondary ground-state OH maser.

\subsubsection{Water masers}
\label{sec:watermaser}
The final maser association considered in this work is with water masers.
These fundamentally differ from the previous maser types in that they are collisionally pumped, found in the regions associated with protostellar outflows and shocks in both high and low-mass star formation. 
This implies that the switch-on of secondary water maser emission in class II methanol maser sources may be at a slightly later evolutionary stage than the onset 6.7\,GHz maser emission as an outflow system must have sufficient time to develop from the protostellar feedback.
However, the presence of a water maser can occur under other conditions aside from outflow activity in a massive star forming region, such as other shocks and PAH emission \citep{Breen201412.2-GHz20}.
To draw any conclusions about the switch-on of water masers in a 6.7\,GHz host clump, the assumption that they occupy a singular, albeit broad, stage of the evolutionary sequence of a massive star forming clump must be made.
\citet{Breen2018TheHOPS} have shown that the inclusion of weak water masers in addition to those detected with HOPS would likely result in a Galactic water maser count in excess of the 6.7\,GHz population, even when accounting for those associated with low-mass star formation and evolved stars.
This implies that this is a strong assumption, and any conclusions drawn about the placement of water masers in the evolutionary timeline for 6.7\,GHz sources is both only approximate and strictly only valid for sources also exhibiting luminous class II methanol maser emission.

\begin{figure}
    \centering
	\includegraphics[width=0.9\columnwidth]{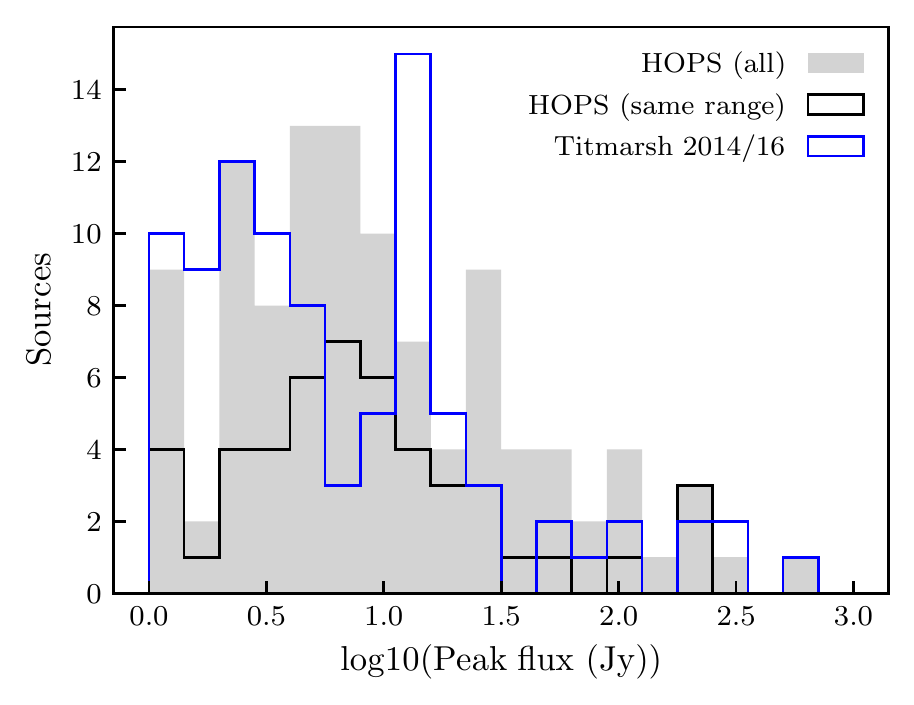}
    \caption{The peak flux density distributions for HOPS and the Titmarsh et al. samples of Galactic water masers that are associated with an MMB maser, showing a significant increase in the number of weak water masers recovered by the Titmarsh et al. samples (blue) relative to the HOPS detections in the same Galactic longitude range (black). The distribution of fluxes for the HOPS masers over the much larger entire survey range is also shown (grey shaded).}
    \label{fig:hops_titmarsh}
\end{figure}

Section \ref{sec:extramaserdata} details the two studies considered in this work for water maser associations.
The HOPS data provides the largest Galactic plane coverage for water maser detections, with 122 of the 501 maser host clumps within the survey range associated with a HOPS maser.
However, the detection of water masers with HOPS is limited by the survey sensitivity of $\sim$5-10\,Jy.
The targeted follow-up study of water masers towards MMB sources presented in \citet{Titmarsh2014A620} and \citet{Titmarsh2016A6} reaches a sensitivity of $\sim$1.6\,Jy between Galactic longitudes $341^{\circ}$ and $20^{\circ}$ through the Galactic centre.
For the 217 Hi-GAL bright methanol maser clumps in this restricted longitude region, the Titmarsh et al. studies recover all 54 HOPS masers associated with an FIR bright maser clump in our sample and provides an additional 56 associations (110 total).
The distribution of peak flux for secondary masers assigned to the MMB host clumps for the two surveys is shown in Figure \ref{fig:hops_titmarsh}, showing the extra water masers that \citet{Titmarsh2014A620,Titmarsh2016A6} find at low flux densities relative to HOPS when comparing the two regions that have been observed with both surveys. 
The distribution of peak flux densities of all HOPS masers found to be associated with a Hi-GAL bright 6.7\,GHz clump is also shown in shaded grey.
Whilst the HOPS sample is likely to return more statistically significant results by providing broader coverage, we make use of the second sample to confirm whether observed differences in the properties of HOPS associated clumps are likely caused by the sensitivity of HOPS.

\begin{figure*}
	\includegraphics[width=\textwidth]{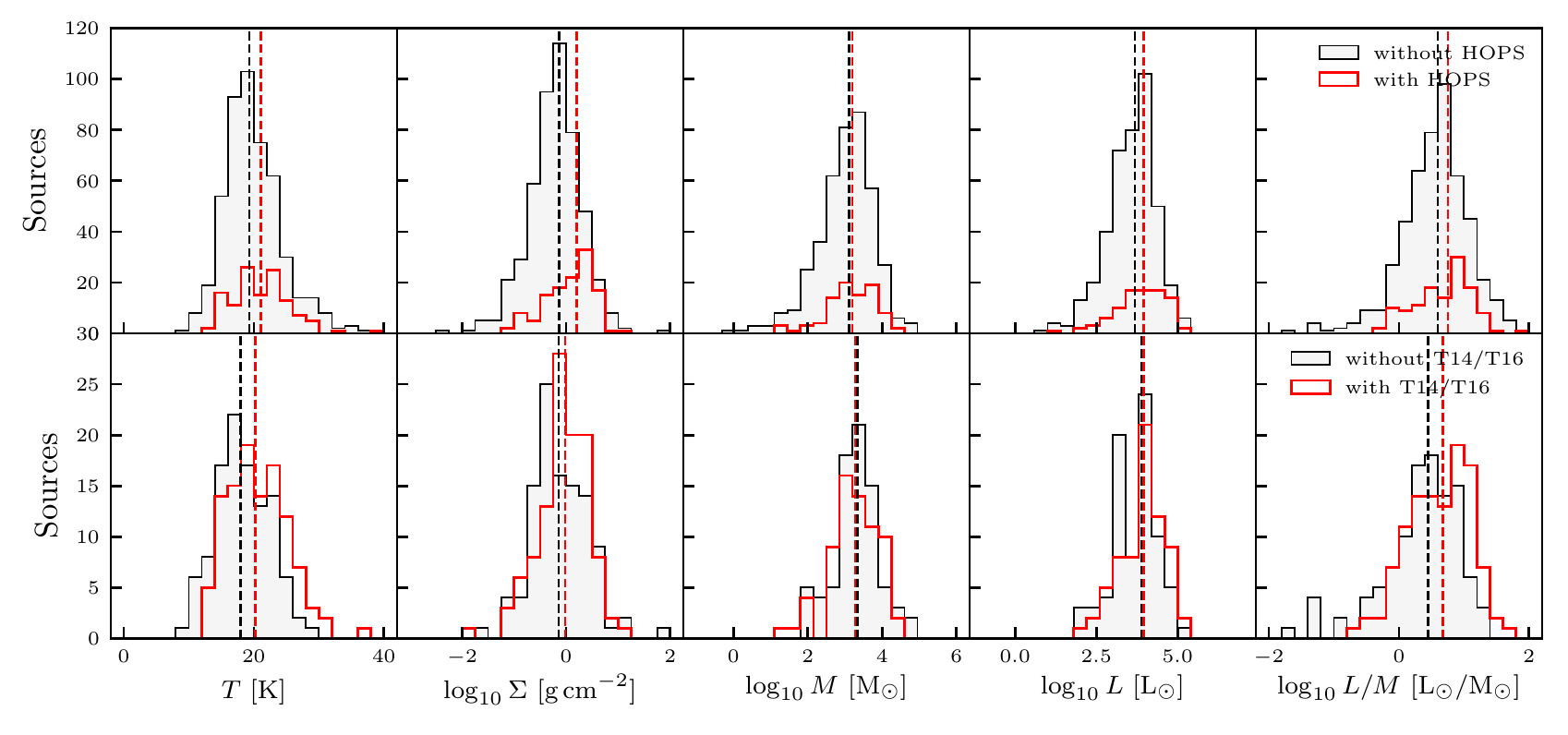}
    \caption{The distributions of properties for methanol maser clumps with (red) and without (black) a water maser detected in top: the HOPS survey and bottom: the Titmarsh et al. MMB follow-up studies. The median values for each distribution are given in Table \ref{tab:maser_assoc}.} 
    \label{fig:h2o}
\end{figure*}

Restricted to the range of the HOPS survey to ensure uniform coverage of the sources for association, the sources with a water maser detected in HOPS are found to be hotter, at higher mass surface density, more luminous and at greater $L/M$ on average with statistical significance, although the median values shown in Table \ref{tab:maser_assoc} are still somewhat lower than those obtained for the ex-OH clumps.
The distributions of clump properties for sources with and without a secondary HOPS maser are shown in Figure \ref{fig:h2o}.

Performing the same analysis with this restricted longitude range as for the HOPS associations, the sample medians and distributions for clumps associated with a Titmarsh et al. water maser are also shown in Table~\ref{tab:maser_assoc} and Figure~\ref{fig:h2o}.
Whilst the statistically significant average increase in temperature and $L/M$ towards the clumps with a secondary water maser is also found for this sample, the clear statistical distinction in mass surface density and luminosity is not. 
This is plausibly due to the sensitivity of HOPS, as the FIR clump luminosity is found to be correlated with the 6.7\,GHz maser luminosity in Section~\ref{sec:IRvsMaser}, and the luminosity of a secondary water maser has in turn found to be correlated with the 6.7\,GHz methanol maser\,\citep{Titmarsh2014A620,Titmarsh2016A6}.
It is often similarly assumed that the water masers hosted in clumps with higher mass surface density are likely to be more luminous and so above the HOPS sensitivity threshold as higher densities enhance the collisional pumping of the maser line, although it should be noted that \citet{Elitzur1989H2ORegions} find that this is not consistent with their model. 

\subsection{Relationship between infrared and maser properties}
\label{sec:IRvsMaser}

The FIR properties of the host clumps may also be compared to the properties of a host maser itself.
For each 6.7\,GHz maser, the MMB catalogue entry contains the peak flux, the flux integrated over the maser line and the maximum and minimum velocity obtained from the spectrum, giving an observed velocity range.
Each of the above maser properties was checked for correlation with the FIR properties obtained, with maser luminosity L$_{6.7}$ derived from the integrated flux of the line where a comparison of intrinsic quantities is required.
We note here that we follow the convention of \citet{Green2017TheMasers} and include a factor of $4\pi$ when calculating maser luminosity.

\begin{figure}
    \centering
    \includegraphics[width=\columnwidth]
    {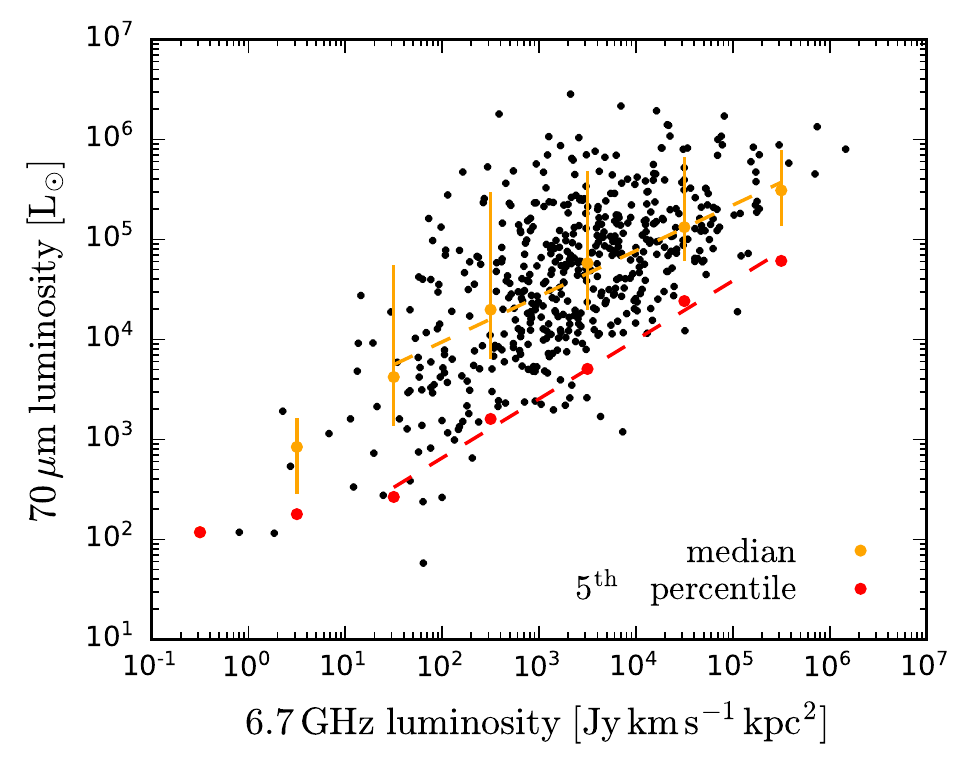}
    \caption{For maser host sources with a distance value available in the MMB catalogue, the 6.7\.GHz methanol maser luminosity (integrated over the maser line) is shown against the 70\microns clump luminosity for each source. Binning in maser luminosity, the general trend between the two properties is characterised by the line through the median 70\microns luminosity (yellow) in each bin, given by $\log_{10}(L_{70}) = (0.46\pm0.04)\log_{10}(L_{6.7}) + (3.1\pm0.1)$ and shown with the standard deviation in each bin. Similarly, an approximate lower threshold on the required 70\microns flux to sustain a class II methanol maser of a given luminosity is given by the 5$^{\mathrm{th}}$ percentile in each bin (red), with a the line of best fit given by $\log_{10}(L_{70}) = (0.59\pm0.03)\log_{10}(L_{6.7}) + (1.6\pm0.1)$. The 70\microns luminosity and maser luminosity are given in units of \Lsun and Jy\,km\,s$^{-1}$\,kpc$^2$.}  
    \label{fig:maser_ir_luminosity}
\end{figure}

No clear correlation is found for any maser property with $T$, $L_{\mathrm{FIR}}$, $M$, $\Sigma$, $L/M$ or any far-infrared or mid-infrared colour shown in Sections \ref{sec:colours} and \ref{sec:MIRresults}.
L$_{6.7}$ is shown against host clump L$_{70}$ in Figure \ref{fig:maser_ir_luminosity} and a general trend of increasing maser luminosity with 70\microns luminosity is found for the sample. 
When logarithmically binned in $L_{6.7}$, the line of best fit through the median $L_{70}$ values returns a relationship of $\log_{10}(L_{70}) = (0.46\pm0.04)\log_{10}(L_{6.7}) + (3.1\pm0.1)$, with 70\microns luminosity given in \Lsun and 6.7\,GHz maser luminosity in units of Jy\,km\,s$^{-1}$\,kpc$^2$.
Perhaps more clearly than the overall trend is the appearance of an apparent minimum 70\microns flux required to sustain a maser of a given luminosity.
An approximation of this limit is found from taking the line of best fit through the 5$^{\mathrm{th}}$ percentile in each bin, returning a steeper line of $\log_{10}(L_{70}) = (0.59\pm0.03)\log_{10}(L_{6.7}) + (1.6\pm0.1)$ in the same units.
As both properties are dependent on the square of the source distance, the slope in logarithmic space would be 1 if this minimum was simply an effect of the flux sensitivity limit of both surveys. 
Given that the slope is significantly less than 1, we conclude that this effect is physical in origin.

Similarly, for the secondary masers present in a 6.7\,GHz host clump, a comparison of IR clump luminosity with maser luminosity can be made.
In addition to the 6.7\,GHz masers, integrated maser luminosities have also been calculated for the 12.2\,GHz class II methanol masers, H$_2$O masers from Titmarsh et al. and the SPLASH OH masers.
As several OH masers detected in the SPLASH survey may be associated with a single MMB 6.7\,GHz maser, the integrated OH luminosity of a clump is taken to be the sum of the integrated luminosities over all of its individual OH masers.
This shows correlation between the clump and maser luminosities in each case, but also that the scatter about the average trend is significant.
We take the average absolute offset from the line of best fit in both maser luminosity (y-direction) and clump luminosity (x-direction) to quantify this scatter, with each of these given alongside the line of best fit parameters in Table \ref{tab:maserL_clumpL}.
In each case, approximately half an order of magnitude of scatter in maser luminosity is found for a given clump luminosity, and vice versa.
This result is unsurprising given that maser luminosity depends on stochastic properties, such as coherence and beaming, and is consistent with the long-term variability found by \citet{Ellingsen2007AFeatures}.
This therefore implies that although a correlation is evident, the use of one given luminosity to predict another in a host clump from the line of best fit relationship is only accurate to an order of magnitude at best, and that a 1:1 correlation of the two properties is not representative of the data.

\begin{table}
\centering
\caption{The lines of best fit and average deviation from this line in each direction for maser luminosity against host clump FIR luminosity. From left to right, the columns are: maser type, number used to calculate the line, the line of best fit parameters of the form $\log_{10}$($L_{\text{m}})= \text{p}_1\*\log_{10}$($L_{\text{FIR}}) + \text{p}_2$, and the average absolute logarithmic offset from this line in clump luminosity and maser luminosity (x- and y-direction respectively in Figure \ref{fig:maserL_clumpL}). }
	\begin{tabular}{l c c c c c}
		\hline
		Maser & N &  \multicolumn{2}{l}{Fit parameters} & \multicolumn{2}{l}{Average offset}\\
		& & p$_1$ & p$_2$ & $\log_{10}$($L_{\text{FIR}}$) &  $\log_{10}$($L_{\text{m}}$)\\ \hline
		6.7\,GHz & 511 & $0.8\pm0.1$ & $0.4\pm0.2$ & 0.7 & 0.6\\
		12.2\,GHz & 219 & $0.7\pm0.1$ & $0.0\pm0.3$ & 0.9 & 0.6 \\ 
		H$_2$O & 67 & $0.7\pm0.1$ & $1.5\pm0.5$ & 0.7 & 0.5\\
		OH & 18 & $0.9\pm0.2$ & $-0.5\pm0.9$ & 0.4 & 0.4\\ \hline
\end{tabular}
\label{tab:maserL_clumpL}
\end{table}

The correlation between the luminosity of the 6.7\,GHz maser and secondary masing species in a host MYSO has also been studied for the targeted MMB follow-ups, so as in \citet{Breen201612.2-GHz2060} and \citet{Titmarsh2014A620,Titmarsh2016A6} for the 12.2\,GHz and water masers respectively, we present this for the host clumps in this sample and the SPLASH OH masers.
As for the correlations between maser and host clump luminosity, the line of best fit is given by $\log_{10}(L_{\text{OH}}) = (0.6\pm0.2)\log_{10}(L_{\text{6.7}}) + (0.5\pm0.7)$, with an average logarithmic scatter of 0.7 in 6.7\,GHz luminosity and 0.4 in total OH luminosity from this line. 

The accompanying plots for these correlations are presented in Appendix~\ref{sec:append_b}

\subsection{Comparison of all associations}

Comparison between the types of secondary maser associations allows the relative evolutionary stage of the protostellar host clump in each category to be assessed through its typical IR properties.
In addition to the above associations, we consider two more categories: `solitary' 6.7\,GHz methanol masers for which no additional masers have been identified, and `associated' masers, in which the host clump has one or more additional masers of any type detected. 

To ensure that all methanol maser host clumps are equally checked for secondary masers, the longitude range of the associated and solitary categories must be equal to the smallest survey range considered.
Due to the small on-sky coverage of the Titmarsh et al. and SPLASH surveys, the sample sizes that would remain by restricting to their common coverage would be challenging to obtain statistically significant results from.
We therefore do not use these surveys when determining whether a 6.7\,GHz host clump also contains a secondary maser and falls within the `associated' category.
As shown in Section \ref{sec:watermaser}, the HOPS survey recovers the strongest water masers.
The `associated' category is therefore characterised by the presence of a second class II methanol maser line at 12.2\,GHz, an excited-state hydroxyl maser or a strong water maser, and the `solitary' category may have weak water masers or ground-state hydroxyl masers present.
The remaining survey with the smallest on-sky coverage is HOPS, and the new number of sources and median properties for each category when restricting to this region are displayed in Table \ref{tab:assoc}.

\begin{figure*}
	\includegraphics{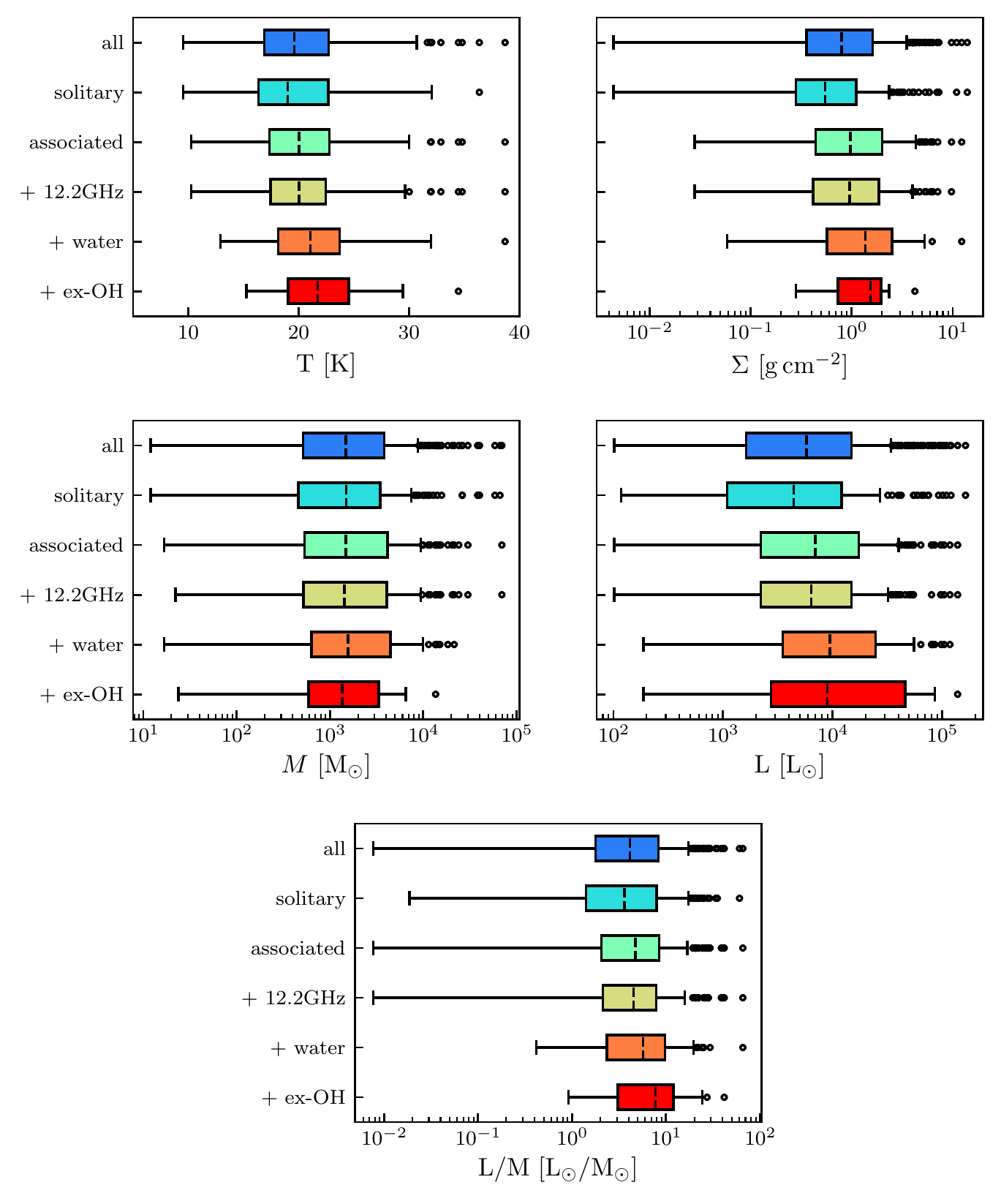}
    \caption{Box-and-whisker plots of the distribution of infrared-derived clump properties for objects hosting a 6.7\,GHz methanol maser, categorised by the presence of any given secondary maser. For each panel from top to bottom: the total maser sample, clumps hosting a methanol maser without any secondary masers found in the surveys used, clumps with at least one secondary maser of any type, and the distributions for clumps with each type of secondary maser (12.2\,GHz CH$_3$OH, H$_2$O from HOPS, and excited-state hydroxyl). Each box is drawn between the lower (25th percentile) to upper (75th percentile) quartile value, with the length of the box defining the interquartile range (IQR) and a vertical dashed line at the median. The median values are given in Table \ref{tab:assoc}. The whiskers (capped solid lines) give the range of the data, defined as 1.5$\times$IQR past each end of the box, and outliers beyond this are plotted as circles.}
    \label{fig:maser_boxes}
\end{figure*}

Figure \ref{fig:maser_boxes} shows all properties as box-and-whisker plots to identify trends between the categories.
We note that the mass distributions of each category are statistically indistinct from one another, implying that the appearance of any given secondary maser is not mass dependent given that a host clump is sufficiently massive to host a class II methanol maser.
In each of the other properties, one or more secondary masers of any type present in a clump traces sources of greater $T$, $\Sigma$, $L$ and $L/M$ on average, relative to solitary 6.7\,GHz masers.
As expected from previously proposed evolutionary schemes such as that of \citet{Breen201012.2-GHzSchemes} and taking $T$, $L$ and $\Sigma$ to increase simultaneously with evolution, clumps with 12.2\,GHz emission are found to be at an earlier evolutionary stage than those with secondary excited hydroxyl masers.
Similarly, these results alone imply that the onset of water masing occurs at a later time than the switch-on of 12.2\,GHz emission.
However, Section \ref{sec:watermaser} shows that the detection limit of HOPS limits this conclusion to the onset of strong secondary water masing.
The use of the more sensitive Titmarsh et al. samples removes the statistically significant difference in the distribution of $\Sigma$ for sources with and without a water maser.
Although the difference in $T$ and $L/M$ remains, constraining the evolutionary state of clumps with 6.7 and 22\,GHz water masers will require a greater longitude coverage with a sensitive survey.
When considering these results, it should still be noted that the overlap in both proposed evolutionary stage and the IR properties of host clumps in each association category is considerable.

\begin{table*}
	\centering
	\caption{Median properties of 6.7\,GHz maser host clumps categorised by secondary maser assessed within the same range of Galactic latitudes, including a `solitary' category for which no secondary masers are found and an `associated' category for which any additional maser is identified.}
	\begin{tabular}{l c c c c c c}
		\hline
		Category & Number (\%) & <T> [K] & <$\Sigma$> [\gpercmsq] & <M> [\Msun] & <L> [\Lsun] & <L/M> [\Lsun/\Msun] \\ \hline
		All 		& 501 (100)  & 19.6 & 0.80 & 1470 & 5340 & 4.14 \\
		Solitary 	& 204 (41) & 19.0 & 0.55 & 1440 & 4190 & 3.61 \\
		Associated 	& 297 (59) & 20.0 & 0.97 & 1470 & 6630 & 4.73 \\
		12.2\,GHz 	& 232 (46) & 20.0 & 0.96 & 1360 & 6380 & 4.53 \\ 
        H$_2$O 		& 122 (24) & 21.0 & 1.36 & 1570 & 9100 & 5.61 \\ 
        ex-OH 		& 29 (6)   & 21.7 & 1.53 & 1360 & 8970 & 7.73 \\ \hline	 
	\end{tabular}
\label{tab:assoc}
\end{table*}

\subsection{Timeline of secondary masers}

As in \citet{Breen201012.2-GHzSchemes}, the statistical lifetime of each secondary maser phase within the 6.7\,GHz lifetime can be estimated from the number of clumps associated with each and a timeline constructed.
The lifetimes estimated here are only for clumps that simultaneously host a class II methanol maser alongside each type of secondary maser, and we do not sample other evolutionary states which may be associated with these masers outside of the 6.7\,GHz lifetime.
We calculate the relative lifetime of each secondary maser species from the percentage of 6.7\,GHz masers and the results are given in Table~\ref{tab:lifetimes}.
As each of the input surveys has different detection limits due to  varying sensitivities, each of these relative lifetimes is a lower bound on the true value as undetected weak secondary masers may be present.
An example of this is  the difference in lifetime estimates from the HOPS survey of only bright water masers and the more sensitive survey by \citet{Titmarsh2014A620,Titmarsh2016A6}, with the detection of weak water masers significantly increasing the estimated lifetime.
Converting the relative timescales requires an estimate of the lifetime 6.7\,GHz maser. 
\citet{vanderWalt2005OnMasers} estimate the lifetime  to be between 2.5$\times10^4$ and 4.5$\times10^4$\,yr, with \citet{Billington2019ATLASGALMasers} recently deriving a consistent lifetime of 3.3$\times10^4$\,yr from analysis of data from the ATLASGAL survey. 

\begin{table*}
	\centering
	\caption{Relative lifetimes of each secondary maser found in clumps hosting a 6.7\,GHz methanol maser. The percentage of clumps associated with each secondary maser have been calculated from the values in Table~\ref{tab:maser_assoc}. }
	\begin{tabular}{l c c}
		\hline
		Species & Percentage & Lifetime (lower-upper) [$10^4$\,yr] \\ \hline
		6.7\,GHz    	    & 100 (reference)   & 2.5 - 4.5   \\
		12.2\,GHz   	    & 45.0              & 1.1 - 2.0   \\
		H$_2$O (HOPS) 	    & 24.4              & 0.6 - 1.1   \\
		H$_2$O (T14/T16) 	& 50.7              & 1.3 - 2.3   \\ 
        OH 		            & 21.7              & 0.5 - 1.0  \\ 
        ex-OH 		        & 4.9               & 0.1 - 0.2 \\ \hline	 
	\end{tabular}
\label{tab:lifetimes}
\end{table*}

An estimate of the relative location of each secondary maser within the 6.7\,GHz lifetime is given by the statistical overlap between each species.
For example, 60\% of the ex-OH masers are found in clumps that also host a 12.2\,GHz maser, and this implies that the two lifetimes should overlap by 60\% of the ex-OH lifetime given in Table~\ref{tab:lifetimes}.
As this method does not identify whether a maser species should overlap with the start or end of the lifetime of another species, the order of appearance within this scheme is given by the sequence of increasing $L_{FIR}/M$ shown in Figure~\ref{fig:maser_boxes}.
The switch-on time of secondary masing is also not constrained by this method, and so the location of this sequence within the lifetime of a 6.7\,GHz maser is not fixed. 

\begin{figure*}
	\includegraphics{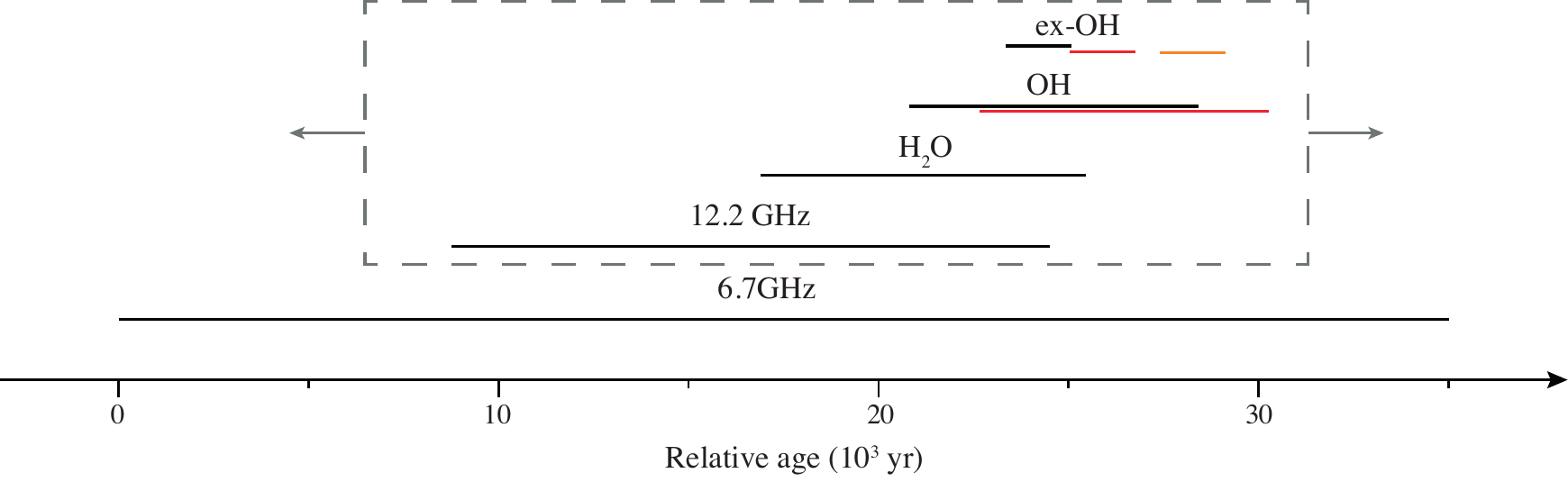}
    \caption{Timeline of the relative duration and evolutionary offset of each secondary maser species found within 6.7\,GHz host clumps. The duration of the class II methanol maser phase is taken to be 3.5$\times10^4$\,yr \citep{vanderWalt2005OnMasers} and length of the line is determined by the fraction of 6.7\,GHz methanol masers with this species. The offset between each species of secondary maser is given by the statistical overlap with the other secondary masers. The black, red and orange lines are found from considering the overlap with 12.2\,GHz class II methanol masers, water masers and ground-state hydroxyl masers respectively. The overall location of the secondary masers within a 6.7\,GHz lifetime is not fixed, as indicated by the grey box.}
    \label{fig:lifetimes2}
\end{figure*}

Considering the overlap between multiple different secondary species also yields different locations within the timeline for both the OH and ex-OH masers. 
The black lines in Figure~\ref{fig:lifetimes2} are the locations calculated from the overlap with 12.2\,GHz masers for each species, and the red and orange from the overlap with water and OH similarly.

It has been assumed that all maser types to form an evolutionary sequence and occupy single sub-phases of the lifetime of 6.7\,GHz masing. 
As we have already selected sources in a constrained evolutionary state through the presence of a 6.7\,GHz maser, it seems unlikely that there are several distinct sub-categories of object within this sample each with the conditions required to host each type of radiative secondary maser. 
However, this may be less true for the water masers as the collisional pumping is dependent on the interaction of a protostar with its environment \citep{Breen201412.2-GHz20}.
For this reason, water masers are much more difficult to reliably place on an evolutionary timeline using only statistical overlaps, and this is seen in incorporating the water masers detected in Titmarsh et al. into Figure~\ref{fig:lifetimes2}.
Similarly, the narrow lifetime and the uncertainty in positioning in the timeline of the ex-OH masers perhaps suggests that their presence is linked to very specific conditions, which may not be simply related to the evolutionary state of a protostar.

The small sample sizes of clumps with OH and ex-OH secondary masers contributes significant uncertainty to the statistical lifetime estimates, and is also a likely source of the uncertainty in their location within the evolutionary timeline when considering different overlaps.
Additionally, the location changes when calculated from the overlap with the HOPS masers as these sample  only bright water maser sources.
It is also important to note that the lifetime of the ex-OH masers is a lower limit as several host clumps are saturated in the Hi-GAL maps and are therefore not included in our sample.
When considering the entire sample of ex-OH masers, \citet{adam2016exohI} estimate a lifetime between $3.3\times10^3$ and $8.3\times10^3$\,yr while the statistical lifetime of $1\times10^3$ to $2\times10^3$\,yr is found in this work.

Determining the switch-on time of a secondary maser species within the 6.7\,GHz lifetime requires a robust evolutionary tracer that can be mapped to time. 
In principle the luminosity to mass ratio, for the appropriate circumstellar mass, could provide such a tracer, but its relationship to time is not well determined and may be a  function of initial mass, and/or the ultimate stellar mass (and luminosity). 

\section{Summary}
We have characterised a complete sample of MYSOs hosting 6.7\,GHz methanol masers through combination of $\sim1000$ of these masers detected with the Methanol MultiBeam survey and compact source catalogues for each of the Hi-GAL wavelengths over the full Galactic plane. 
We associate 96\% of all masers with a compact source in at least one Hi-GAL wavelength.
We find 73 per cent of all MMB class II methanol masers to be associated with a compact source in four Hi-GAL bands, yielding a sample of 647 maser host clumps for further analysis following removal of sources with unreliable flux estimates.
For maser-bearing clumps of far-infrared luminosity $\geq100$\Lsun, we find the MMB to be complete to 2.7\,kpc, taking both weak methanol masers and beaming into consideration.

Our key results are as follows:

\begin{enumerate}[label=\roman*)]
\item We have derived the typical properties of the massive protostellar clumps hosting MMB masers by fitting to their far-infrared spectral energy distributions.
The maser host clumps on the near and far side of the Galactic centre (taken to be at a heliocentric distance of 8.0\,kpc) represent different populations due to the selection angular size used to create the Hi-GAL compact source catalogues.
The typical mass, radius and far-infrared luminosity of nearby sources (median distance 3.8\,kpc) are 630\Msun, 0.2\,pc and 2500\Lsun, whereas the same properties for the maser hosts on the far side of the Galaxy (median distance 11.4\,kpc) are 3200\Msun, 0.6\,pc and 10\,000\Lsun respectively.
We do not detect small clumps on the far side of the Galaxy, confirmed by the lack of a clear multi-wavelength detection of a host clump for all of the MMB sources.
Despite the difference in typical size, we do not find these two populations to be distinct in luminosity-to-mass ratio, and thus are in the same relative evolutionary state.
The median $L_{FIR}/M$ for maser host clumps is 4.2\Lsun/\Msun, and the median temperature and mass surface density are 19.5\, and 0.76\gpercmsq respectively. 

\item We find that 6.7\,GHz methanol masers, and therefore high-mass protostars, are present in clumps down to mass surface densities of approximately 0.08\gpercmsq.
This value is similar to thresholds of $\sim$0.1\gpercmsq for massive star formation reported by other work, despite derivation from significantly different methods.

\item We have identified a correlation between the 6.7\,GHz integrated luminosity of a maser and the luminosity of its host clump at 70\microns, finding the minimum 70\microns clump luminosity required to be associated with a maser of a given luminosity to be characterised by $L_{70}\propto L_{6.7}^{0.6}$.

\item Class II methanol maser hosts are found to be at greater values in all clump properties relative to the population of all protostellar clumps visible with Hi-GAL. The range of $L_{FIR}/M$ covered by the maser sample is narrower relative to the spread in $L_{bol}/M$ seen for the protostellar sample.
Identifying approximate thresholds on each property and through the use of PCA, we identify 896 protostellar objects, from a sample of 5497, that match the properties of the methanol maser host clumps but lack a detection of a strong 6.7\,GHz maser with the MMB survey.
Accounting for completeness in both samples, we find a factor of 2.5 more maser sources than matched maserless clumps.

Finding a 70\microns deficiency in the sources lacking a methanol maser, a comparison of the mid-infrared 8\microns properties of these sources suggests that the objects lacking a 6.7\,GHz methanol maser may be in a later evolutionary stage.
However, this conclusion is limited by the small number of sources in each sample with a detected GLIMPSE counterpart (19 and 47 per cent of sources in the non-maser and maser samples respectively). The origin of these low values needs further investigation.
It is unlikely that the majority of these sources host weak methanol masers (\S~\ref{sec:nonmaser_interp}). 
However, confirming this requires more sensitive observations while observations on smaller, core-scales will allow us to more fully understand the  properties of the maser-less, embedded protostars and their apparent lack of methanol masers. 
This is the subject of future work.

\item Considering secondary water, hydroxyl and 12.2\,GHz class II methanol masers in each MMB maser host clump, we find that clumps associated with any secondary maser to be more evolved than those hosting a solitary 6.7\,GHz maser.
Sources with a hydroxyl maser are found to be significantly more advanced in evolution.
We find sources with water masers detected in HOPS to be at an intermediate evolutionary stage between the solitary 6.7\,GHz and hydroxyl-bearing clumps, although we do not find a statistically significant difference in luminosity or mass surface density relative to clumps associated with a water maser relative to those that are not when considering higher-sensitivity surveys.
We find a 1\,K median temperature increase of borderline statistical significance towards sources with a second class II methanol maser at 12.2\,GHz.
We present a maser timeline summarising these results.

 \item Although trends between FIR clump luminosity and maser luminosity are found for each type of secondary maser in a 6.7\,GHz host clump, these relationships have order of magnitude scatter and are not 1:1 relationships. 
 Any predictions of maser luminosity from infrared clump luminosities should account for this. \end{enumerate}

In summary, comparison of the sources identified in the Methanol MultiBeam (MMB) survey and the Herschel Infrared Galactic Plane Survey (Hi-GAL) has allowed the most comprehensive, Galaxy-wide, study of the properties and evolutionary status of the sources which host 6.7 GHz methanol masers to date. 
The maser sources have median luminosity of 5600,\Lsun and are associated with clumps of median radius 0.42\,pc with median mass 1400\,\Msun, although these are influenced by the bi-modal distribution of distances to the sources. 
On the near side of the Galaxy the typical source has a luminosity of 2500,\Lsun, lower than typically previously assumed for these sources. 
However, the clumps in which these sources are found have high surface densities, consistent with them eventually becoming high mass, and high luminosity, sources.  
Our analysis has identified for the first time a sample of sources which are similar to the maser sources in a range of properties but do not host masers. 
We show that these are consistent with being at a more evolved phase of evolution than the maser sources, but confirmation of this requires further comparative study of these objects. 

\section*{Acknowledgements}
BMJ acknowledges the support of a studentship grant from the UK Science and Technology Facilities Council (STFC). 
AA is funded by the STFC at the UK ALMA Regional Centre Node through grant ST/P000827/1.
SPE acknowledges the support of an Australian Research Council Discovery Project (project number DP180101061).
MM acknowledges support from the grant 2017/23708-0, S\~ao Paulo Research Foundation (FAPESP).
This research has made use of the NASA/ IPAC Infrared Science Archive, which is operated by the Jet Propulsion Laboratory, California Institute of Technology, under contract with the National Aeronautics and Space Administration.
We would like to thank Andr\'es Guzm\'an and Yanett Contreras for their very helpful discussion about the temperature of the maser sources. 





\bibliographystyle{mnras}
\bibliography{main}

\begin{thebibliography}{}
\makeatletter
\relax
\def\mn@urlcharsother{\let\do\@makeother \do\$\do\&\do\#\do\^\do\_\do\%\do\~}
\def\mn@doi{\begingroup\mn@urlcharsother \@ifnextchar [ {\mn@doi@}
  {\mn@doi@[]}}
\def\mn@doi@[#1]#2{\def\@tempa{#1}\ifx\@tempa\@empty \href
  {http://dx.doi.org/#2} {doi:#2}\else \href {http://dx.doi.org/#2} {#1}\fi
  \endgroup}
\def\mn@eprint#1#2{\mn@eprint@#1:#2::\@nil}
\def\mn@eprint@arXiv#1{\href {http://arxiv.org/abs/#1} {{\tt arXiv:#1}}}
\def\mn@eprint@dblp#1{\href {http://dblp.uni-trier.de/rec/bibtex/#1.xml}
  {dblp:#1}}
\def\mn@eprint@#1:#2:#3:#4\@nil{\def\@tempa {#1}\def\@tempb {#2}\def\@tempc
  {#3}\ifx \@tempc \@empty \let \@tempc \@tempb \let \@tempb \@tempa \fi \ifx
  \@tempb \@empty \def\@tempb {arXiv}\fi \@ifundefined
  {mn@eprint@\@tempb}{\@tempb:\@tempc}{\expandafter \expandafter \csname
  mn@eprint@\@tempb\endcsname \expandafter{\@tempc}}}

\bibitem[\protect\citeauthoryear{Avison, Peretto, Fuller, Duarte-Cabral,
  Traficante  \& Pineda}{Avison et~al.}{2015}]{adam2015sdc335}
Avison A.,  Peretto N.,  Fuller G.~A.,  Duarte-Cabral A.,  Traficante A.,
  Pineda J.~E.,  2015, \mn@doi [Astronomy {\&} Astrophysics]
  {10.1051/0004-6361/201425041}, 577, A30

\bibitem[\protect\citeauthoryear{Avison et~al.,}{Avison
  et~al.}{2016}]{adam2016exohI}
Avison A.,  et~al., 2016, \mn@doi [Monthly Notices of the Royal Astronomical
  Society] {10.1093/mnras/stw1101}, 461, 136

\bibitem[\protect\citeauthoryear{Bartkiewicz, Szymczak, van Langevelde,
  Richards  \& Pihlstr{\"{o}}m}{Bartkiewicz
  et~al.}{2009}]{Bartkiewicz2009TheObservations}
Bartkiewicz A.,  Szymczak M.,  van Langevelde H.~J.,  Richards A. M.~S.,
  Pihlstr{\"{o}}m Y.~M.,  2009, \mn@doi [Astronomy {\&} Astrophysics]
  {10.1051/0004-6361/200912250}, 502, 155

\bibitem[\protect\citeauthoryear{Bartkiewicz, Szymczak  \& van
  Langevelde}{Bartkiewicz et~al.}{2014}]{Bartkiewicz2014EuropeanObjects}
Bartkiewicz A.,  Szymczak M.,   van Langevelde H.~J.,  2014, \mn@doi [Astronomy
  {\&} Astrophysics] {10.1051/0004-6361/201322629}, 564, A110

\bibitem[\protect\citeauthoryear{Bartkiewicz, Szymczak  \& van
  Langevelde}{Bartkiewicz et~al.}{2016}]{Bartkiewicz2016EuropeanMasers}
Bartkiewicz A.,  Szymczak M.,   van Langevelde H.~J.,  2016, \mn@doi [Astronomy
  {\&} Astrophysics] {10.1051/0004-6361/201527541}, 587, A104

\bibitem[\protect\citeauthoryear{Benjamin et~al.,}{Benjamin
  et~al.}{2003}]{benjamin2003glimpse}
Benjamin R.~A.,  et~al., 2003, Publications of the Astronomical Society of the
  Pacific, 115, 953

\bibitem[\protect\citeauthoryear{Billington et~al.,}{Billington
  et~al.}{2019}]{Billington2019ATLASGALMasers}
Billington S.~J.,  et~al., 2019, preprint (arXiv:1907.00564)

\bibitem[\protect\citeauthoryear{Breen, Ellingsen, Caswell  \& Lewis}{Breen
  et~al.}{2010}]{Breen201012.2-GHzSchemes}
Breen S.~L.,  Ellingsen S.~P.,  Caswell J.~L.,   Lewis B.~E.,  2010, \mn@doi
  [Monthly Notices of the Royal Astronomical Society]
  {10.1111/j.1365-2966.2009.15831.x}, 401, 2219

\bibitem[\protect\citeauthoryear{Breen, Ellingsen, Caswell, Green, Voronkov,
  Fuller, Quinn  \& Avison}{Breen et~al.}{2012a}]{breen201212}
Breen S.~L.,  Ellingsen S.~P.,  Caswell J.~L.,  Green J.~A.,  Voronkov M.~A.,
  Fuller G.~A.,  Quinn L.~J.,   Avison A.,  2012a, Monthly Notices of the Royal
  Astronomical Society, 421, 1703

\bibitem[\protect\citeauthoryear{Breen, Ellingsen, Caswell, Green, Voronkov,
  Fuller, Quinn  \& Avison}{Breen et~al.}{2012b}]{Breen201212.2-GHz186-330}
Breen S.~L.,  Ellingsen S.~P.,  Caswell J.~L.,  Green J.~A.,  Voronkov M.~A.,
  Fuller G.~A.,  Quinn L.~J.,   Avison A.,  2012b, \mn@doi [Monthly Notices of
  the Royal Astronomical Society] {10.1111/j.1365-2966.2012.21759.x}, 426, 2189

\bibitem[\protect\citeauthoryear{Breen, Ellingsen, Contreras, Green, Caswell,
  Stevens, Dawson  \& Voronkov}{Breen
  et~al.}{2013}]{shari2013exclusivemethanol}
Breen S.~L.,  Ellingsen S.~P.,  Contreras Y.,  Green J.~A.,  Caswell J.~L.,
  Stevens J.~B.,  Dawson J.~R.,   Voronkov M.~A.,  2013, \mn@doi [Monthly
  Notices of the Royal Astronomical Society] {10.1093/mnras/stt1315}, 435, 524

\bibitem[\protect\citeauthoryear{Breen et~al.,}{Breen
  et~al.}{2014}]{Breen201412.2-GHz20}
Breen S.~L.,  et~al., 2014, \mn@doi [Monthly Notices of the Royal Astronomical
  Society] {10.1093/mnras/stt2447}, 438, 3368

\bibitem[\protect\citeauthoryear{Breen et~al.,}{Breen
  et~al.}{2015}]{Breen2015The2060}
Breen S.~L.,  et~al., 2015, \mn@doi [Monthly Notices of the Royal Astronomical
  Society] {10.1093/mnras/stv847}, 450, 4109

\bibitem[\protect\citeauthoryear{Breen, Ellingsen, Caswell, Green, Voronkov,
  Avison, Fuller  \& Quinn}{Breen et~al.}{2016}]{Breen201612.2-GHz2060}
Breen S.~L.,  Ellingsen S.~P.,  Caswell J.~L.,  Green J.~A.,  Voronkov M.~A.,
  Avison A.,  Fuller G.~A.,   Quinn L.~J.,  2016, \mn@doi [Monthly Notices of
  the Royal Astronomical Society] {10.1093/mnras/stw965}, 459, 4066

\bibitem[\protect\citeauthoryear{Breen et~al.,}{Breen
  et~al.}{2018}]{Breen2018TheHOPS}
Breen S.~L.,  et~al., 2018, \mn@doi [Monthly Notices of the Royal Astronomical
  Society] {10.1093/mnras/stx3051}, 474, 3898

\bibitem[\protect\citeauthoryear{Butler \& Tan}{Butler \&
  Tan}{2012}]{Butler2012MID-INFRAREDCLUMPS}
Butler M.~J.,  Tan J.~C.,  2012, \mn@doi [The Astrophysical Journal]
  {10.1088/0004-637X/754/1/5}, 754, 5

\bibitem[\protect\citeauthoryear{Carey et~al.,}{Carey
  et~al.}{2009}]{carey2009mipsgal}
Carey S.~J.,  et~al., 2009, Publications of the Astronomical Society of the
  Pacific, 121, 76

\bibitem[\protect\citeauthoryear{Caswell}{Caswell}{2003}]{Caswell2003SpectraMHz}
Caswell J.~L.,  2003, \mn@doi [Monthly Notices of the Royal Astronomical
  Society] {10.1046/j.1365-8711.2003.06418.x}, 341, 551

\bibitem[\protect\citeauthoryear{Caswell \& Vaile}{Caswell \&
  Vaile}{1995}]{Caswell1995Excited-stateGHz}
Caswell J.~L.,  Vaile R.~A.,  1995, \mn@doi [Monthly Notices of the Royal
  Astronomical Society] {10.1093/mnras/273.2.328}, 273, 328

\bibitem[\protect\citeauthoryear{Caswell, Kramer  \& Reynolds}{Caswell
  et~al.}{2009}]{Caswell2009Maser300.969+1.147}
Caswell J.~L.,  Kramer B.~H.,   Reynolds J.~E.,  2009, \mn@doi [Monthly Notices
  of the Royal Astronomical Society] {10.1111/j.1365-2966.2009.14952.x}, 398,
  528

\bibitem[\protect\citeauthoryear{Caswell et~al.,}{Caswell
  et~al.}{2010}]{Caswell2010The6}
Caswell J.~L.,  et~al., 2010, \mn@doi [Monthly Notices of the Royal
  Astronomical Society] {10.1111/j.1365-2966.2010.16339.x}, 404, 1029

\bibitem[\protect\citeauthoryear{Caswell et~al.,}{Caswell
  et~al.}{2011}]{Caswell2011The345}
Caswell J.~L.,  et~al., 2011, \mn@doi [Monthly Notices of the Royal
  Astronomical Society] {10.1111/j.1365-2966.2011.19383.x}, 417, 1964

\bibitem[\protect\citeauthoryear{Cesaroni}{Cesaroni}{2005}]{Cesaroni2005HotCores}
Cesaroni R.,  2005, \mn@doi [Proceedings of the International Astronomical
  Union] {10.1017/S1743921305004369}, 1, 59

\bibitem[\protect\citeauthoryear{Chanapote, Asanok, Dodson, Rioja, Green  \&
  Hutawarakorn~Kramer}{Chanapote et~al.}{2019}]{Chanapote2019TracingArray}
Chanapote T.,  Asanok K.,  Dodson R.,  Rioja M.,  Green J.~A.,
  Hutawarakorn~Kramer B.,  2019, \mn@doi [Monthly Notices of the Royal
  Astronomical Society] {10.1093/mnras/sty2767}, 482, 1670

\bibitem[\protect\citeauthoryear{Chen, Gan, Ellingsen, He, Shen  \&
  Titmarsh}{Chen et~al.}{2013}]{Chen2013NEWLYCATALOG}
Chen X.,  Gan C.-G.,  Ellingsen S.~P.,  He J.-H.,  Shen Z.-Q.,   Titmarsh A.,
  2013, \mn@doi [The Astrophysical Journal Supplement Series]
  {10.1088/0067-0049/206/1/9}, 206, 9

\bibitem[\protect\citeauthoryear{Churchwell et~al.,}{Churchwell
  et~al.}{2009}]{churchwell2009spitzer}
Churchwell E.,  et~al., 2009, Publications of the Astronomical Society of the
  Pacific, 121, 213

\bibitem[\protect\citeauthoryear{Cragg, Johns, Godfrey  \& Brown}{Cragg
  et~al.}{1992}]{Cragg1992PumpingMasers}
Cragg D.~M.,  Johns K.~P.,  Godfrey P.~D.,   Brown R.~D.,  1992, \mn@doi
  [Monthly Notices of the Royal Astronomical Society]
  {10.1093/mnras/259.1.203}, 259, 203

\bibitem[\protect\citeauthoryear{Cragg, Sobolev, Ellingsen, Caswell, Godfrey,
  Salii  \& Dodson}{Cragg et~al.}{2001}]{Cragg2001MultitransitionMasers}
Cragg D.~M.,  Sobolev A.~M.,  Ellingsen S.~P.,  Caswell J.~L.,  Godfrey P.~D.,
  Salii S.~V.,   Dodson R.~G.,  2001, \mn@doi [Monthly Notices of the Royal
  Astronomical Society] {10.1046/j.1365-8711.2001.04294.x}, 323, 939

\bibitem[\protect\citeauthoryear{Cragg, Sobolev  \& Godfrey}{Cragg
  et~al.}{2002}]{Cragg2002ModellingRegions}
Cragg D.~M.,  Sobolev A.~M.,   Godfrey P.~D.,  2002, \mn@doi [Monthly Notices
  of the Royal Astronomical Society] {10.1046/j.1365-8711.2002.05226.x}, 331,
  521

\bibitem[\protect\citeauthoryear{Cyganowski et~al.,}{Cyganowski
  et~al.}{2008}]{cyganowski2008catalog}
Cyganowski C.~J.,  et~al., 2008, The Astronomical Journal, 136, 2391

\bibitem[\protect\citeauthoryear{Cyganowski, Brogan, Hunter  \&
  Churchwell}{Cyganowski et~al.}{2009}]{Cyganowski2009ASURVEY}
Cyganowski C.~J.,  Brogan C.~L.,  Hunter T.~R.,   Churchwell E.,  2009, \mn@doi
  [The Astrophysical Journal] {10.1088/0004-637X/702/2/1615}, 702, 1615

\bibitem[\protect\citeauthoryear{Cyganowski, Brogan, Hunter, Churchwell  \&
  Zhang}{Cyganowski et~al.}{2011}]{Cyganowski2011BIPOLAREGOs}
Cyganowski C.~J.,  Brogan C.~L.,  Hunter T.~R.,  Churchwell E.,   Zhang Q.,
  2011, \mn@doi [The Astrophysical Journal] {10.1088/0004-637X/729/2/124}, 729,
  124

\bibitem[\protect\citeauthoryear{Dawson et~al.,}{Dawson
  et~al.}{2014}]{Dawson2014SPLASH:Region}
Dawson J.~R.,  et~al., 2014, \mn@doi [Monthly Notices of the Royal Astronomical
  Society] {10.1093/mnras/stu032}, 439, 1596

\bibitem[\protect\citeauthoryear{De~Buizer \& Vacca}{De~Buizer \&
  Vacca}{2010}]{DeBuizer2010DIRECTREGIONS}
De~Buizer J.~M.,  Vacca W.~D.,  2010, \mn@doi [The Astronomical Journal]
  {10.1088/0004-6256/140/1/196}, 140, 196

\bibitem[\protect\citeauthoryear{Deacon, Chapman, Green  \& Sevenster}{Deacon
  et~al.}{2007}]{Deacon2007HSources}
Deacon R.~M.,  Chapman J.~M.,  Green A.~J.,   Sevenster M.~N.,  2007, \mn@doi
  [The Astrophysical Journal] {10.1086/511383}, 658, 1096

\bibitem[\protect\citeauthoryear{Dunham, Crapsi, Evans~II, Bourke, Huard, Myers
   \& Kauffmann}{Dunham et~al.}{2008}]{Dunham2008IdentifyingCores}
Dunham M.~M.,  Crapsi A.,  Evans~II N.~J.,  Bourke T.~L.,  Huard T.~L.,  Myers
  P.~C.,   Kauffmann J.,  2008, \mn@doi [The Astrophysical Journal Supplement
  Series] {10.1086/591085}, 179, 249

\bibitem[\protect\citeauthoryear{Egan, Price  \& Kraemer}{Egan
  et~al.}{2003}]{Egan2003The2.3}
Egan M.~P.,  Price S.~D.,   Kraemer K.~E.,  2003, American Astronomical Society
  Meeting Abstracts, 203, 57.08

\bibitem[\protect\citeauthoryear{Elia et~al.,}{Elia
  et~al.}{2013}]{Elia2013THE225.5}
Elia D.,  et~al., 2013, \mn@doi [The Astrophysical Journal]
  {10.1088/0004-637X/772/1/45}, 772, 45

\bibitem[\protect\citeauthoryear{Elia et~al.,}{Elia
  et~al.}{2017}]{Elia2017The67_.circ0}
Elia D.,  et~al., 2017, \mn@doi [Monthly Notices of the Royal Astronomical
  Society] {10.1093/mnras/stx1357}, 471, 100

\bibitem[\protect\citeauthoryear{Elitzur, Hollenbach  \& McKee}{Elitzur
  et~al.}{1989}]{Elitzur1989H2ORegions}
Elitzur M.,  Hollenbach D.~J.,   McKee C.~F.,  1989, \mn@doi [The Astrophysical
  Journal] {10.1086/168080}, 346, 983

\bibitem[\protect\citeauthoryear{Ellingsen}{Ellingsen}{2006}]{ellingsen2006methanol}
Ellingsen S.~P.,  2006, The Astrophysical Journal, 638, 241

\bibitem[\protect\citeauthoryear{Ellingsen}{Ellingsen}{2007}]{Ellingsen2007AFeatures}
Ellingsen S.~P.,  2007, \mn@doi [Monthly Notices of the Royal Astronomical
  Society] {10.1111/j.1365-2966.2007.11615.x}, 377, 571

\bibitem[\protect\citeauthoryear{Ellingsen, Voronkov, Cragg, Sobolev, Breen  \&
  Godfrey}{Ellingsen et~al.}{2007}]{Ellingsen2007InvestigatingSurveys}
Ellingsen S.~P.,  Voronkov M.~A.,  Cragg D.~M.,  Sobolev A.~M.,  Breen S.~L.,
  Godfrey P.~D.,  2007, \mn@doi [Astrophysical Masers and their Environments,
  Proceedings of the International Astronomical Union, IAU Symposium, Volume
  242, p. 213-217] {10.1017/S1743921307012999}, 242, 213

\bibitem[\protect\citeauthoryear{Field \& Gray}{Field \&
  Gray}{1988}]{Field1988TheCase}
Field D.,  Gray M.~D.,  1988, \mn@doi [Monthly Notices of the Royal
  Astronomical Society] {10.1093/mnras/234.2.353}, 234, 353

\bibitem[\protect\citeauthoryear{Fish}{Fish}{2007}]{Fish2007Expanded1}
Fish V.~L.,  2007, \mn@doi [The Astrophysical Journal] {10.1086/524136}, 669,
  L81

\bibitem[\protect\citeauthoryear{Forster \& Caswell}{Forster \&
  Caswell}{1989}]{Forster1989TheRelation}
Forster J.~R.,  Caswell J.~L.,  1989, Astronomy and Astrophysics, 213, 339

\bibitem[\protect\citeauthoryear{Fujisawa et~al.,}{Fujisawa
  et~al.}{2014}]{Fujisawa2014ObservationsObservations}
Fujisawa K.,  et~al., 2014, \mn@doi [Publications of the Astronomical Society
  of Japan] {10.1093/pasj/psu015}, 66, 31

\bibitem[\protect\citeauthoryear{Gallaway et~al.,}{Gallaway
  et~al.}{2013}]{gallaway2013mid}
Gallaway M.,  et~al., 2013, Monthly Notices of the Royal Astronomical Society,
  430, 808

\bibitem[\protect\citeauthoryear{Green \& McClure-Griffiths}{Green \&
  McClure-Griffiths}{2011}]{Green2011DistancesSelf-absorption}
Green J.~A.,  McClure-Griffiths N.~M.,  2011, \mn@doi [Monthly Notices of the
  Royal Astronomical Society] {10.1111/j.1365-2966.2011.19418.x}, 417, 2500

\bibitem[\protect\citeauthoryear{Green et~al.,}{Green
  et~al.}{2009}]{green20096}
Green J.~A.,  et~al., 2009, Monthly Notices of the Royal Astronomical Society,
  392, 783

\bibitem[\protect\citeauthoryear{Green et~al.,}{Green
  et~al.}{2010}]{Green2010The20}
Green J.~A.,  et~al., 2010, \mn@doi [Monthly Notices of the Royal Astronomical
  Society] {10.1111/j.1365-2966.2010.17376.x}, 409, 913

\bibitem[\protect\citeauthoryear{Green et~al.,}{Green
  et~al.}{2012}]{Green2012TheRegion}
Green J.~A.,  et~al., 2012, \mn@doi [Monthly Notices of the Royal Astronomical
  Society] {10.1111/j.1365-2966.2011.20229.x}, 420, 3108

\bibitem[\protect\citeauthoryear{Green et~al.,}{Green
  et~al.}{2017}]{Green2017TheMasers}
Green J.~A.,  et~al., 2017, \mn@doi [Monthly Notices of the Royal Astronomical
  Society] {10.1093/mnras/stx887}, 469, 1383

\bibitem[\protect\citeauthoryear{Gutermuth \& Heyer}{Gutermuth \&
  Heyer}{2015}]{Gutermuth2015A/MIPSGAL}
Gutermuth R.~A.,  Heyer M.,  2015, \mn@doi [The Astronomical Journal]
  {10.1088/0004-6256/149/2/64}, 149, 64

\bibitem[\protect\citeauthoryear{Guzm{\'{a}}n, Sanhueza, Contreras, Smith,
  Jackson, Hoq  \& Rathborne}{Guzm{\'{a}}n
  et~al.}{2015}]{Guzman2015FAR-INFRAREDSAMPLE}
Guzm{\'{a}}n A.~E.,  Sanhueza P.,  Contreras Y.,  Smith H.~A.,  Jackson J.~M.,
  Hoq S.,   Rathborne J.~M.,  2015, \mn@doi [The Astrophysical Journal]
  {10.1088/0004-637X/815/2/130}, 815, 130

\bibitem[\protect\citeauthoryear{Juvela et~al.,}{Juvela
  et~al.}{2015}]{Juvela2015GalacticCores}
Juvela M.,  et~al., 2015, \mn@doi [Astronomy {\&} Astrophysics]
  {10.1051/0004-6361/201425269}, 584, A94

\bibitem[\protect\citeauthoryear{Juvela et~al.,}{Juvela
  et~al.}{2018}]{Juvela2018iHerschel/isup/sup}
Juvela M.,  et~al., 2018, \mn@doi [Astronomy {\&} Astrophysics]
  {10.1051/0004-6361/201731921}, 612, A71

\bibitem[\protect\citeauthoryear{Kauffmann \& Pillai}{Kauffmann \&
  Pillai}{2010}]{Kauffmann2010HOWCLUSTERS}
Kauffmann J.,  Pillai T.,  2010, \mn@doi [The Astrophysical Journal]
  {10.1088/2041-8205/723/1/L7}, 723, L7

\bibitem[\protect\citeauthoryear{Kauffmann, Pillai, Shetty, Myers  \&
  Goodman}{Kauffmann et~al.}{2010}]{Kauffmann2010THECLOUDS}
Kauffmann J.,  Pillai T.,  Shetty R.,  Myers P.~C.,   Goodman A.~A.,  2010,
  \mn@doi [The Astrophysical Journal] {10.1088/0004-637X/712/2/1137}, 712, 1137

\bibitem[\protect\citeauthoryear{Krumholz}{Krumholz}{2014}]{Krumholz2014TheFunction}
Krumholz M.~R.,  2014, \mn@doi [Physics Reports]
  {10.1016/j.physrep.2014.02.001}, 539, 49

\bibitem[\protect\citeauthoryear{Krumholz \& McKee}{Krumholz \&
  McKee}{2008}]{Krumholz2008AFormation}
Krumholz M.~R.,  McKee C.~F.,  2008, \mn@doi [Nature] {10.1038/nature06620},
  451, 1082

\bibitem[\protect\citeauthoryear{Kudritzki}{Kudritzki}{2002}]{Kudritzki2002LinedrivenStars}
Kudritzki R.~P.,  2002, \mn@doi [The Astrophysical Journal] {10.1086/342178},
  577, 389

\bibitem[\protect\citeauthoryear{Kurtz}{Kurtz}{2005}]{Kurtz2005HypercompactRegions}
Kurtz S.,  2005, \mn@doi [Proceedings of the International Astronomical Union]
  {10.1017/S1743921305004424}, 1, 111

\bibitem[\protect\citeauthoryear{Lada \& Lada}{Lada \&
  Lada}{2003}]{Lada2003EmbeddedClouds}
Lada C.~J.,  Lada E.~A.,  2003, \mn@doi [Annual Review of Astronomy and
  Astrophysics] {10.1146/annurev.astro.41.011802.094844}, 41, 57

\bibitem[\protect\citeauthoryear{Mainzer et~al.,}{Mainzer
  et~al.}{2011}]{mainzer2011preliminary}
Mainzer A.,  et~al., 2011, The Astrophysical Journal, 731, 53

\bibitem[\protect\citeauthoryear{Minier, Ellingsen, Norris  \& Booth}{Minier
  et~al.}{2003}]{Minier2003TheMasers}
Minier V.,  Ellingsen S.~P.,  Norris R.~P.,   Booth R.~S.,  2003, \mn@doi
  [Astronomy {\&} Astrophysics] {10.1051/0004-6361:20030465}, 403, 1095

\bibitem[\protect\citeauthoryear{Molinari et~al.,}{Molinari
  et~al.}{2010a}]{molinari2010hi}
Molinari S.,  et~al., 2010a, Publications of the Astronomical Society of the
  Pacific, 122, 314

\bibitem[\protect\citeauthoryear{Molinari et~al.,}{Molinari
  et~al.}{2010b}]{molinari2010clouds}
Molinari S.,  et~al., 2010b, Astronomy {\&} Astrophysics, 518, L100

\bibitem[\protect\citeauthoryear{Molinari, Schisano, Faustini, Pestalozzi,
  Di~Giorgio  \& Liu}{Molinari et~al.}{2011}]{molinari2011source}
Molinari S.,  Schisano E.,  Faustini F.,  Pestalozzi M.,  Di~Giorgio A.~M.,
  Liu S.,  2011, Astronomy {\&} Astrophysics, 530, A133

\bibitem[\protect\citeauthoryear{Molinari et~al.,}{Molinari
  et~al.}{2016a}]{molinari2016dr1}
Molinari S.,  et~al., 2016a, Astronomy {\&} Astrophysics, 591, A149

\bibitem[\protect\citeauthoryear{Molinari, Merello, Elia, Cesaroni, Testi  \&
  Robitaille}{Molinari et~al.}{2016b}]{Molinari2016CALIBRATIONFORMATION}
Molinari S.,  Merello M.,  Elia D.,  Cesaroni R.,  Testi L.,   Robitaille T.,
  2016b, \mn@doi [The Astrophysical Journal] {10.3847/2041-8205/826/1/L8}, 826,
  L8

\bibitem[\protect\citeauthoryear{Moscadelli, Xu, Chen, Moscadelli, Xu  \&
  Chen}{Moscadelli et~al.}{2010}]{Moscadelli2010RevisingW3OH}
Moscadelli L.,  Xu Y.,  Chen X.,  Moscadelli L.,  Xu Y.,   Chen X.,  2010,
  \mn@doi [ApJ] {10.1088/0004-637X/716/2/1356}, 716, 1356

\bibitem[\protect\citeauthoryear{Motte et~al.,}{Motte
  et~al.}{2010}]{motte2010initial}
Motte F.,  et~al., 2010, Astronomy {\&} Astrophysics, 518, L77

\bibitem[\protect\citeauthoryear{Nguy{\^{e}}n~Luong et~al.,}{Nguy{\^{e}}n~Luong
  et~al.}{2011}]{NguyenLuong2011TheMini-starburst}
Nguy{\^{e}}n~Luong Q.,  et~al., 2011, \mn@doi [Astronomy {\&} Astrophysics]
  {10.1051/0004-6361/201117831}, 535, A76

\bibitem[\protect\citeauthoryear{Pandian, Goldsmith  \& Deshpande}{Pandian
  et~al.}{2007}]{Pandian2007TheData}
Pandian J.~D.,  Goldsmith P.~F.,   Deshpande A.~A.,  2007, \mn@doi [The
  Astrophysical Journal] {10.1086/510512}, 656, 255

\bibitem[\protect\citeauthoryear{Pandian, Momjian, Xu, Menten  \&
  Goldsmith}{Pandian et~al.}{2011}]{Pandian2011THEMORPHOLOGIES}
Pandian J.~D.,  Momjian E.,  Xu Y.,  Menten K.~M.,   Goldsmith P.~F.,  2011,
  \mn@doi [The Astrophysical Journal] {10.1088/0004-637X/730/1/55}, 730, 55

\bibitem[\protect\citeauthoryear{Persi, Tapia, Roth, Elia  \&
  L{\'{o}}pez-V{\'{a}}zquez}{Persi et~al.}{2016}]{Persi2016The15507-5359}
Persi P.,  Tapia M.,  Roth M.,  Elia D.,   L{\'{o}}pez-V{\'{a}}zquez J.~A.,
  2016, \mn@doi [Monthly Notices of the Royal Astronomical Society]
  {10.1093/mnras/stw741}, 459, 1946

\bibitem[\protect\citeauthoryear{Preibisch, Ossenkopf, Yorke  \&
  Henning}{Preibisch et~al.}{1993}]{preibisch1993influence}
Preibisch T.,  Ossenkopf V.,  Yorke H.~W.,   Henning T.,  1993, Astronomy and
  Astrophysics, 279, 577

\bibitem[\protect\citeauthoryear{Qiao et~al.,}{Qiao
  et~al.}{2016}]{Qiao2016ACCURATEREGION}
Qiao H.-H.,  et~al., 2016, \mn@doi [The Astrophysical Journal Supplement
  Series] {10.3847/1538-4365/227/2/26}, 227, 26

\bibitem[\protect\citeauthoryear{Qiao et~al.,}{Qiao
  et~al.}{2018}]{Qiao2018AccurateRegion}
Qiao H.-H.,  et~al., 2018, \mn@doi [The Astrophysical Journal Supplement
  Series] {10.3847/1538-4365/aae580}, 239, 15

\bibitem[\protect\citeauthoryear{Reach et~al.,}{Reach
  et~al.}{2006}]{Reach2006AGalaxy}
Reach W.~T.,  et~al., 2006, \mn@doi [The Astronomical Journal]
  {10.1086/499306}, 131, 1479

\bibitem[\protect\citeauthoryear{Reid, Dame, Menten  \& Brunthaler}{Reid
  et~al.}{2016}]{Reid2016ASOURCES}
Reid M.~J.,  Dame T.~M.,  Menten K.~M.,   Brunthaler A.,  2016, \mn@doi [The
  Astrophysical Journal] {10.3847/0004-637X/823/2/77}, 823, 77

\bibitem[\protect\citeauthoryear{Sanna, Moscadelli, Surcis, van Langevelde,
  Torstensson  \& Sobolev}{Sanna et~al.}{2017}]{Sanna2017PlanarCepheusAHW2}
Sanna A.,  Moscadelli L.,  Surcis G.,  van Langevelde H.~J.,  Torstensson K.
  J.~E.,   Sobolev A.~M.,  2017, \mn@doi [Astronomy {\&} Astrophysics]
  {10.1051/0004-6361/201730773}, 603, A94

\bibitem[\protect\citeauthoryear{Slysh, Kalenskii, Val'tts  \& Otrupcek}{Slysh
  et~al.}{1994}]{Slysh1994TheGHz}
Slysh V.~I.,  Kalenskii S.~V.,  Val'tts I.~E.,   Otrupcek R.,  1994, \mn@doi
  [Monthly Notices of the Royal Astronomical Society]
  {10.1093/mnras/268.2.464}, 268, 464

\bibitem[\protect\citeauthoryear{Sobolev, Ostrovskii, Kirsanova, Shelemei,
  Voronkov  \& Malyshev}{Sobolev et~al.}{2005}]{Sobolev2005MethanolFormation}
Sobolev A.~M.,  Ostrovskii A.~B.,  Kirsanova M.~S.,  Shelemei O.~V.,  Voronkov
  M.~A.,   Malyshev A.~V.,  2005, \mn@doi [Proceedings of the International
  Astronomical Union] {10.1017/S1743921305004503}, 1, 174

\bibitem[\protect\citeauthoryear{Svoboda et~al.,}{Svoboda
  et~al.}{2016}]{Svoboda2016THECLUMPS}
Svoboda B.~E.,  et~al., 2016, \mn@doi [The Astrophysical Journal]
  {10.3847/0004-637X/822/2/59}, 822, 59

\bibitem[\protect\citeauthoryear{Tan et~al.,}{Tan
  et~al.}{2014}]{Tan2014MassiveFormation}
Tan J.~C.,  et~al., 2014, {Massive Star Formation}.
University of Arizona Press

\bibitem[\protect\citeauthoryear{Titmarsh, Ellingsen, Breen, Caswell  \&
  Voronkov}{Titmarsh et~al.}{2014}]{Titmarsh2014A620}
Titmarsh A.~M.,  Ellingsen S.~P.,  Breen S.~L.,  Caswell J.~L.,   Voronkov
  M.~A.,  2014, \mn@doi [Monthly Notices of the Royal Astronomical Society]
  {10.1093/mnras/stu1346}, 443, 2923

\bibitem[\protect\citeauthoryear{Titmarsh, Ellingsen, Breen, Caswell  \&
  Voronkov}{Titmarsh et~al.}{2016}]{Titmarsh2016A6}
Titmarsh A.~M.,  Ellingsen S.~P.,  Breen S.~L.,  Caswell J.~L.,   Voronkov
  M.~A.,  2016, \mn@doi [Monthly Notices of the Royal Astronomical Society]
  {10.1093/mnras/stw636}, 459, 157

\bibitem[\protect\citeauthoryear{Traficante et~al.,}{Traficante
  et~al.}{2011}]{alessio2011romagal}
Traficante A.,  et~al., 2011, Monthly Notices of the Royal Astronomical
  Society, 416, 2932

\bibitem[\protect\citeauthoryear{Traficante, Fuller, Peretto, Pineda  \&
  Molinari}{Traficante et~al.}{2015}]{traficante2015initial}
Traficante A.,  Fuller G.~A.,  Peretto N.,  Pineda J.~E.,   Molinari S.,  2015,
  Monthly Notices of the Royal Astronomical Society, 451, 3089

\bibitem[\protect\citeauthoryear{Traficante, Fuller, Billot, Duarte-Cabral,
  Merello, Molinari, Peretto  \& Schisano}{Traficante
  et~al.}{2017}]{alessio2017quietclumpsI}
Traficante A.,  Fuller G.~A.,  Billot N.,  Duarte-Cabral A.,  Merello M.,
  Molinari S.,  Peretto N.,   Schisano E.,  2017, \mn@doi [Monthly Notices of
  the Royal Astronomical Society] {10.1093/mnras/stx1375}, 470, 3882

\bibitem[\protect\citeauthoryear{Traficante, Fuller, Smith, Billot,
  Duarte-Cabral, Peretto, Molinari  \& Pineda}{Traficante
  et~al.}{2018}]{alessio2018darkclumpsII}
Traficante A.,  Fuller G.~A.,  Smith R.~J.,  Billot N.,  Duarte-Cabral A.,
  Peretto N.,  Molinari S.,   Pineda J.~E.,  2018, \mn@doi [Monthly Notices of
  the Royal Astronomical Society] {10.1093/mnras/stx2672}, 473, 4975

\bibitem[\protect\citeauthoryear{Urquhart et~al.,}{Urquhart
  et~al.}{2015}]{Urquhart2015TheDust}
Urquhart J.~S.,  et~al., 2015, \mn@doi [Monthly Notices of the Royal
  Astronomical Society] {10.1093/mnras/stu2300}, 446, 3461

\bibitem[\protect\citeauthoryear{Urquhart et~al.,}{Urquhart
  et~al.}{2018}]{urquhart2018atlasgal}
Urquhart J.~S.,  et~al., 2018, \mn@doi [Monthly Notices of the Royal
  Astronomical Society] {10.1093/mnras/stx2258}, 473, 1059

\bibitem[\protect\citeauthoryear{Wall \& Jenkins}{Wall \&
  Jenkins}{2012}]{Wall2012PracticalAstronomers}
Wall J.~V.,  Jenkins C.~R.,  2012, {Practical statistics for astronomers}.
Cambridge University Press

\bibitem[\protect\citeauthoryear{Walsh et~al.,}{Walsh
  et~al.}{2011}]{Walsh2011TheData}
Walsh A.~J.,  et~al., 2011, \mn@doi [Monthly Notices of the Royal Astronomical
  Society] {10.1111/j.1365-2966.2011.19115.x}, 416, 1764

\bibitem[\protect\citeauthoryear{Walsh, Purcell, Longmore, Breen, Green,
  Harvey-Smith, Jordan  \& Macpherson}{Walsh
  et~al.}{2014}]{Walsh2014AccurateHOPS}
Walsh A.~J.,  Purcell C.~R.,  Longmore S.~N.,  Breen S.~L.,  Green J.~A.,
  Harvey-Smith L.,  Jordan C.~H.,   Macpherson C.,  2014, \mn@doi [Monthly
  Notices of the Royal Astronomical Society] {10.1093/mnras/stu989}, 442, 2240

\bibitem[\protect\citeauthoryear{Wright et~al.,}{Wright
  et~al.}{2010}]{wright2010wide}
Wright E.~L.,  et~al., 2010, The Astronomical Journal, 140, 1868

\bibitem[\protect\citeauthoryear{Zinnecker \& Yorke}{Zinnecker \&
  Yorke}{2007}]{Zinnecker2007TowardFormation}
Zinnecker H.,  Yorke H.~W.,  2007, \mn@doi [Annual Review of Astronomy and
  Astrophysics] {10.1146/annurev.astro.44.051905.092549}, 45, 481

\bibitem[\protect\citeauthoryear{van~der Walt}{van~der
  Walt}{2005}]{vanderWalt2005OnMasers}
van~der Walt J.,  2005, \mn@doi [Monthly Notices of the Royal Astronomical
  Society] {10.1111/j.1365-2966.2005.09026.x}, 360, 153

\makeatother
\end{thebibliography}




\appendix

\section{Principal component analysis}
\label{append:pca}

\begin{figure*}
	\includegraphics[width=\textwidth]{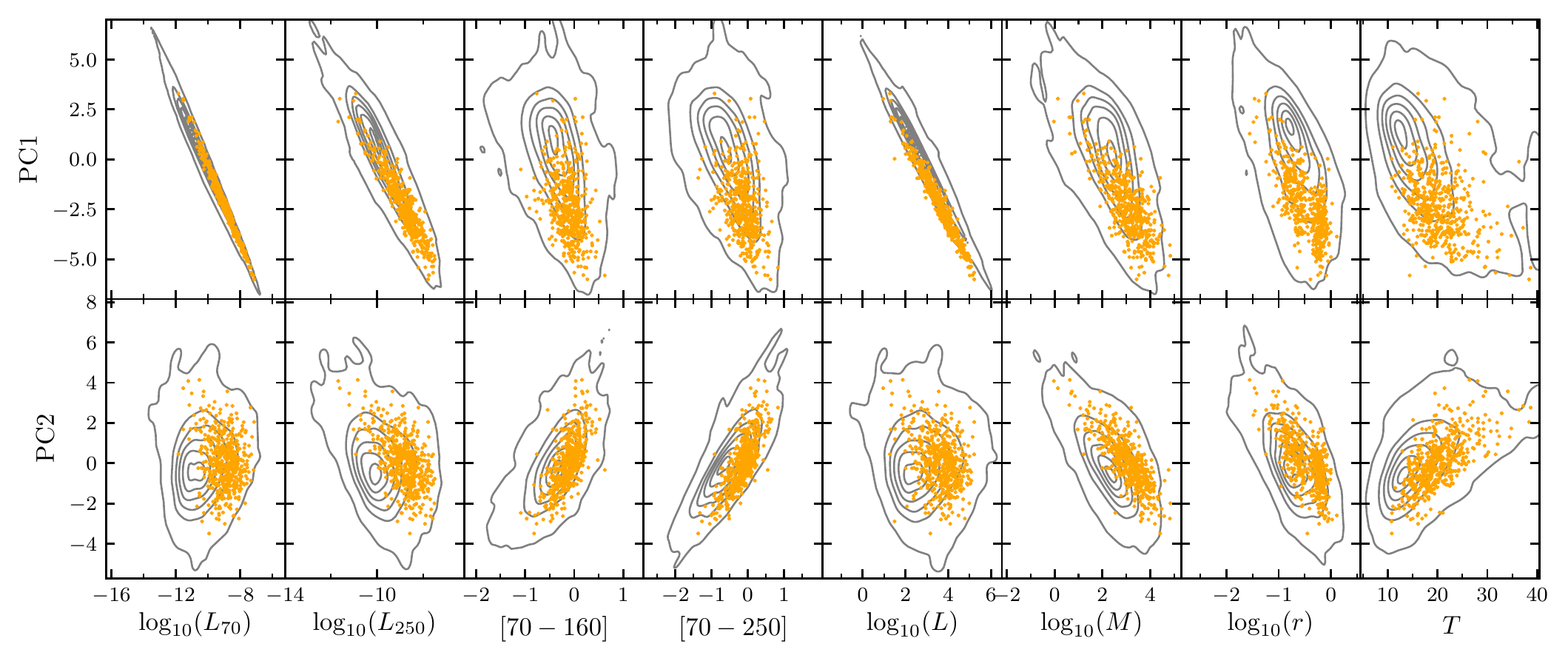}
    \caption{The decomposition of the first two principal components, PC1 (top row) and PC2(bottom row) into the physical source parameters used as input to the PCA. The correlation of each principal component with the source properties shows the direction of the principal component in the 8-dimensional infrared parameter space. The protostellar sources are shown as grey contours at density levels of 1\% to 90\% in intervals of 15\%, with the maser hosts marked in orange. The original infrared property units are [L:R] \Lsun, \Lsun, dimensionless, dimensionless, \Lsun, \Msun, pc and K respectively, and the magnitudes of PC1 and PC2 are arbitrary.}
    \label{fig:pca}
\end{figure*}

PCA is a statistical technique with which to reduce the dimensionality to only a few new dimensions.
The new variables, the principal components, are defined along mutual trends in the N-dimensional parameter space, where N is the original number of variables, each of which can be mapped back to the original physical variables \citep{Wall2012PracticalAstronomers}.
The direction of a principal component is identified as the direction of maximum variation in the N-D parameter space.
For example, if three parameters are all mutually correlated with a linear relationship when plotted in 3D, the first principal component will be in the direction along this correlation as the scatter, and therefore variation, about this line is minimised.
Each subsequent principal component is identified in the remaining direction of maximum correlation after the previous trend has been removed, and therefore each accounts for a decreasing fraction of the total variance in the original parameter space.
Only the first few components accounting for the majority of the original variance require further consideration as the components beyond this are only very weak correlations.

In addition to correlations, PCA is often applied to identify clustering in data sets with which to separate categories of objects.
For our case of maser and non-maser objects, if the two samples cluster into separate regions in the parameter space defined by all of their properties, PCA will identify a principal component in the direction of separation to remove the variance caused by their separation.
The two samples will display a maximum offset along this principal component, and physical trend responsible for the offset can be identified.
For example, although we have already found that the maser objects are more massive, luminous and hotter than generic protostellar hosts on average, there remains a large overlap between the two types of object in these distributions.
It is also not clear whether the same non-maser clumps overlap with the maser sample in all three properties.
However, if each methanol maser host is \textit{simultaneously} more massive, luminous and hotter than a general protostellar object, separation between maser and non-maser objects will be seen along a principal component in this direction.

In our analysis, the returned explained variance ratios (the fraction of total variance accounted for by each component) indicate that only the first two components should be considered further (0.51 and 0.32 respectively).
Figure \ref{fig:pca} shows the correlation of each principal component with the individual parameters.
A single principal component might describe a mutual change in several properties with evolution of an object or with the size of an object.
This decomposition of a principal component into the original parameters allows the direction of the new component to be interpreted as a physical trend.
The first principal component identifies the strong correlation between $L_{70}$, $L_{250}$ and $L_{\mathrm{FIR}}$, but also the trend of increasing luminosity with the mutual increase of the mass and radius of a clump.
The second component is in the direction of increasing $[70-160]$ and $[70-250]$ colour with temperature alongside decreasing mass and radius.
Although principal component analysis has identified that these trends are independent, the variance between the two samples is not sufficiently significant in comparison to the intra-sample variance in each property to identify a direction of separation between the two. 
PCA therefore selects the trends in parameter associated with evolutionary or scaling effects of the clumps.
Non-linear clustering methods may be more appropriate for identifying small offsets, but these are much more challenging to map back to physical properties.

Making use of the independent clump property trends that have been identified through PCA, there are two possible explanations for the trend towards greater infrared colours seen in PC2.
An increase in the amount of 70\microns flux relative to longer wavelength fluxes for a source, indicated by an increase in [70-160]\microns and [70-250]\microns colours, may indicate an increase in the ability of short wavelength photons to escape from the inner regions of the clump, such as through holes in the envelope.
The second possibility is that there is a greater amount of hot dust emitting at short wavelengths within the clump.

An enhancement in the ability of 70\microns photons to escape from the inner regions should therefore correlate with an increase in the ability of photons to escape at 8\microns if this is the underlying cause for this trend.
In the case of removal of envelope material, the percentage of the inner surface area exposed by lines of sight with little or no intervening material determines the ability of photons to escape. 
We therefore expect a similar enhancement in the 70\microns and 8\microns emission through this effect and in such a case, the maser sources with 8\microns counterparts would be expected to occupy the high IR colour regions of PC2, and therefore low-mass and radius regions.
An increase in hot dust would also cause an increase in the amount of emission produced at 8\microns by a protostar, although without the removal of envelope material, the 8\microns emission is still much more strongly attenuated by the envelope material than emission at 70\microns.
The correlation between increasing in 70\microns emission and 8\microns emission would be less pronounced in this case.
When plotted the 8\microns sources show no preferential location and are scattered throughout the entire regions occupied by the maser sources.
This therefore supports the conclusion that an increase in the hot dust content of the clump is more likely responsible for this trend.

\section{Correlation of maser and clump luminosities}
\label{sec:append_b}

\begin{figure*}
    \centering
    \includegraphics[width=0.7\textwidth]
    {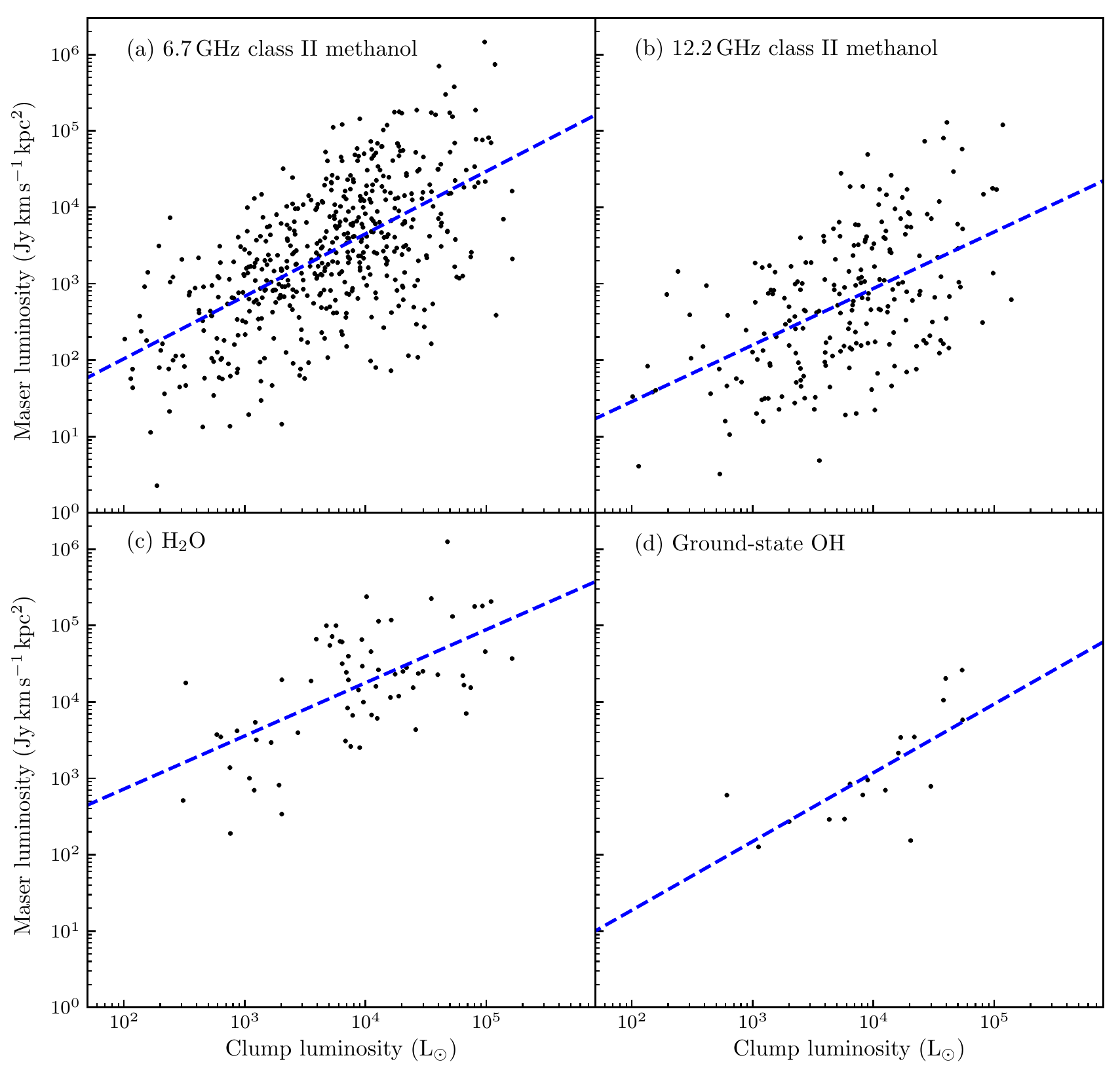}
    \caption{The correlation between integrated maser luminosity and host clump FIR luminosity for the 6.7\,GHz and other secondary masers within the clump. The panels show: (a) 6.7\,GHz luminosity and (b) 12.2\,GHz luminosity from the follow-up MMB study by \citet{breen201212,Breen201212.2-GHz186-330,Breen201412.2-GHz20,Breen201612.2-GHz2060}, (c) the water maser luminosity from the study by \citet{Titmarsh2014A620,Titmarsh2016A6} and (d) the OH maser luminosity from the SPLASH survey \citep{Qiao2016ACCURATEREGION,Qiao2018AccurateRegion}, where the total integrated OH maser luminosity is calculated as the sum of the integrated luminosities of all individual OH masers in a clump. The lines of best fit are shown in blue dashed lines and the properties are given in Table \ref{tab:maserL_clumpL}.}
    \label{fig:maserL_clumpL}
\end{figure*}

\begin{figure}
    \centering
    \includegraphics[width=\columnwidth]
    {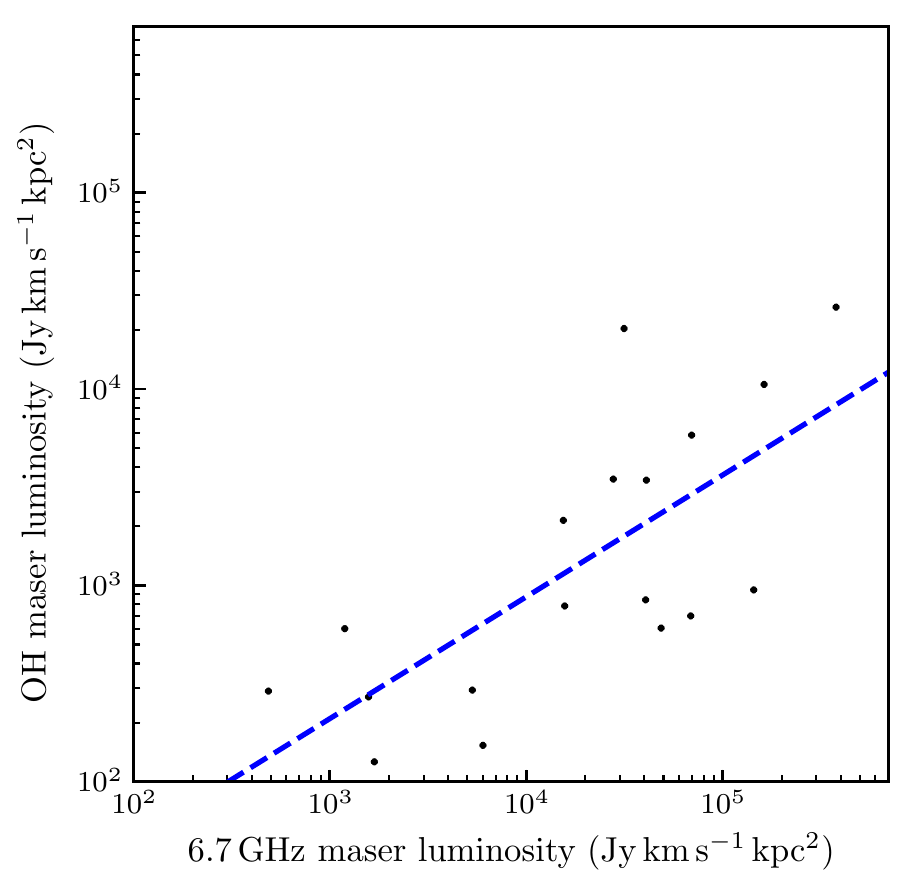}
    \caption{The correlation between integrated 6.7\,GHz maser luminosity and total integrated OH luminosity for the 18 clumps with a secondary OH maser and a reliable distance. The line of best fit (blue dashed) is given by  $\log_{10}(L_{\text{OH}}) = 0.62\log_{10}(L_{\text{6.7}}) + 0.46$.}
    \label{fig:mmbL_splashL}
\end{figure}

The figures to accompany Section~\ref{sec:IRvsMaser} are presented here.
Figure~\ref{fig:maserL_clumpL} displays the correlations between the luminosity of each 6.7\,GHz and secondary masers considered.
Figure~\ref{fig:mmbL_splashL} shows the correlation and line of best fit between 6.7\,GHz and secondary OH maser luminosity.

\bsp	
\label{lastpage}
\end{document}